\documentclass[10pt,letterpaper]{article}

\def\m{$\mu$m}
\def\f{$^\star$}
\def\g{$^\times$}
\def\wl{$\lambda$}
\def\dc{$^\circ$C}
\def\nu{$^\dagger$}
\def\nd{$^\ddagger$}
\def\nt{$^\diamondsuit$}
\def\nq{$^\clubsuit$}
\def\nc{$^\spadesuit$}
\def\ns{$^\sharp$}
 \usepackage{float}
\usepackage{subcaption}
\usepackage[top=0.85in,left=2.75in,footskip=0.75in]{geometry}
\usepackage{changepage}
\usepackage[utf8]{inputenc}
\usepackage{textcomp,marvosym}
\usepackage{fixltx2e}
\usepackage{amsmath,amssymb}
\usepackage[noadjust]{cite}
\usepackage{nameref,hyperref}
\usepackage[right]{lineno}
\usepackage{microtype}
\DisableLigatures[f]{encoding = *, family = * }
\usepackage{rotating}
\usepackage{pdflscape}
\usepackage{pbox} 
\usepackage{changepage}
\usepackage{rotfloat}
\usepackage[usenames, dvipsnames]{color}
\usepackage{soul}
\usepackage{multirow}
\usepackage{subeqnarray}
\usepackage{xfrac}
\usepackage{mathrsfs}

\raggedright
\usepackage[aboveskip=1pt,labelfont=bf,labelsep=period,justification=raggedright,singlelinecheck=off]{caption}

\makeatletter
\renewcommand{\@biblabel}[1]{\quad#1.}
\makeatother

\date{}

\usepackage{lastpage,fancyhdr,graphicx}
\usepackage{epstopdf}
\fancypagestyle{std}{
\pagestyle{myheadings}
\pagestyle{fancy}
\fancyhf{}
\lhead{\includegraphics[width=2.0in]{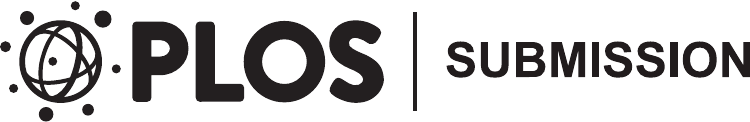}}
\rfoot{\thepage/\pageref{LastPage}}

\fancyheadoffset[L]{2.25in}
\fancyfootoffset[L]{2.25in}
\lfoot{\sf PLOS}
}

\fancypagestyle{tab}{
\pagestyle{myheadings}
\pagestyle{fancy}
\fancyhf{}
\lhead{\includegraphics[width=2.0in]{PLOS-Submission.eps}}
\rfoot{\thepage/\pageref{LastPage}}

\fancyheadoffset[L]{-0.03in}
\fancyfootoffset[L]{-0.03in}
\lfoot{\sf PLOS}
}

\begin{document}
\vspace*{0.35in}

\begin{flushleft}
{\Large
\textbf\newline{The Bank Of Swimming Organisms at the Micron Scale (BOSO-Micro)}
}
\newline
\\
Marcos F. Velho Rodrigues\textsuperscript{1}, Maciej Lisicki\textsuperscript{2},
Eric Lauga\textsuperscript{1,*}
\\
\bigskip
\textbf{1} Department of Applied Mathematics and Theoretical Physics, University of Cambridge, Cambridge CB3 0WA, United Kingdom.
\\
\textbf{2} Faculty of Physics, University of Warsaw, Warsaw, Poland.
\bigskip

*Email: \href{mailto:e.lauga@damtp.cam.ac.uk}{e.lauga@damtp.cam.ac.uk}

\end{flushleft}

\section*{Abstract}
Unicellular microscopic organisms living in aqueous environments outnumber all other creatures on Earth. A large proportion of them are able to self-propel in fluids with  a vast diversity of swimming gaits and motility patterns. In this paper we  present  a {biophysical}
 survey of the available experimental data produced to date on the characteristics of motile behaviour in unicellular microswimmers. We assemble from the available literature empirical data on the motility of four broad categories of organisms: bacteria (and archaea), flagellated eukaryotes, spermatozoa and ciliates. Whenever possible, we gather  the following   {biological, morphological, kinematic and dynamical  parameters}: species, geometry and size of the organisms, swimming speeds, actuation frequencies, actuation amplitudes, number of flagella and   properties of the surrounding fluid.  We then {organise} the
{data} using the established fluid mechanics principles for propulsion at low Reynolds number. {Specifically, we} {use  theoretical {biophysical} models for the locomotion of cells within the same taxonomic groups of organisms as a means of rationalising the raw material we have assembled,} while demonstrating the variability for organisms of different species within the same group.
 {The material  gathered in our work  is an attempt to  summarise the available experimental data in the {field}, providing a convenient and practical reference point for future studies.}

\tableofcontents


\section{\label{sec:intro} Introduction}

Swimming microorganisms were first observed almost 350 years  ago by Antonie van Leeuwenhoek~\cite{VLee}. Since then, extensive knowledge has been obtained on their form, function, genetics and behaviour~\cite{braybook}. We now also understand the  vital  role they play in ecosystems~{\cite{andersen2016characteristic}}
 as well as  in the individual organisms they can inhabit, and whose health they influence {\cite{zhang2015impacts}}. Their ubiquity demonstrates {an} astonishing diversity and adaptability to the most extreme conditions. Furthermore, the involvement of swimming microorganisms in biological processes, irrespective of habitat, is invariably and directly linked to their motility.  The chance of a ciliate escaping a predator~\cite{L173, L152}, the capacity of a spermatozoon to enter and fertilise an egg~\cite{L226}, and the virulent spreading of pathogenic bacteria~\cite{L108} are but a few examples of how cell motility can be decisive for survival.

Swimming in a  fluid on  small, cellular length scales is subject to the physical constraints imposed by the viscosity of the fluid. With typical lengths of the order of microns, and speeds of a few to hundreds of microns per second, the fluid flows set up by microswimmers are characterised by  negligibly small Reynolds numbers. The world in which their locomotion takes place is therefore dominated by viscous {friction} and the effects of inertia are unimportant~\cite{LaLRN,lauga2009,lauga_book}. {As a result,}  the propulsion strategies employed by larger organisms such as fish, mammals, insects and birds are ineffective on cellular length and time scales~\cite{childress81,vogel96,triantafyllou00,alexander02,dudley02,fish06,videler2012fish}.

Swimming microorganisms have thus developed physical {mechanisms} to successfully overcome, and in fact exploit, viscous drag  by  actuating slender tail-like appendages called flagella~\cite{BandW}. Somewhat confusingly, the same name is used to refer to either the  polymeric filaments of prokaryotes or the more complex, muscle-like {flexible} organelles of eukaryotes. In the former case, the filaments are semi-rigid and helical, and they are rotated passively by molecular motors embedded in the cell wall~\cite{lauga16}. For the latter, the flagella undergo three-dimensional active motion resulting from the action of internally-distributed motor proteins~\cite{braybook}.  Despite the variation in structure, distribution and beating pattern of flagella between species, the actuation of flagella in a viscous fluid provides the unifying {biophysical}  picture through which the locomotion of all microorganisms can be understood.

Assessing how fast a certain microorganism can swim is not a simple task. Motility is strongly dependent on  temperature~\cite{L118,L120,L190,149} and on the viscosity of the medium in which the cells swim~\cite{183, L129,L221,L267,149}. Absolute pressure~\cite{L136}, pH~\cite{L135} and even magnetic field~\cite{L232} have {also} been shown to {influence} the  motility of certain species. The motile behaviour of microorganisms may also change depending on whether they are undertaking the role of prey or predator~\cite{L142,L236,L173,L152}.
Furthermore, {cellular propulsion}   also  depends on  biochemical factors~\cite{manson1992bacterial,Belas1849}. Swimming speeds for different species within the same genus (e.g.~\textit{Vibrio}, \textit{Ceratium}, \textit{Peridinium} and \textit{Paramecium})
 and even different strains of the same species (e.g.~\textit{Escherichia coli}~\cite{L123,L251}, \textit{Campylobacter jejuni}~\cite{L129} and \textit{Pseudomonas aeruginosa}~\cite{L79})
   are available in the literature but  little information is given on the variability of the swimming speed within a species or even for an individual organism.  Overall, data on the swimming speed variability of different organisms are rather scarce.
 {Our} recent study for eukaryotic microswimmers has shown that some of the swimming speed distributions have a universal character when appropriately re-scaled~\cite{elife} but the lack of data limits a more detailed analysis. {Since {motility} may be the key factor distinguishing between the {regimes of} cell feeding  (i.e.~advective vs diffusive) or   sensing  (e.g.~spatial  vs~temporal)~\cite{wan2020origins}, extensive {data on} swimming  might aid elucidating the physical mechanisms affecting the cell behaviour.}

{The {biophysical} description} of cellular propulsion  was pioneered in the last century with the work{s} of Gray  {(from the biology side)}~\cite{L208} and Taylor {(mathematics)}~\cite{AotSoMO}, and  it has  now grown into a mature field of research~\cite{jahn72,pedley92,fauci06,lauga2009,gaffney11,stocker,goldstein14,lauga16}.   Despite {many} theoretical advances, the  difficulties of observation and measurement on small scales, as well as the complexity of the fluctuating fluid flows continue to offer outstanding challenges for detailed studies. In addition, the  locomotion of cells links to  the rapidly growing field of artificial active matter, addressing the question of how microbiology, medicine and robotics could work together for the creation and manipulation of artificial swimmers, {some of which are} inspired by flagellated organisms~\cite{RmpBechinger}. These {laboratory} swimmers have a promising potential to perform site-specific drug deliveries, or chemical sensing,  and to assist micro-manipulations in advanced surgery, enhancing the effectiveness of medical treatments~\cite{BJNelson2010,Chen2012,Loghin2017, Kim2014}.

Motivated by the combination of current activity in the research field and its rich scientific history, we carry out in this paper a {biophysical} survey of the available experimental data produced to date ({13 April 2021}) on the characteristics of motile behaviour in unicellular microswimmers. Specifically, we   assemble from the available published  literature empirical data on the motility of four broad categories of organisms, namely bacteria (and archaea), flagellated eukaryotes, spermatozoa and ciliates. Whenever possible, we  gather  a broad set of parameters related to  {biolog}{ical}{, morpholog}{ical}{, kinematic and  dynamical} aspects of the swimming cells: species, geometry and size of the organisms, swimming speeds, actuation frequencies and amplitudes, number of flagella and   properties of the surrounding fluid. We assemble our results in a large downloadable database that we call BOSO-Micro, with BOSO standing for
``Bank Of Swimming Organisms'' and ``Micro'' emphasising their microscopic scale.

We then analyse the data from the database in light of the established fluid mechanics principles for propulsion at low Reynolds number {in order to sort and organise the assembled raw material}.
We reproduce classical scalings for the locomotion of cells within the same taxonomic groups, while demonstrating the variability between different species within the same group. The resulting database, which is made available with this paper {and downloadable from the Center for Open Science  (OSF) repository,}
provides a convenient and practical reference point for future studies~{\cite{Rodrigues_Lisicki_Lauga_2020}}. {{Despite our best efforts}, some species and studies may have been left out of our dataset, {and since research in the field is active and ongoing}, it is important to also allow   our {database} to be easily}
{and continuously}
{extended}.
{{To allow future}  collaborative effort of the community, we have {also} organised an open source version}
{of the database}
{on GitHub~\cite{BOSOgit}, which can be supplemented with new data while retaining a version control.}

The paper is organised as follows. In Section \ref{sec:MM}, we describe in detail the structure of the database, its sources, and the  procedures used for data selection, extraction and processing. We also briefly outline the  theoretical basis of locomotion at low Reynolds number that serves as a guide for the exploration of our data. We then present {and discuss} the collected data,  separating them according to the different taxonomic groups: bacteria and archaea (Sec.~\ref{Sec:bacteria}), flagellated eukaryotes (Sec.~\ref{Sec:flageuk}), spermatozoa (Sec.~\ref{Sec:sperms}) and ciliates (Sec.~\ref{Sec:cil}).
We summarise the findings in Sec.~\ref{sec:disc}, where we {also comment on the potential caveats and  limitations of our work}. We conclude the paper by {displaying} the complete database in Appendix~\ref{sec:data}.

\section{\label{sec:MM} Methods}

 \subsection{Propulsion at low Reynolds number}

Cellular swimming is invariably coupled to the fluid mechanics of the surrounding environment. Biological locomotion in aqueous media happens on a wide range of spatial scales, from sub-micrometre bacteria to whales measuring  tens of metres. In all cases, steady swimming results {from} balancing the propulsive forces generated by the moving swimmer {with}  the frictional (drag) forces {from the surrounding environment}~\cite{LaLRN,lauga2009}. {Propulsion} results from  the {biological} actuation, which always involves motion of the body relative to the fluid. This in turn generates flow, which dissipates energy and thus resists the motion.

{For biological locomotion in} Newtonian fluids, the fluid flow around a swimming organism is governed by the Navier-Stokes equations. However, in the  regime of interest for this work, the effects of viscosity on the motion typically dominate   inertial effects, as {classically} quantified by the dimensionless Reynolds number. Assuming $U$ to be the typical {speed} scale of a swimmer of a characteristic size $B$, moving through a fluid of mass density $\rho$ and dynamic viscosity $\eta$, the ratio of inertial to viscous forces is defined as the {(steady)   Reynolds number},
$\text{Re} = \rho U B /\eta$.   {Because the} propulsion mechanism   often involves the periodic motion of {biological} organelles of   characteristic length $\ell$ and angular frequency $\omega$, another dimensionless number can be constructed, termed the  oscillatory Reynolds number {and defined as} $\text{Re}_{\omega}=\rho \omega \ell^2/\eta$.

In Table \ref{tab:Re} we  estimate both values of $\text{Re}$ and
$\text{Re}_{\omega}$  for a number of representative organisms from the database assuming their environment to be water at $25$\dc. In the majority of cases, {these} estimates suggest that it is appropriate to  neglect  all inertial effects when compared to   viscous forces, as {both} $\mathrm{Re}\ll 1$ {and} $\text{Re}_{\omega}\ll 1$, or at most {just below} one. To {interpret} the dynamics of microswimmers, it is thus appropriate to consider the over-damped  limit, when the fluid dynamics are governed by the {steady} Stokes equations.  For a   detailed overview of the fluid dynamics of locomotion at low Reynolds we refer to classical work in Refs.~\cite{BandW,lauga2009,FlagHyd,55,CiliaBL}.

\begin{table}[t]
    \centering
    \begin{tabular}{p{0.34\textwidth}|p{0.05\textwidth}p{0.08\textwidth}p{0.08\textwidth}p{0.05\textwidth}|p{0.1\textwidth}p{0.1\textwidth}}
         \textbf{Species}  	&	$B$ [$\mu$m]	&	$U$	[$\mu$ms$^{-1}$] &	$\omega$ [rad\,s$^{-1}$]	&	$\ell$ [$\mu$m]	&	Re	&	Re$_{\omega}$	\\\hline\hline
        \textit{E.~coli} (bacteria)	&	2.5	&	24.1	&	823.1	&	8.3	&	$6.75\, 10^{-5}$	&	$6.35\, 10^{-2}$	\\
       \textit{H.~salinarum} (archaea)	&	2.6	&	3.3	&	144.5	&	4.3	&	$9.61\, 10^{-6}$	&	$2.99\, 10^{-3}$	\\
        \textit{G.~lamblia} (flag. eukaryote)	&	11.3	&	26	&	81.7	&	11.6	&	$3.28\, 10^{-4}$	&	$1.22\, 10^{-2}$	\\
       Bull spermatozoon (Metazoa)	&	8.9	&	97	&	129.2	&	54.0	&	$9.64\, 10^{-4}$	&	$4.22\, 10^{-1}$	\\
       \textit{P.~caudatum} (ciliate)	&	242	&	1476.5	&	197.3	&	12	&	$4.00\, 10^{-1}$	&	$3.18\, 10^{-2}$	\\\hline
    \end{tabular}
    \caption{Steady (Re) and oscillatory (Re$_{\omega}$) Reynolds numbers  for five representative organisms from the database. The values of the mass density ($\rho$) and dynamic viscosity ($\eta$) used correspond to water at $25$\dc.
    }
    \label{tab:Re}
\end{table}

\subsection{Data collection and processing}

In this paper we focus on unicellular microorganisms that can swim on their own, either using the actuation of flagella and cilia or by periodic deformations of their cell bodies,  so that they generate net   displacements via interactions with the surrounding fluid. We therefore do not {include} gliding and twitching motility, nor amoeboid displacement. Swarming bacteria were however included, because swarmer cells are also swimmer cells.

In order to identify in the available literature the swimming characteristics of multiple organisms, we selected six seminal  {biophysical} papers in the field of biological fluid dynamics {of} microscale locomotion (ordered by year of publication):
(i)  an early analysis of microscale swimming by Taylor~\cite{AotSoMO};
(ii) the work of Gray and Hancock on the swimming of spermatozoa~\cite{93};
(iii) the lecture on the theory of flagellar hydrodynamics by Lighthill~\cite{FlagHyd};
(iv) the  introduction to life at low Reynolds number by Purcell~\cite{LaLRN};
(v) the classical review paper on locomotion by cilia and flagella by Brennen and Winet~\cite{BandW}; and
(vi) the study on bacterial locomotion in viscous environments by Berg and Turner~\cite{L91}. These papers {are commonly viewed by the community as} groundbreaking {biophysical}  contributions to the field of microswimmer hydrodynamics, which is reflected in the number of citations of these works, summing up to over
 5300\footnote{The respective numbers of citations are:
 614~\cite{BandW};
 240~\cite{L91};
 733~\cite{93};
 541~\cite{FlagHyd};
 2461~\cite{LaLRN};
 736~\cite{AotSoMO}.  Source: Web of Knowledge,  13 April 2021.  }.

In order to construct the database, we first {used}  the Web of Knowledge database {to assemble} two lists of published references: (a) papers that are cited by any of the six source papers, (b) papers that cite any of the six source papers. Each of the resulting  {references} was then examined to determine whether it contained any measurements or reports on the swimming characteristics of any unicellular microswimmer{, or if it led   to other useful references.} {We acknowledge that our selection of six initial papers is clearly biased towards the fluid mechanics and {biophysical} aspects, yet we hope that by a thorough query of the cited and citing papers we managed to {sufficiently} extend the scope of the search  to construct a {comprehensive and relevant} dataset. In order to allow further extension of the database to include new and possibly omitted studies, we refer to the open GitHub version of it~\cite{BOSOgit}.} Note that we reproduce all {the collected} information in the form of tables in  Appendix~\ref{sec:data}, {in which we list all  relevant material  in a} concise form.

\begin{figure}[t]
    \centering
    \includegraphics[width=\textwidth]{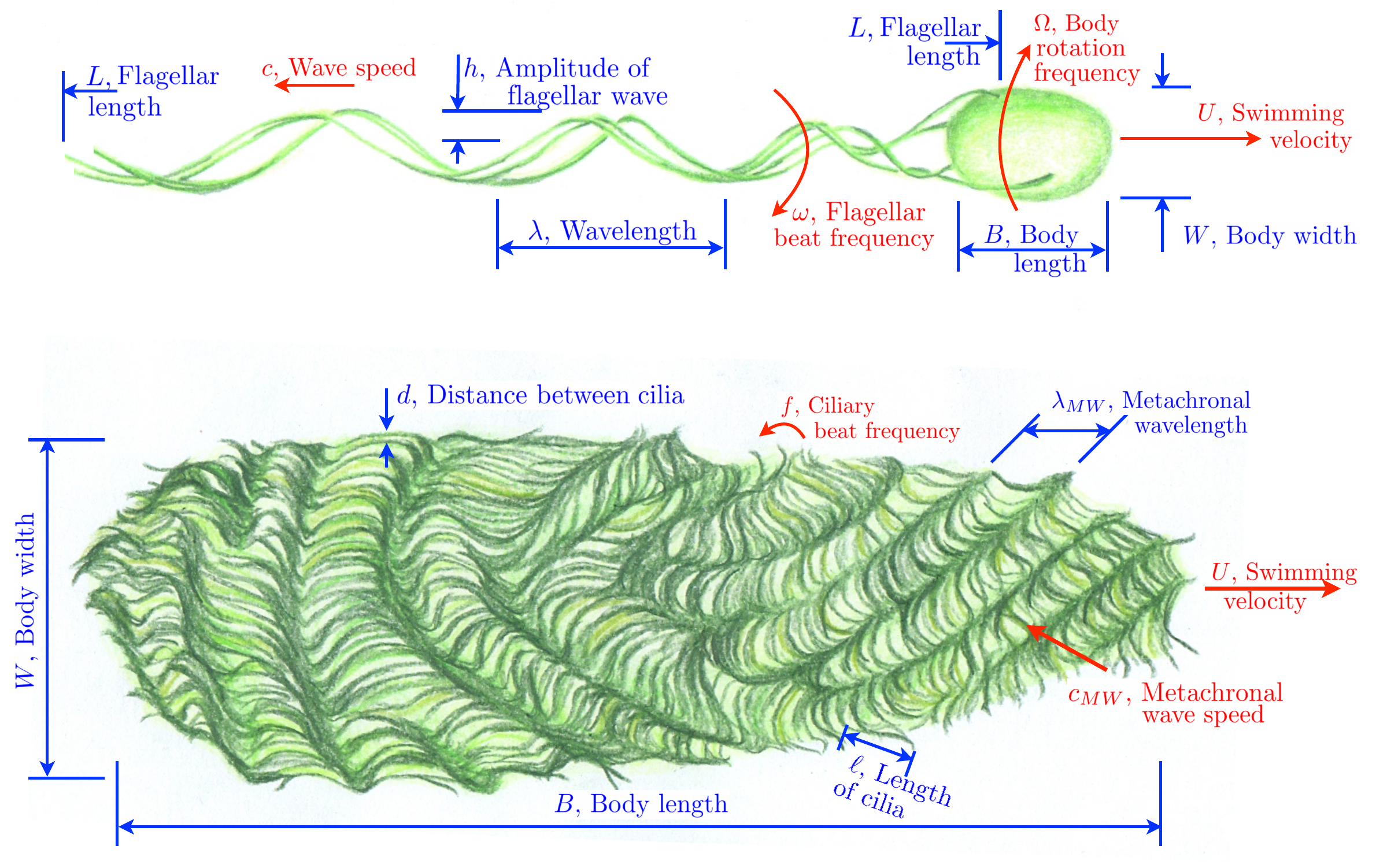}
    \caption{Top: {G}eometrical and kinematic parameters of flagellated swimmers, {illustrated} here for a bacterium; {we use the same symbols for cells employing} planar or helical waves  for simplicity.
    Bottom: Geometrical and kinematic parameters of ciliated swimmers.
    Drawings by Marcos F. Velho Rodrigues.}
    \label{fig:flag}
    \label{fig:cilia}
\end{figure}

In addition to the cell swimming speed, we extracted other geometrical and kinematic characteristics of the organisms when available in experimental studies. These parameters are summarised {on the sketches in} Fig.~\ref{fig:flag}  for cells with a small number of  flagella (top) and  for cells with many appendages (bottom):
 dimensions of cell bodies, swimming speeds, lengths and beat frequencies of cilia and flagella, wavelengths, wave speeds, amplitudes and form of the propagated waves (two or three-dimensional, sinusoidal, helicoidal or complex patterns for flagella, and metachrony for cilia~\cite{162}). Note that  several works  exist that {review solely}  the morphological features of swimming microorganisms~\cite{gage1998,L112,anderson2005sperm}. As the focus of our paper is on the relationship between geometry, kinematics and locomotion, we {chose} not to include in our database any study that does not report any swimming speeds.

\begin{figure}[t]
    \centering
    \includegraphics[width=0.85\textwidth]{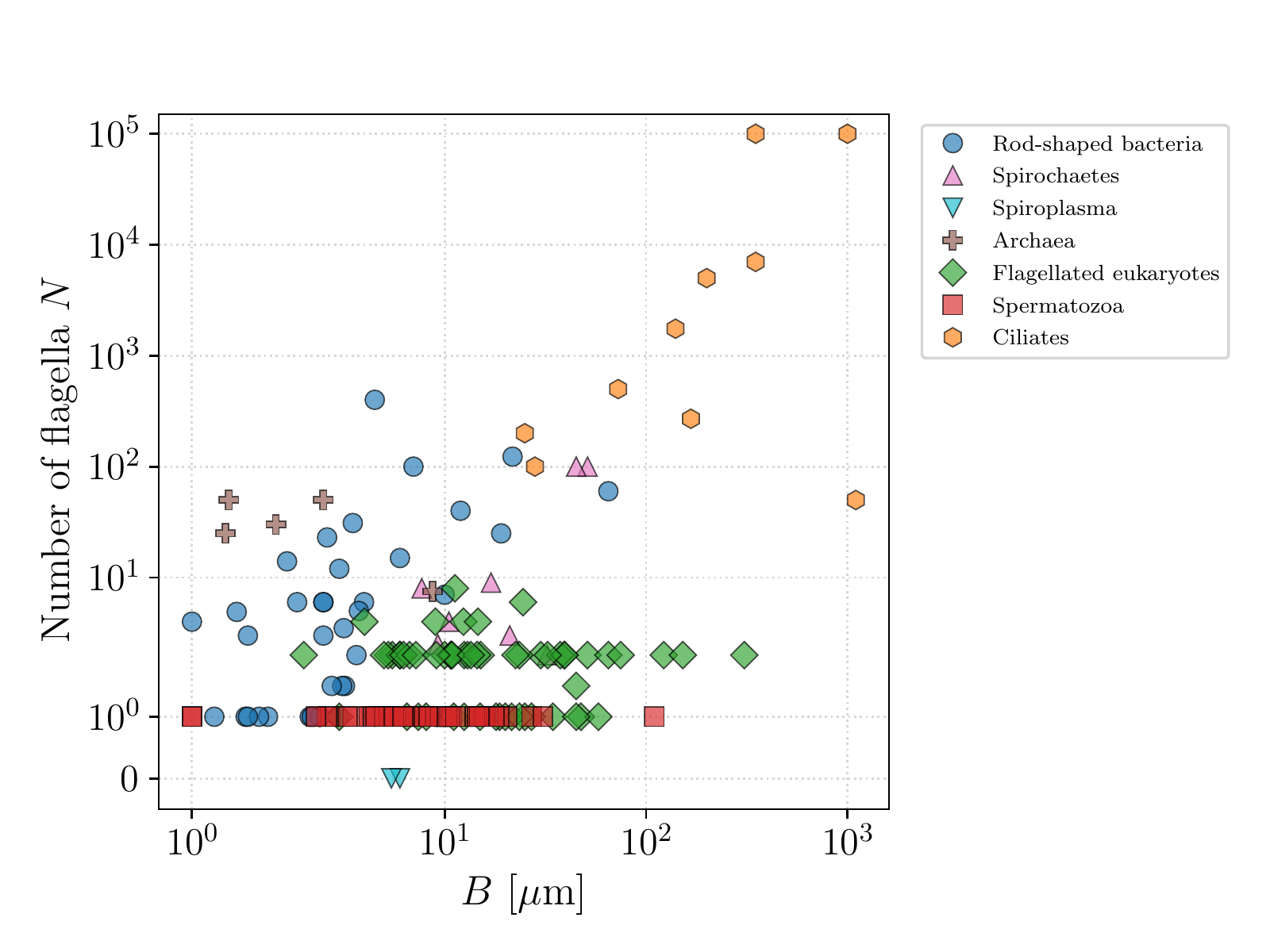}
    \caption{Number of appendages, {i.e.}~cilia or flagella, of each organism (whenever available) plotted against the cell body length. Both characteristics span  orders of magnitude but the data cluster within taxonomic groups.}
    \label{fig:Nflag}
\end{figure}

 In all, the database contains {a} total of {382} species for which we were able to find {at least one measurement on swimming speed along with} other characteristics. Within the tree of life, microswimmers of these species are present in all domains:  Bacteria {and} Archaea (together encompassing prokaryotic organisms), and Eukaryota {({including}   flagellated and ciliated cells and the spermatozoa of multicellular organisms). Members of these {different} groups clearly differ in size, propulsion modes and other physical characteristics. In particular, we plot in Fig.~\ref{fig:Nflag} the number of flagella {(or cilia)} of each organism against the typical cell body length, demonstrating the partial clustering of organisms within their taxonomic groups. On top of variability within taxa, there is a considerable diversity even within groups, and both parameters can span several orders of magnitude. Bearing this   in mind, we analyse each taxonomic group separately in what follows.}

\begin{sidewaysfigure}
 \centering
  \includegraphics[width=\textwidth]{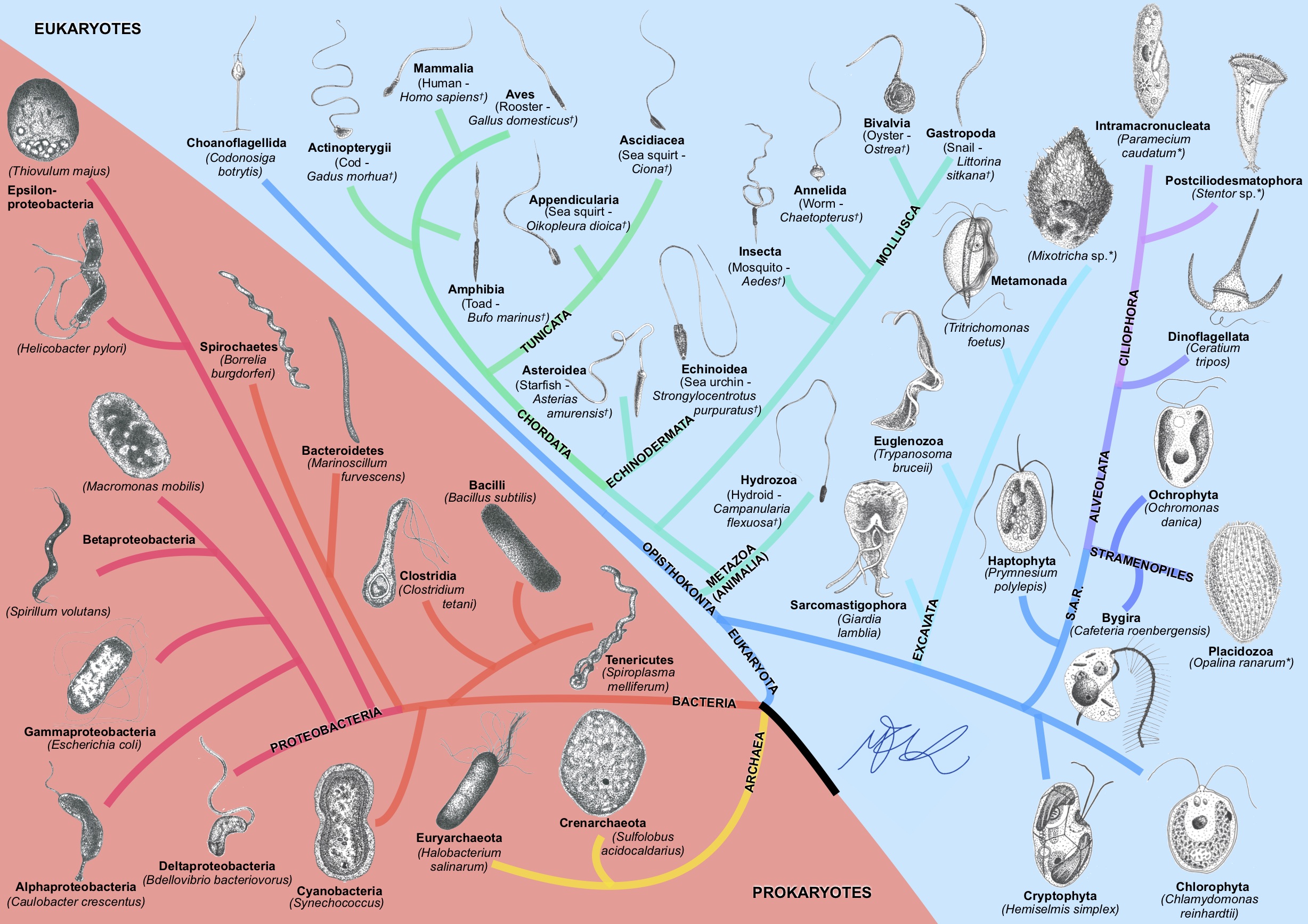}
  \caption{BOSO-Micro Tree of Life. The taxonomy was obtained from the Open Tree of Life~\cite{opentreeoflife}. Ciliates are indicated by an asterisk *, and spermatozoa by a dagger \nu\ beside their species' names. The drawings are not to scale and were inspired by real {microscopy images} or by illustrations. All drawings by Marcos F. Velho Rodrigues.}
  \label{fig:Tree}
 \end{sidewaysfigure}

In order to {help visualise} the {range}  of the present study, we {also} {follow} taxonomy as presented in the  Open Tree of Life~\cite{opentreeoflife} and sketch in Fig.~\ref{fig:Tree}   the various phylogenetic branches {included in our work} together with a drawing of one representative organism within each phylum covered.

\newgeometry{top=0.85in,left=0.5in,right=0.5in, footskip=0.35in, bottom=1in}
\pagestyle{tab}

\newgeometry{top=0.85in,left=2.75in,right=0.5in, footskip=0.35in, bottom=1in}
\pagestyle{std}

\section{\label{Sec:bacteria} Bacteria and Archaea}

We start our journey through swimming microorganisms with prokaryotes, namely the domains Bacteria and Archaea.
Bacteria constitute the bulk of the biomass on Earth, inhabiting the soil, water reservoirs, and the guts of larger organisms. They are simple cells without a nucleus, yet they display a remarkable diversity of shapes~\cite{madigan2017brock}. Motility is a crucial feature for many species of bacteria, in particular for nutrition purposes, and to this end bacteria have developed various propulsion strategies~\cite{jarrell2008}.

 {Two broad  categories of swimming bacteria  exist. In the first one},
  propulsion is enabled by the actuated motion of  flagella located in the fluid outside the cell body~\cite{lauga16}. Unlike their {active} eukaryotic analogues, prokaryotic flagellar filaments are passive organelles \cite{berganderson1973} of typical length of a few microns, attached to a flexible hook that acts as a joint connected to a molecular motor embedded in the  cell wall. The word flagellum (plural flagella) is used to refer to the motor\textendash hook\textendash  filament complex. The bacterial rotary motor,  driven internally by ion fluxes, exerts a torque on the hook, which transmits it to  the filament thereby inducing its rotational motion. Because the flagellar filaments have helical shapes,  their rotation in a viscous fluid induces  a hydrodynamic propulsive force and leads to the motion of the organism~\cite{lauga2009}.

{Flagellated} bacteria can {be equipped} with anything from one flagellum (monotrichous cells) to a few flagella originating from different points on the cell body {\cite{schuhmacher2015how}}. Polar bacteria have their flagella positioned in the vicinity of the pole of the cell. Other arrangements are seen in lophotrichous (a tuft of flagella at the pole) and amphitrichous (flagella  at each pole) cells, while for peritrichous species {(including the well-studied model organism {\it \textit{Escherichia coli}})} the rotary motors are located approximately randomly on the cell body.

Some species of {flagellated bacteria}  can also display a {mode} of motility named swarming, where  cells undergo changes in morphology and rely on intercellular interactions to move {near surfaces}~\cite{L70}. Some species can transition from swimming to swarming behaviours by relying on   polar flagella for swimming, while  {exploiting} several   flagella distributed along the sides of their bodies {for swarming}~\cite{L146}.   {The data for most bacteria in our database is presented in Table~\ref{tab:rod}.}

\begin{figure}[t!]
\centering
\includegraphics[width=0.5\columnwidth]{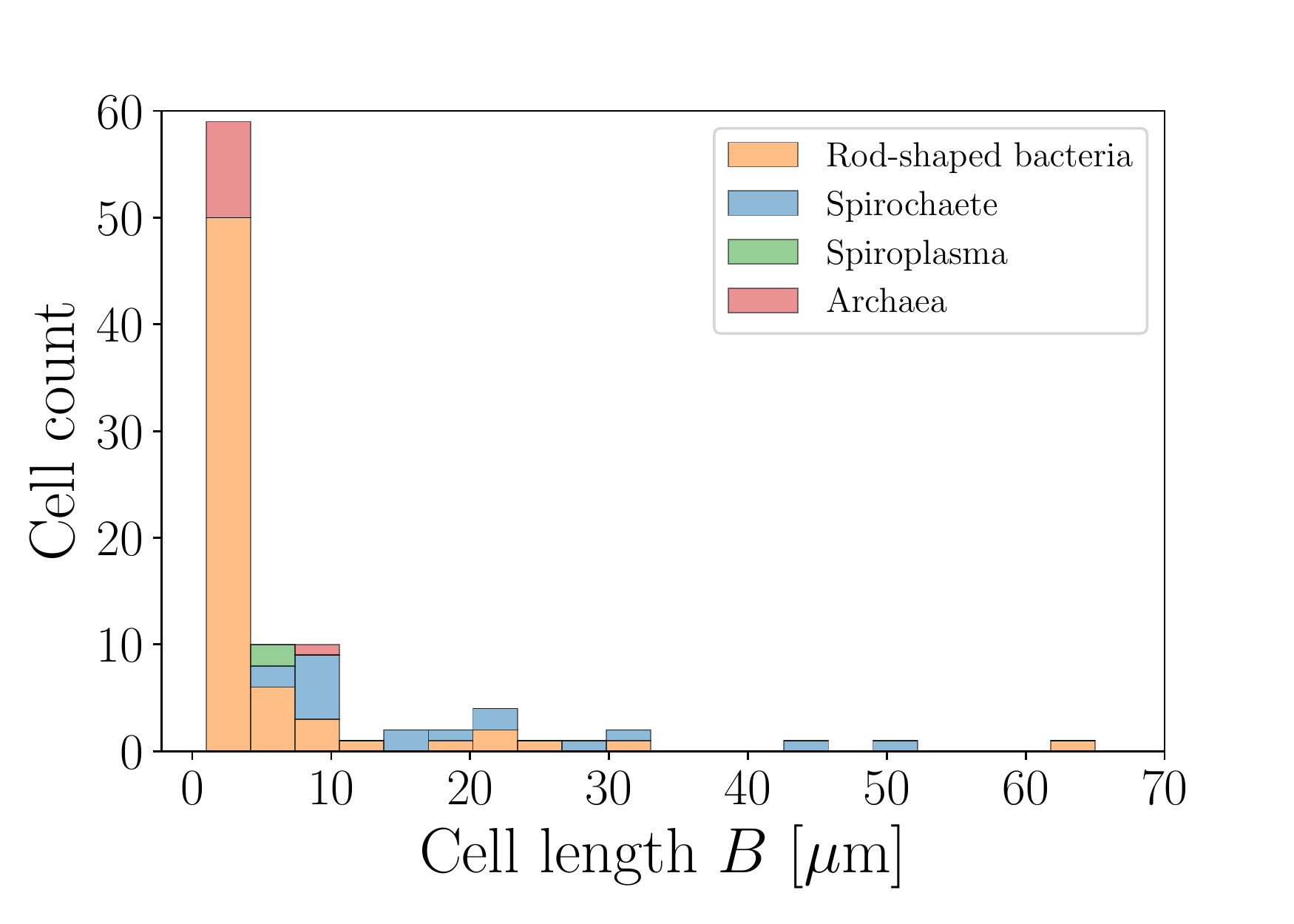}\includegraphics[width=0.5\columnwidth]{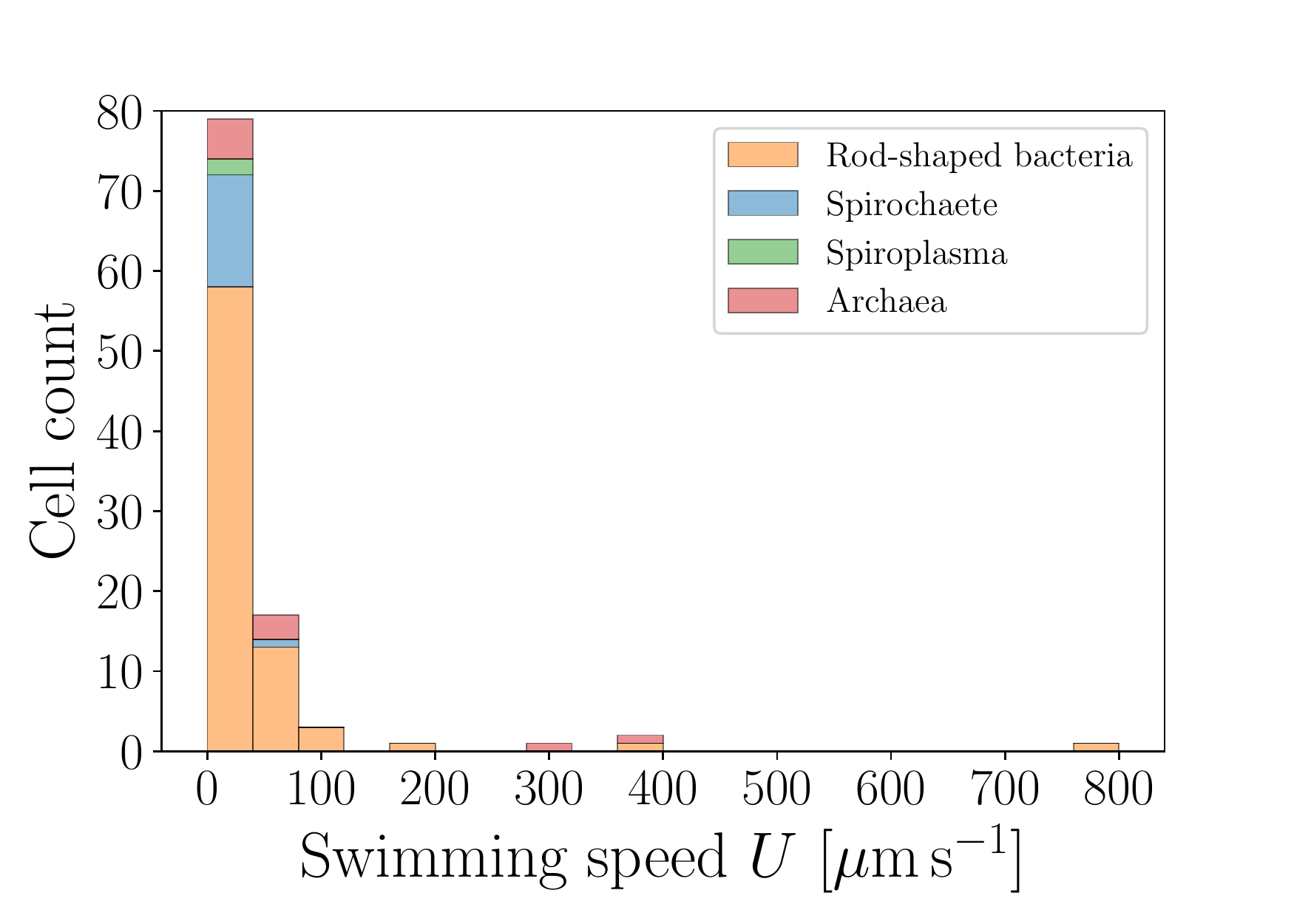}
\caption{Histograms of body lengths, $B$ ($\mu$m, left), and swimming speeds, $U$ ($\mu\text{m}\,\text{s}^{-1}$, right),  for rod-shaped bacteria (excluding spirochaetes and \textit{Spiroplasma})
({$\langle B\rangle=5.79\pm9.33\ \mu\textrm{m}\ (n=66),\ \langle U\rangle=48.33\pm98.47\ \mu\textrm{m}\,\textrm{s}^{-1}\ (n=77)$}), spirochaetes
({$\langle B\rangle=18.59\pm13.02\ \mu\textrm{m}\ (n=17),\ \langle U\rangle=17.94\pm18.84\ \mu\textrm{m}\,\textrm{s}^{-1}\ (n=15)$}), \textit{Spiroplasma}
({$\langle B\rangle=5.72\pm0.28\ \mu\textrm{m}\ (n=2),\ \langle U\rangle=1.69\pm0.81\  \mu\textrm{m}\,\textrm{s}^{-1}\ (n=2)$}) and archaea
({$\langle B\rangle=2.71\pm2.12\ \mu\textrm{m}\ (n=10),\ \langle U\rangle=89.18\pm126.57\ \mu\textrm{m}\,\textrm{s}^{-1}\ (n=10)$}) from our database. Most organisms have sizes below 10~$\mu$m {($\langle B\rangle=7.75\pm10.85\ \mu\textrm{m}\ (n=95)$)} and swimming speeds below 100~$\mu\text{m}\,\text{s}^{-1}$ {($\langle U\rangle=46.98\pm95.42\ \mu\textrm{m}\,\textrm{s}^{-1}\ (n=104)$)}. \label{histBnL_rsb}}
\end{figure}

 {In the second type of bacterial swimming, cells  move via a time-dependent deformation of their   body}. {{Famously, cells in }the phylum Spirochaetes are morphologically distinguished by   having internal axial flagellar filaments running lengthwise between the inner and outer membrane of their periplasmatic space}, producing helical waves in the cell body with no apparent slippage with respect to the surrounding fluid \cite{L280}. Unlike typical rod-shaped bacteria, this particular configuration allows them to swim in extremely viscous gel-like media.

{{Finally, cells} in the genus \textit{Spiroplasma} do not present axial flagellar filaments. Instead, they swim by propagating kink pairs along their {helical} body using the motion of its cytoskeleton. {This creates} a processive change in the helicity of the body, which also allows them to move through extremely viscous fluids~\cite{L195}.} {Our data for spirochaetes and \textit{Spiroplasma} is presented in Table~\ref{tab:spiro}.}

Relatively less studied are the species in the {prokaryotic} domain Archaea.
Archaea  {have morphologies}   similar to bacteria but, {equipped} with a   different molecular organisation, {they are} able to live under conditions that are extreme and hostile to other forms of life. {Other differences exist;} for example, some species of archaea have square-shaped bodies, unlike any bacterium or eukaryote {\cite{L211,oren1999haloarcula}}. {Although the actuation of archaeal flagella  has been characterised in detail~\cite{Kinosita2016}, the motile behaviour of only about 10 species in the whole domain has been studied so far, {with all data summarised in Table~\ref{tab:archaea}}.}

\subsection{Geometry and swimming speeds of the cells}

\begin{figure}[t]
\centering
\includegraphics[width=0.5\columnwidth]{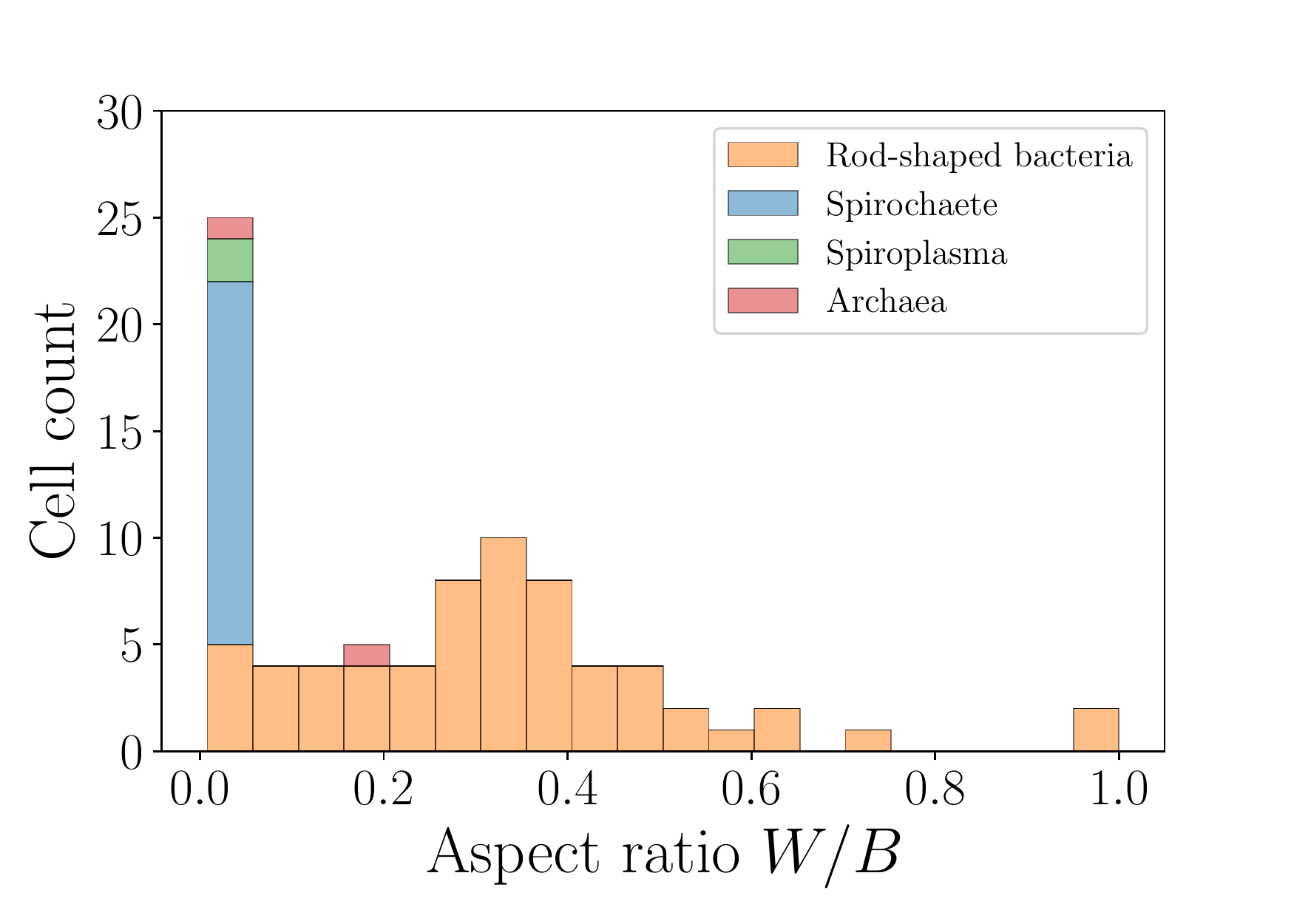}\includegraphics[width=0.5\columnwidth]{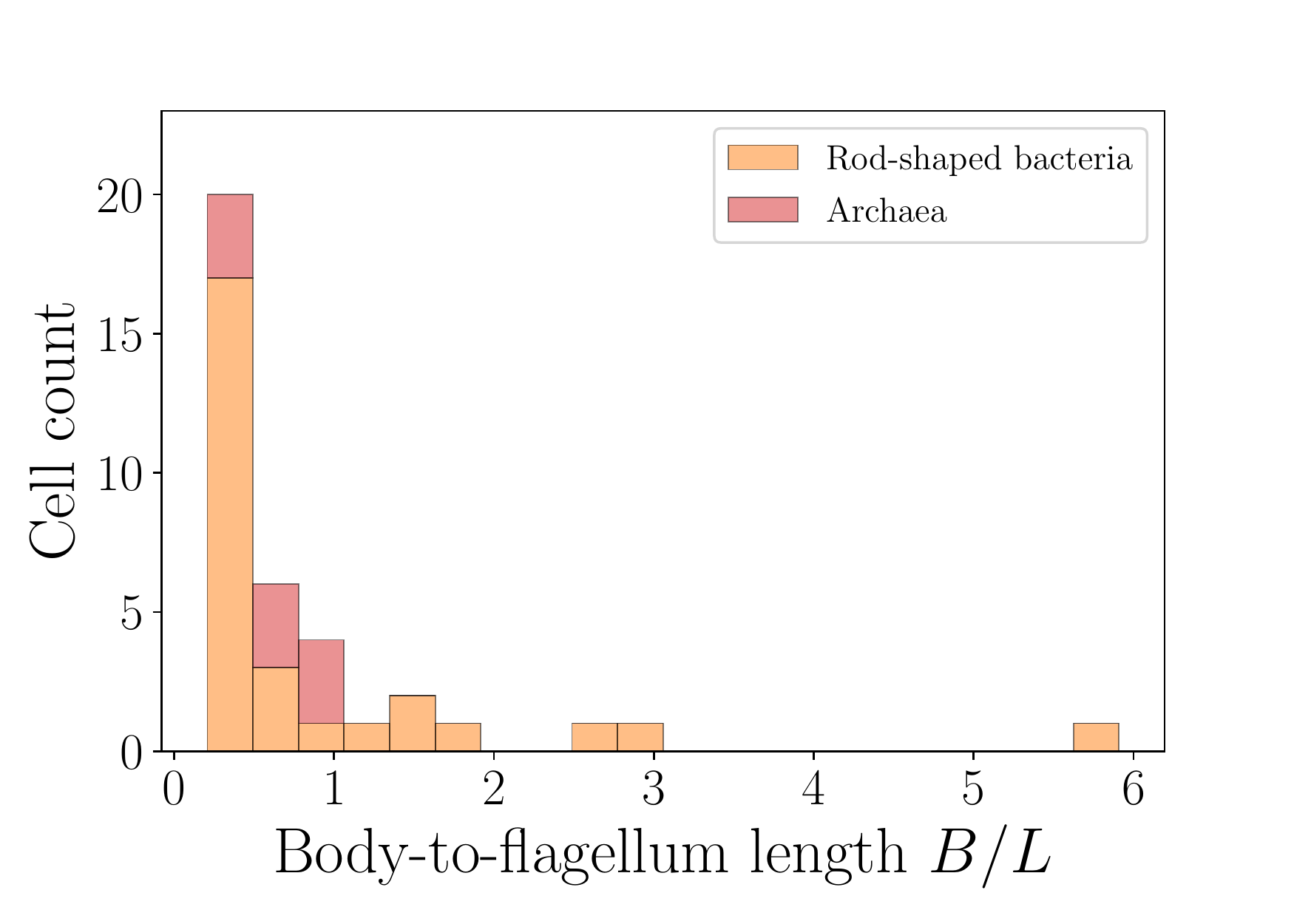}
\caption{Histograms of aspect ratios $W/B$ (left) and body-to-flagellum length $B/L$ (right) for rod-shaped bacteria (excluding spirochaetes and \textit{Spiroplasma})
({$\langle W/B\rangle=0.33\pm0.20\ (n=63), \langle B/L\rangle=0.93\pm1.19 \ (n=28)$}), spirochaetes
({$\langle W/B\rangle=0.02\pm0.01\ (n=17)$}), \textit{Spiroplasma}
({$\langle W/B\rangle=0.03\pm0.00\ (n=2)$}) and archaea
({$\langle W/B\rangle=0.11\pm 0.06\ (n=2), \langle B/L\rangle=0.63\pm0.24 \ (n=9)$}). All bacteria in our study are prolate, with a{n average}
aspect ratio {$\langle W/B\rangle=0.25\pm 0.22\ (n=84)$}, with a notable slenderness of spirochaetes and \textit{Spiroplasma}. If the prokaryotes possess freely rotating flagella, their length often exceeds the body size {$\langle B/L\rangle=0.86\pm1.05\ (n=37)$} (both spirochaetes and {\it Spiroplasma} are not included in the $B/L$ graph).  \label{histBnL_rsb_AR}}
\end{figure}{}

The distribution of sizes and speeds of the prokaryotes from Tables \ref{tab:rod}, \ref{tab:spiro} and \ref{tab:archaea} are {shown} in Fig.~\ref{histBnL_rsb}. The characteristic length of the cell bodies does not exceed 10~$\mu\text{m}$ while the typical swimming speeds are of the order of tens of $\mu\text{m}\,\text{s}^{-1}$.

The shapes of the prokaryotes are {next} quantified in the {distributions shown in} Fig.~\ref{histBnL_rsb_AR} (left). {The cells are close to ellipsoidal}, with an aspect ratio $W/B$ (body width to length) not exceeding 1 and an average of about 0.25. {In contrast}, spirochaetes and \textit{Spiroplasma} are  slender, with the aspect ratio not exceeding 0.05. We also {plot} in Fig.~\ref{histBnL_rsb_AR} (right) the distribution of   body-to-flagellum {lengths} for
{cells with external flagellar filaments (i.e.~excluding  spirochaetes and \textit{Spiroplasma})}. This is typically smaller than unity, indicating that {the helical filaments} are longer than the cell body in most cases.

\begin{figure}[t!]
\centering
\includegraphics[width=0.85\columnwidth]{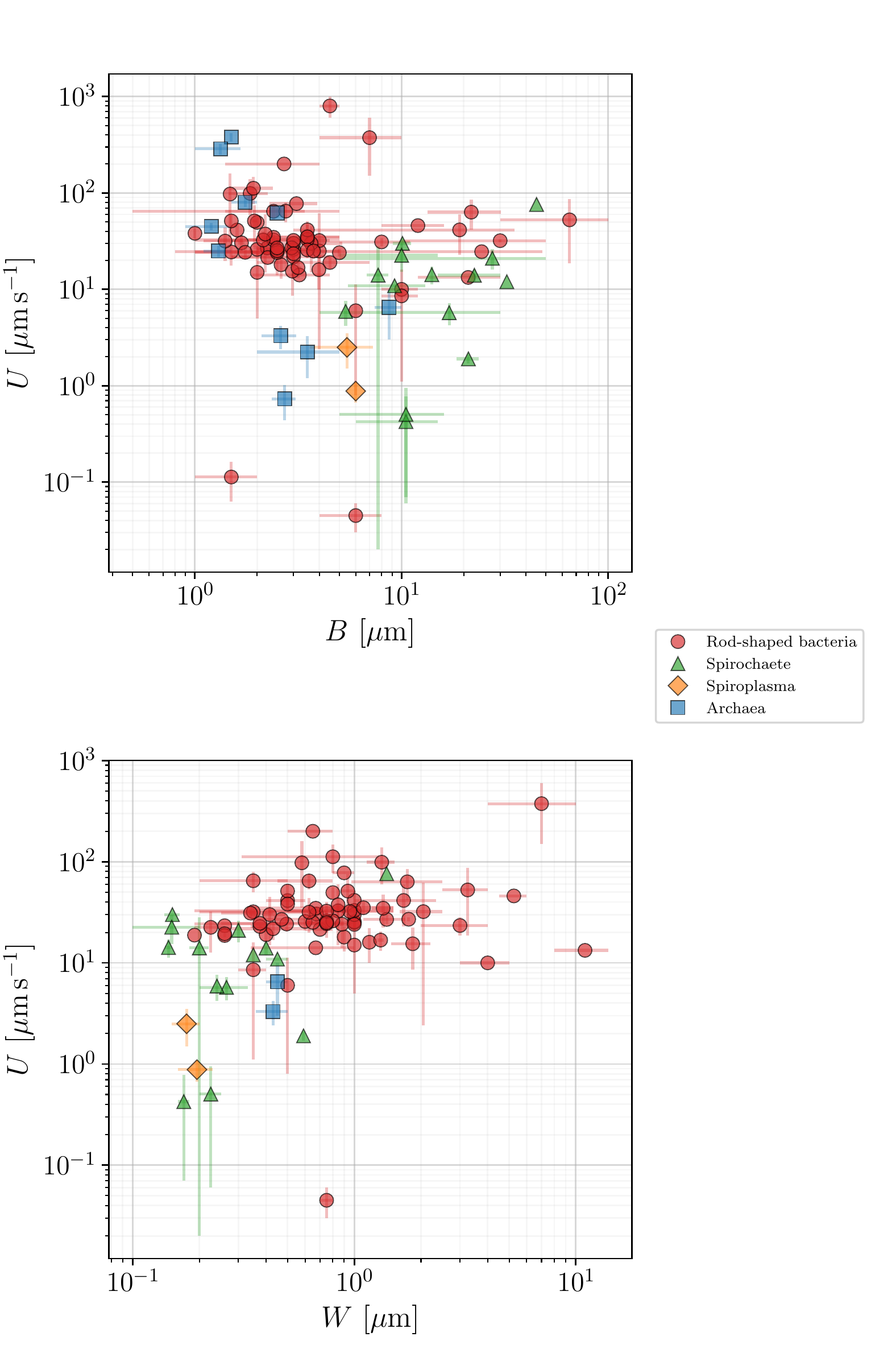}
\caption{Swimming speed, $U$ ($\mu$m\,s$^{-1}$), as function of the cell body length, $B$ ($\mu$m, top), and body width, $W$ ($\mu$m, bottom), for all our registered prokaryotes. Error bars represent standard deviations, whenever available, or the span between the recorded maximum and minimum values.
}\label{prok_UvsB}
\end{figure}

The swimming speed for all  prokaryotes in our database is plotted in Fig.~\ref{prok_UvsB} against the cell body length (top {panel}) and width (bottom {panel}), {with  colours used} to divide our dataset into the specific taxonomic groups. Clearly,  a wide spread of values {exist} for the swimming speeds and  in the next section we {use} a mathematical model for bacterial locomotion in order {to gain further insight into the data}.

 \subsection{Modelling of swimming for flagellated prokaryotes}

We  focus {in what follows} on the case of rod-shaped prokaryotes. The flagellar locomotion of bacteria relies on the motor rotation being transmitted to the passive {flagellar filament} via the flexible hook~\cite{lauga16}.
The rotation of the motor is  generated by ion fluxes {and} in the forward propulsion mode the rotary motor works at approximately constant torque~\cite{chenberg2000}. The value of this torque, however, has been under some debate. Berry and Berg estimated the stall torque in an optical tweezers experiment to be of the order of $4600\ \text{pN}\,\text{nm}$ \cite{Berry1997}, while Reid {\it et al.} attached micrometer beads to flagella to measure  the motor torque to be $1260\pm 190\ \text{pN}\,\text{nm}$ {\cite{Reid2006}}. In magnetic tweezers experiments involving the attachment of paramagnetic beads, the corresponding torque amounted to $874\pm206\ \text{pN}\,\text{nm}$ \cite{VanOene2017}. In contrast, a simplified theoretical model predicts a lower value of $370\pm100\ \text{pN}\,\text{nm}$ \cite{M45} while recent numerical simulations reported  values in the range  $440-820\ \text{pN}\,\text{nm}$ \cite{Das2018}. {Kinosita {\it et al.}~\cite{Kinosita2016} managed to observe in detail the flagellar rotation  of the archaeon \textit{Halobacterium salinarum} and estimated {its} motor torque to be as low as 50\,pN\,nm.} {However, different species of bacteria can have very different motor structures \cite{carroll2020}, which leads to a wide range of possible values for the propulsive torque~\cite{beeby2016}.}

In order to {estimate} the  motor torque of various species in our dataset, we consider a generalised mathematical model  for the swimming of flagellated prokaryotes. For simplicity we focus on the case of a {cell} rotating a single helical filament {\cite{Purcell1997}}. {{However}, the resulting model should remain valid for a prokaryote with many flagella, since during  swimming all   {flagellar filaments} bundle together and rotate as if they formed a single helix}~{\cite{lauga_book}}. {Furthermore, as we show later, the model can be easily adapted to cope with the impact of bundled flagella}.

A prokaryotic flagellar filament of length $L$ is well approximated by a rigid helix of pitch $\lambda$ and radius $h$ (as shown in Fig.~\ref{fig:flag}, top), rotating at an angular velocity $\omega=2\pi f$ relative to the cell body, where $f$ is the frequency of rotation of the flagellum. {In turn}, the cell body  rotates at an angular velocity $\Omega$ relative to the fluid to maintain an overall torque balance {on the cell}. At low Reynolds number, the {helical filament is subject to}  a hydrodynamic thrust $F$ and a viscous torque $T$ which depend linearly with the axial speed $U$ and the rotation rate of the {filament} relatively to the fluid $\Omega+\omega$.
This may be written as
\begin{equation}
  \left(\begin{array}{c}
    F\\T \end{array}\right)_{\text{flagellum}}=-\left(\begin{array}{cc}
      f_{11} & f_{12} \\ f_{12} & f_{22}\end{array}\right) \left(\begin{array}{c}
        U\\ \Omega+\omega\end{array}\right).\label{eq:fric}
\end{equation}

Prokaryotic flagella{r filaments} are very thin, with typical radii of 0.02~$\mu$m and average lengths a thousand times greater (the mean value of all lengths in our database is $\langle L \rangle=8.28~\mu\text{m}$), so that the coefficients of the symmetric matrix $f_{ij}$ can be evaluated using the resistive-force theory of viscous hydrodynamics valid for very slender filaments \cite{93}. Integrating the local hydrodynamic forces on each small segment of the flagellum using the viscous drag coefficients per unit length, $c_\bot$ and $c_\parallel$,  in the directions normal and tangential to each segment respectively (see details below),  yields the classical result that the {resistance} coefficients in Eq.~\eqref{eq:fric} are given by
\begin{subeqnarray}\label{eq:rfcf}
f_{11}&=&(c_\parallel \cos^2\theta+c_\bot\sin^2\theta)\,L,\\
f_{12}&=&(c_\bot-c_\parallel)\sin\theta\cos\theta\,hL,\\
f_{22}&=&(c_\bot \cos^2\theta+c_\parallel\sin^2\theta)\,h^2 L.
\end{subeqnarray}
where $\theta=\arctan(2\pi {h}/{\lambda})$ is the helix pitch angle~\cite{RodenbornE338,lauga_book}.

The cell body,  modelled as a prolate spheroid of length $B$ and diameter $W$ (Fig.~\ref{fig:flag}, top),  is  subject to a hydrodynamic force $F$ proportional to the swimming speed $U$ and a hydrodynamic torque $T$ proportional to its rotation rate $\Omega$. {Assuming the cell to rotate about its principal axis} {leads to}
\begin{equation}\label{eq:fric2}
  \left(\begin{array}{c}
    F\\T \end{array}\right)_{\text{body}}=-\left(\begin{array}{cc}
      b_{11} & 0 \\ 0 & b_{22}\end{array}\right) \left(\begin{array}{c}
        U\\ \Omega\end{array}\right),
\end{equation}
where the off-diagonal coefficients   vanish due to the symmetry of the body.

During steady, straight swimming, the sum of forces and torques on the swimming bacterium must vanish, and thus {combining Eqs.~\eqref{eq:fric} and \eqref{eq:fric2}} we obtain a linear system of equations for the swimming speed and angular rotation as a function of the rotation rate of the filament as
\begin{equation}
    \left(\begin{array}{cc}
      b_{11}+f_{11} & f_{12} \\ f_{12} & b_{22}+f_{22}\end{array}\right) \left(\begin{array}{c}
        U\\ \Omega\end{array}\right)=-\left(\begin{array}{c}
          f_{12}\\ f_{22}\end{array}\right)\omega.
\end{equation}
Solving for $U$ and $\Omega$ as functions of $\omega$ leads to the relations
\begin{subeqnarray}\label{eq:UnOm}
  U&=&\frac{f_{12}b_{22}}{f_{12}^2-(b_{11}+f_{11})(b_{22}+f_{22})}\omega,\\
  \Omega&=&\frac{f_{22}(f_{11}+b_{11})-f_{12}^2}{f_{12}^2-(b_{11}+f_{11})(b_{22}+f_{22})}\omega.
\end{subeqnarray}

The torque $T_m$ exerted by the flagellar motor is, by definition, given by $T_m=f_{12}U+f_{22}(\Omega+\omega)$, which after substitution into
Eq.~\eqref{eq:UnOm} yields
\begin{equation}
  T_m=\frac{b_{22}(f_{12}^2-f_{22}(b_{11}+f_{11}))}{f_{12}^2-(b_{11}+f_{11})(b_{22}+f_{22})}\omega,
\end{equation}
and therefore  the ratio between the swimming speed and the torque exerted by the motor is only a function of the various resistance coefficients, as
\begin{equation}\label{eq:UovTm1}
  \frac{U}{T_m}=\frac{f_{12}}{f_{12}^2-f_{22}(b_{11}+f_{11})}.
\end{equation}

The ratio between $f_{12}^2$ and $f_{11}f_{22}$ can be computed using the expressions given by Eq.~\eqref{eq:rfcf} and we obtain
\begin{equation}\label{eq:ratio}
  \frac{f_{12}^2}{f_{11}f_{22}}=\frac{(c_\bot-c_\parallel)^2\sin^2\theta \cos^2\theta}{(c_\parallel\cos^2\theta+c_\bot\sin^2\theta)(c_\parallel\sin^2\theta+c_\bot\cos^2\theta)}.
\end{equation}
The right hand side of Eq.~\eqref{eq:ratio} is always positive (since $c_\bot,c_\parallel>0$). Its derivative with respect to $\theta$ is given by
\begin{equation}
  \frac{2c_\bot c_\parallel (c_\bot - c_\parallel)^2\sin\theta\cos\theta(\cos^2\theta-\sin^2\theta)}{(c_\bot \sin^2\theta +c_\parallel \cos^2\theta)^2(c_\bot \cos^2\theta +c_\parallel \sin^2\theta)^2},
\end{equation}
which has $\theta = \{k\pi/4, k\in \mathbb{Z}\}$ as roots, for all values of $c_\parallel$ and $c_\bot$. Since $\theta=\{0,{\pi}/{2}\}$ are zeros of Eq.~\eqref{eq:ratio} themselves, $\theta={\pi}/{4}$ gives the maximum possible value for the ratio as
\begin{equation}
\frac{f_{12}^2}{f_{11}f_{22}}\leq \frac{(c_\bot-c_\parallel)^2}{(c_\bot+c_\parallel)^2}=\left(\frac{\displaystyle 1-{c_\parallel}/{c_\bot}}
{\displaystyle 1+ {c_\parallel}/{c_\bot}} \right) ^2.
\end{equation}
It is usually a good approximation to take $c_\parallel/c_\bot\approx {1}/{2}$, so that the ratio $f_{12}^2/(f_{11}f_{22})$ is bound from above by $ {1}/{9}$, and one may thus {approximately} neglect the contribution of $f_{12}^2$ in the denominator of Eq.~\eqref{eq:UovTm1}, yielding the simplified result
\begin{equation}\label{eq:UovTm2}
  \left|\frac{U}{T_m}\right|\approx\frac{f_{12}}{f_{22}(b_{11}+f_{11})}.
\end{equation}

The drag coefficient $b_{11}$  for a prolate spheroid of length $B$ and diameter $W$ depends on a 
 {geometric factor $C_{FB}$}  that involves the eccentricity $e$ of the  spheroid, given by $e=\sqrt{1-({W}/{B})^2}$ ($0 \leq e < 1$),  as
 \cite{chwang_wu_1975}
\begin{equation}
	b_{11}=3\pi\eta B C_{FB}(W/B),\quad   C_{FB} =\frac{8}{3}e^3 \left[-2e+(1+e^2)\log\frac{1+e}{1-e}\right]^{-1}.
	\label{eq:Cfb}
\end{equation} 
The asymptotic limit of very slender spheroids, evaluated in Ref.  \cite{chwang_wu_1975},  also gives  the friction coefficients for the motion of a rod of length $L$ and maximal thickness $2b$ as
\begin{equation}
    c_\bot=\frac{4\pi\eta}{\log({L}/{b})+ {1}/{2}},\ \ \  c_\parallel=\frac{2\pi\eta}{\log({L}/{b})- {1}/{2}},\ \ \ (b/L\ll 1),
\label{eq:cbot}
\end{equation}
which, for large aspect ratios, yield the approximation $c_\parallel/c_\bot\approx {1}/{2}$, as above. Assuming for simplicity the pitch angle to be $\theta\approx {\pi}/{4}$, and using the friction coefficients as in
Eq.~\eqref{eq:cbot}, Eq.~\eqref{eq:UovTm2} takes the final explicit form

  \begin{equation}
\label{eq:rsbfit}
    U=\frac{T_m}{\eta \xi^2},
    \quad     \xi=\sqrt{9\pi h\left(BC_{FB}\left({W}/{B}\right)+ \left[L/\left(\log({L}/{b})+ {1}/{2}\right)\right]\right)},
  \end{equation}
{where} the characteristic length $\xi$  depends solely on the morphology of the swimmer and results   from the interplay of body and flagellum size.
{The   result in Eq.~\eqref{eq:rsbfit} relates therefore the swimming speed $U$ to the flagellar motor torque $T_m$ via the viscosity of the fluid ($\eta$) and a morphological factor ($\xi$).} {Note that by adjusting the helix thickness $2b$, the model can {address}  the impact of having several filaments inside the flagellar bundle~{\cite{kim04_PIV}}. Since the effect of $b$ in Eq.~\eqref{eq:rsbfit} is logarithmic, its {impact on our results is minimal}.}

  \subsection{{Insights from data}}

\begin{figure}[t]
\centering
\includegraphics[width=0.85\columnwidth]{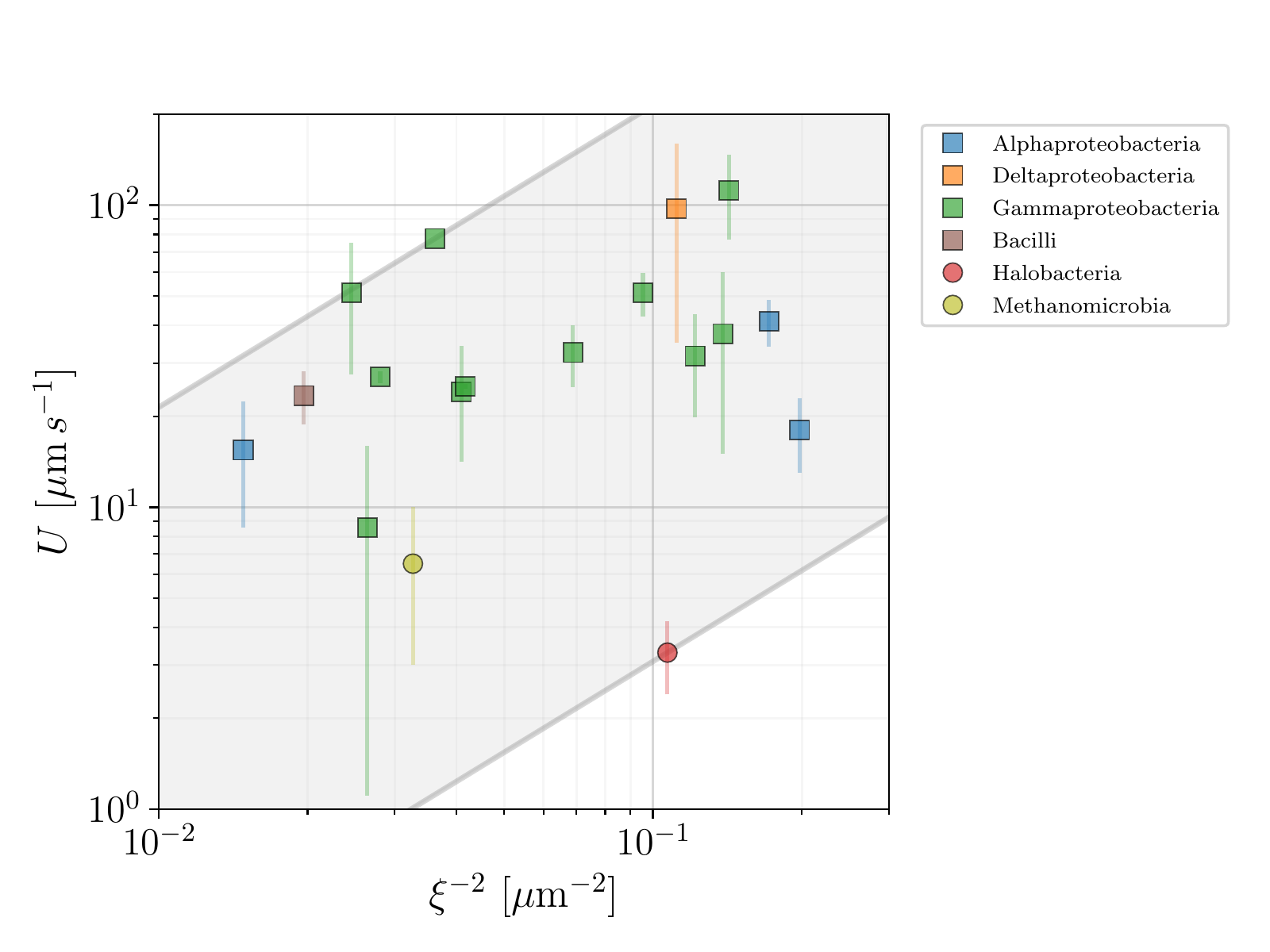}
\caption{{Propulsion speed of rod-shaped prokaryotes vs morphological factor $1/\xi^2$. Bacteria are plotted in squares and archaea in circles with colours used to distinguish between the different taxonomic classes. The plot, along with Eq.~\eqref{eq:rsbfit}, allows to {estimate} the range of bacterial motor torques $27.48 - 1907$ pN\,nm, represented by the shaded area.}
\label{fig:rsp_fit}}
\end{figure}

{We can now use the model introduced above in order to help organise  our database and    provide a simple estimate of the range of motor torques in the available data.} In Fig.~\ref{fig:rsp_fit} we plot the  swimming speed, $U$, for rod-shaped bacteria and archaea as a function of the {morphological factor} $1/\xi^{2}$
for all the species for which our database gives access to the parameters {involved in the definition of $\xi$} in Eq.~\eqref{eq:rsbfit} (we assumed the thickness of the flagella to be $2b=0.02\ \mu$m in all cases).
{The ratio between $U$ and $1/\xi^2$ {should yield} an estimate of the {effective} flagellar motor torque, $T_m$. {An important limitation  is that the value of the viscosity is, alas,  rarely given directly in the studies gathered in our database. We thus assume  the viscosity $\eta$ in Eq.~\eqref{eq:rsbfit} to be that of water at 25\dc~and
in Fig.~\ref{fig:rsp_fit} we display the  range of motor torques so} obtained using   parallel lines enclosing the shaded area. The   {lower and upper bounds of the motor torque $T_m$ are obtained to be} $27.48$~pN\,nm (for \textit{Halobacterium salinarum}) and $1907$~pN\,nm (\textit{Pseudomonas fluorescens}).} {This large range highlights} {the intrinsic variability within this group, corresponding to the observed scatter of the data.
}

\section{\label{Sec:flageuk}Flagellated eukaryotes (excluding spermatozoa and ciliates)}

Eukaryotic cells are not just morphologically distinct from prokaryotes, they also have different important biological features, {including} the  presence of a cellular nucleus. {Their propulsion is enabled  by an} {internal mechanism that is fundamentally different from}, and    more complex than,  {that of prokaryotes}.  The central structure of eukaryotic flagella and cilia is termed the axoneme and is usually composed of nine  microtubule doublet filaments surrounding a tenth central pair of microtubules. Cross-linking dynein motors allow the relative sliding of the microtubules, which results in the propagation of bending deformations along the flexible  flagellum~\cite{M101} {that can take the form of}  {travelling waves, either planar or helical, as well as of complex  two- (2D) and three-dimensional (3D)} {kinematics}.

The {eukaryotic flagellar} waves usually propagate from the flagellum base to its distal end, but some species have waves travelling {in the other direction}. Similarly, while most species swim with flagella trailing, {some, such as} {the alga \textit{Ochromonas danica},   self-propel   with}
their flagella leading the cell. {We refer to} {Jahn \& Votta  for an extensive overview of the observed beating patterns \cite{JahnVotta}.}  One of the most fundamental  beating patterns displayed by eukaryotic cells  is a simple planar sine wave, and we will use it as the basis for the modelling introduced below. {Note that flagella of some eukaryotic species display perpendicularly attached rigid structures, termed mastigonemes, which give a hairy microstructure to the flagellum and allow the cells to generate propulsion in the same direction as the propagating wave~\cite{BandW,116}.} Some algae such as \textit{Chlamydomonas} do not even rely on waves to swim, but do so by swinging a pair of short flagella in a breaststroke way.

Eukaryotic cells are generally one or two orders of magnitude larger {in size} than prokaryotes and are therefore more easily observed experimentally. {A number of} past review papers gathered  swimming speeds and body lengths for tens of organisms~\cite{163,L160,47,L144}. Our database builds on, and extends, these datasets by introducing a number of new important cellular parameters {and new organisms}. Note that parts of our data for eukaryotic cells, particularly the average sizes and swimming speeds have been published elsewhere~\cite{elife}.

Among swimming unicellular eukaryotes, three families with different morphology can be distinguished:  flagellates, spermatozoa, and ciliates. Flagellates -- {the focus o{f} this section} -- typically possess a few long flagella  attached to their bodies, which they actuate in order to achieve propulsion ({for organisms in this section, the typical number of flagella rarely exceeds 10}). Spermatozoa are also remarkable flagellated swimmers but they lack the ability to reproduce, thus {are not} considered living organisms.
Lastly, ciliates are  much larger in size and are covered by dense arrays of cilia, which are short flagella {that move} collectively to create flow along the cell surfaces. The qualitative difference in their swimming speeds, as well as their geometric characteristics such as their  size and their number of flagella, warrants separate statistical analysis for each group~\cite{elife}; {spermatozoa are therefore addressed in Sec.~\ref{Sec:sperms}  and ciliates in    Sec.~\ref{Sec:cil}.}

 \subsection{Geometry and swimming speeds of the cells}

\begin{figure}[t!]
\centering
\includegraphics[width=0.5\columnwidth]{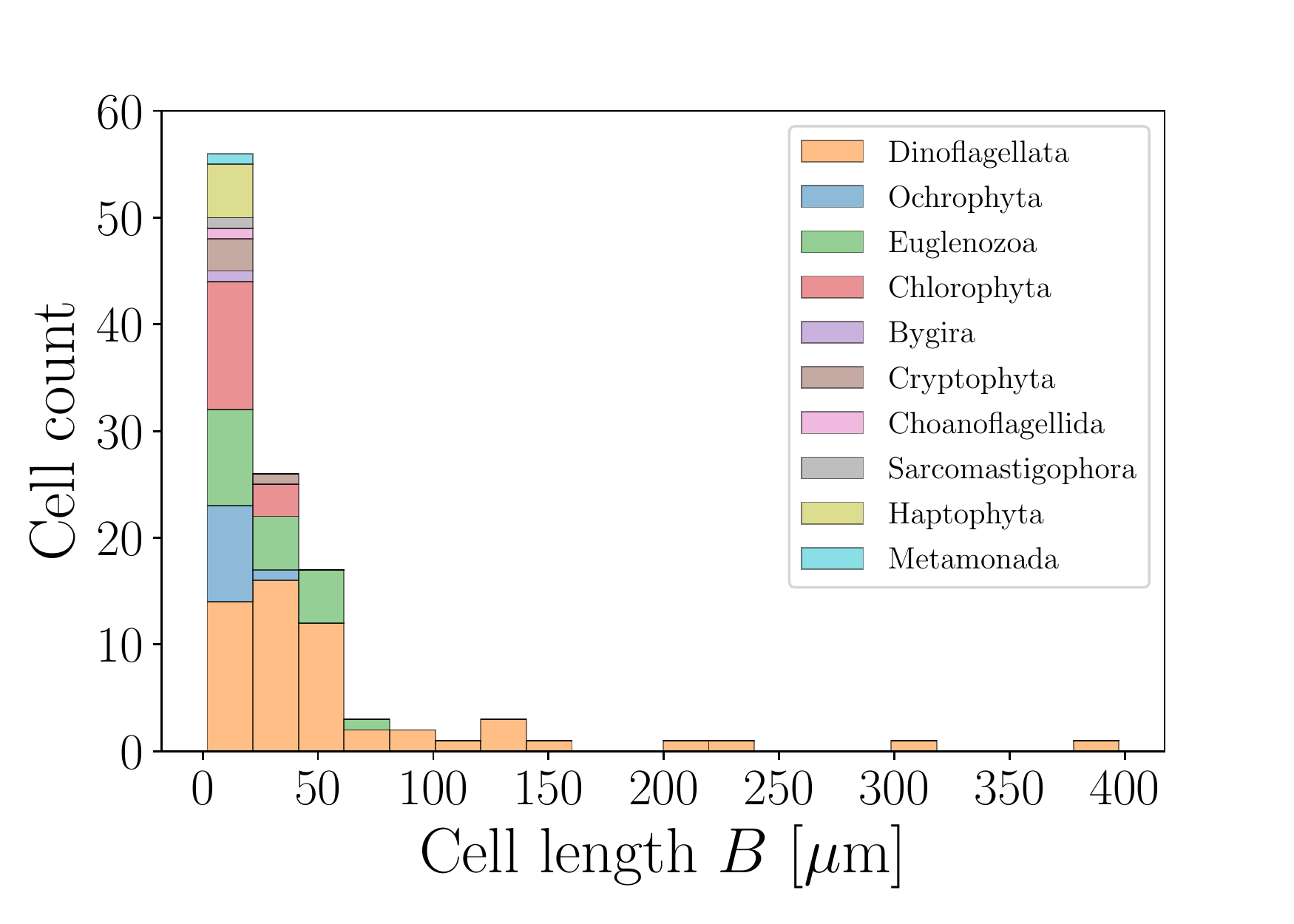}\includegraphics[width=0.5\columnwidth]{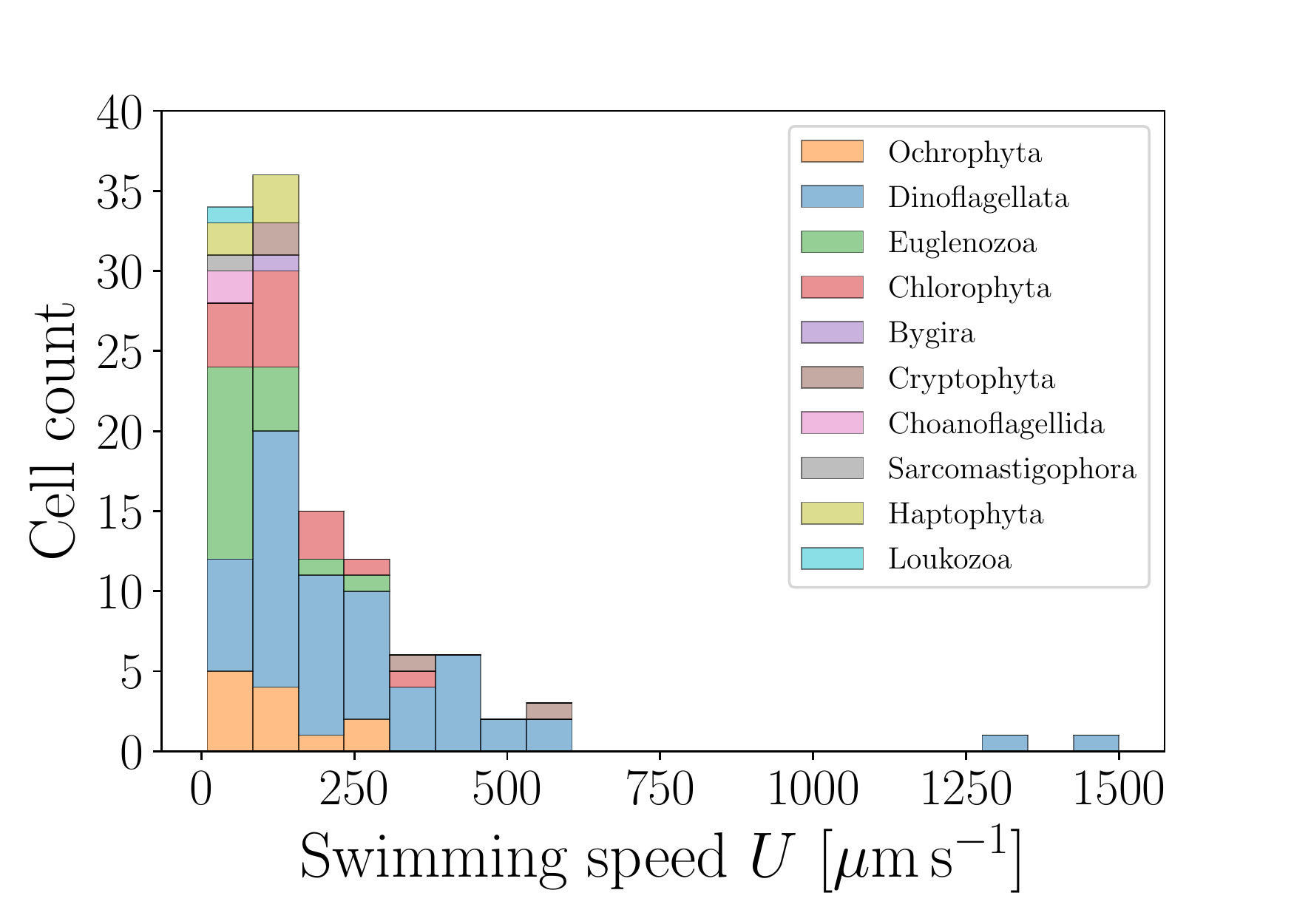}
\caption{Histograms of body lengths, $B$ ($\mu$m, left),  and swimming speeds,  $U$ ($\mu$m$\,$s$^{-1}$, right),  for
flagellated eukaryotic swimmers (excluding spermatozoa and ciliates) in our dataset. The average cell length is {$\langle B\rangle = 38.87\pm56.64\ \mu\textrm{m}\ (n=113)$} and the average swimming speed {$\langle U \rangle = 186.70\pm208.77\ \mu\textrm{m}\,\textrm{s}^{-1}\ (n=116)$}. \label{histBnL_flageuk}}
\end{figure}

The typical sizes and swimming speeds of eukaryotic flagellates are presented in Fig.~\ref{histBnL_flageuk}, based on {the data from} Table \ref{tab:flageuk}. {Significantly} larger and faster than prokaryotic cells, the distributions are dominated by the low-values tails.
 \begin{figure}[t]
 \centering
 \includegraphics[width=0.5\columnwidth]{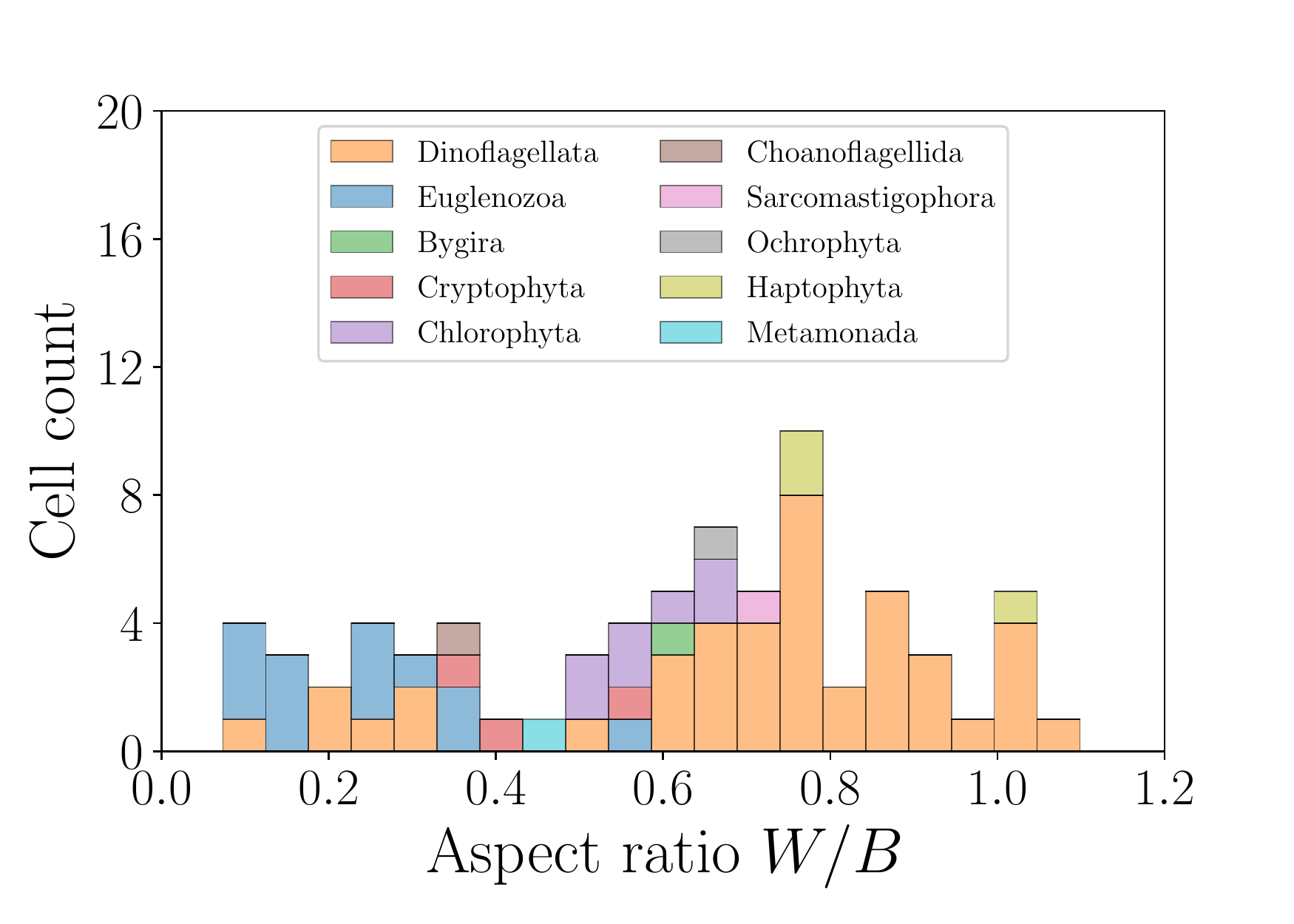}\includegraphics[width=0.5\columnwidth]{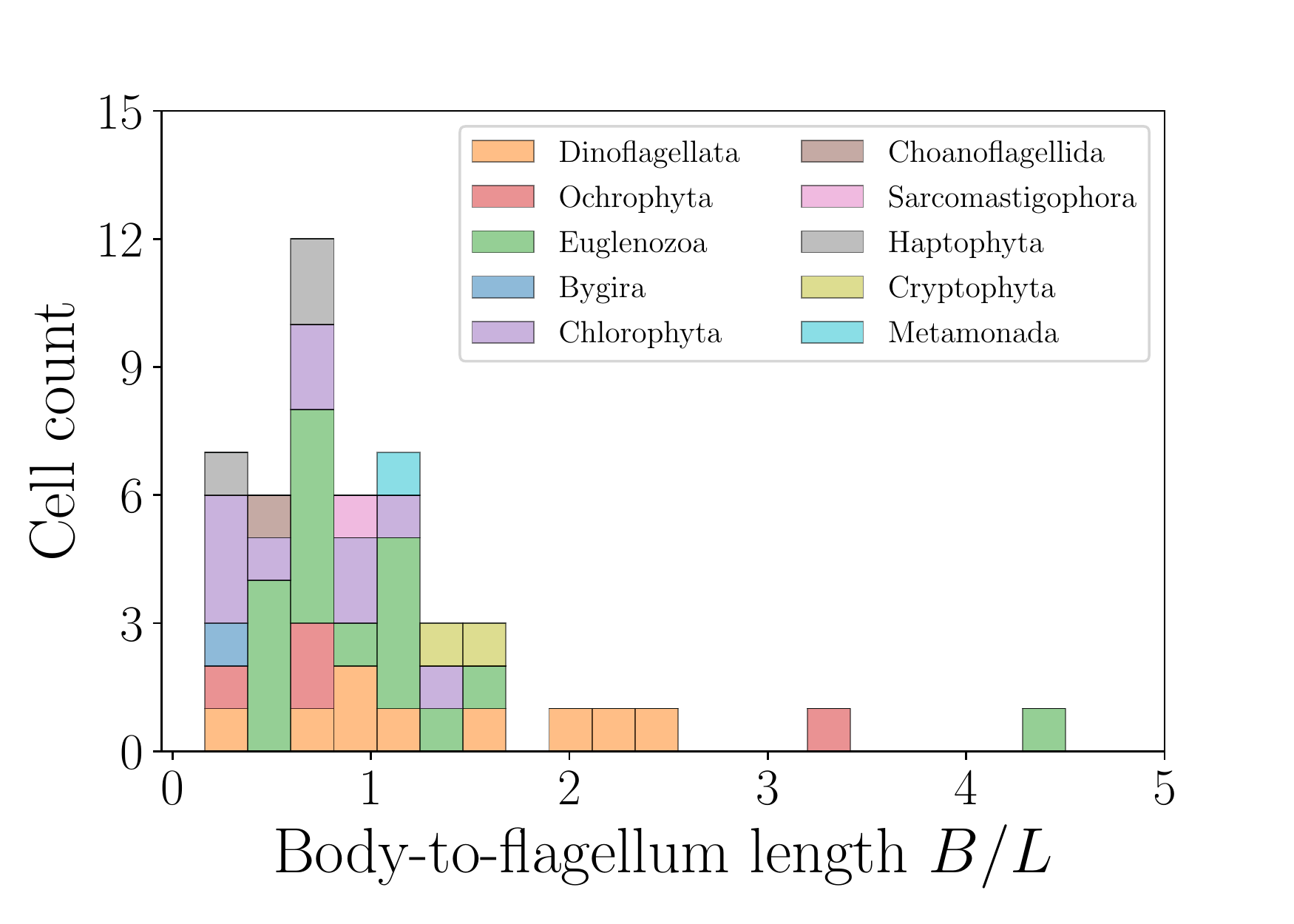}
 \caption{Histograms of aspect ratios $W/B$ (left) and body-to-flagellum length ratios $B/L$ (right) for flagellated eukaryotic swimmers. For all organisms in this category, the aspect ratios do not exceed $\approx 1.1$, and the shape distribution {indicates a} slightly prolate shape on average, {with} {$\langle W/B\rangle=0.60\pm0.27\ (n=73)$}. The distribution of body-to-flagellum length ratios  shows that  flagella {tend to be} of   length comparable  to the {cell body}, with a few exceptions {$\langle B/L\rangle=1.03\pm0.79\ (n=49)$}. \label{histBnL_flageuk_AR}}
 \end{figure}{}

  {Most cells are close to the average values, with several outliers in the large size and speed ranges.} The statistical properties of these distributions have been discussed in detail in our previous {work}~\cite{elife}. {We {may} gain further insight by considering  the distribution of  aspect ratios for the cell bodies, $W/B$, and the relative cell body-to-flagella lengths, $B/L$,  {both of which are shown} in Fig.~\ref{histBnL_flageuk_AR}. The  {wide} distribution of aspect ratios   confirms that most flagellates are slightly prolate, although several more elongated swimmers are also reported. In addition, for most cells the ratio of body to flagella length does not exceed 1, confirming that the length of the flagella is  comparable to the cell size.} This  {feature allows to    distinguish flagellated eukaryotes from} spermatozoa and ciliates.

  \begin{figure}[t]
 \centering
 \includegraphics[width=0.75\columnwidth]{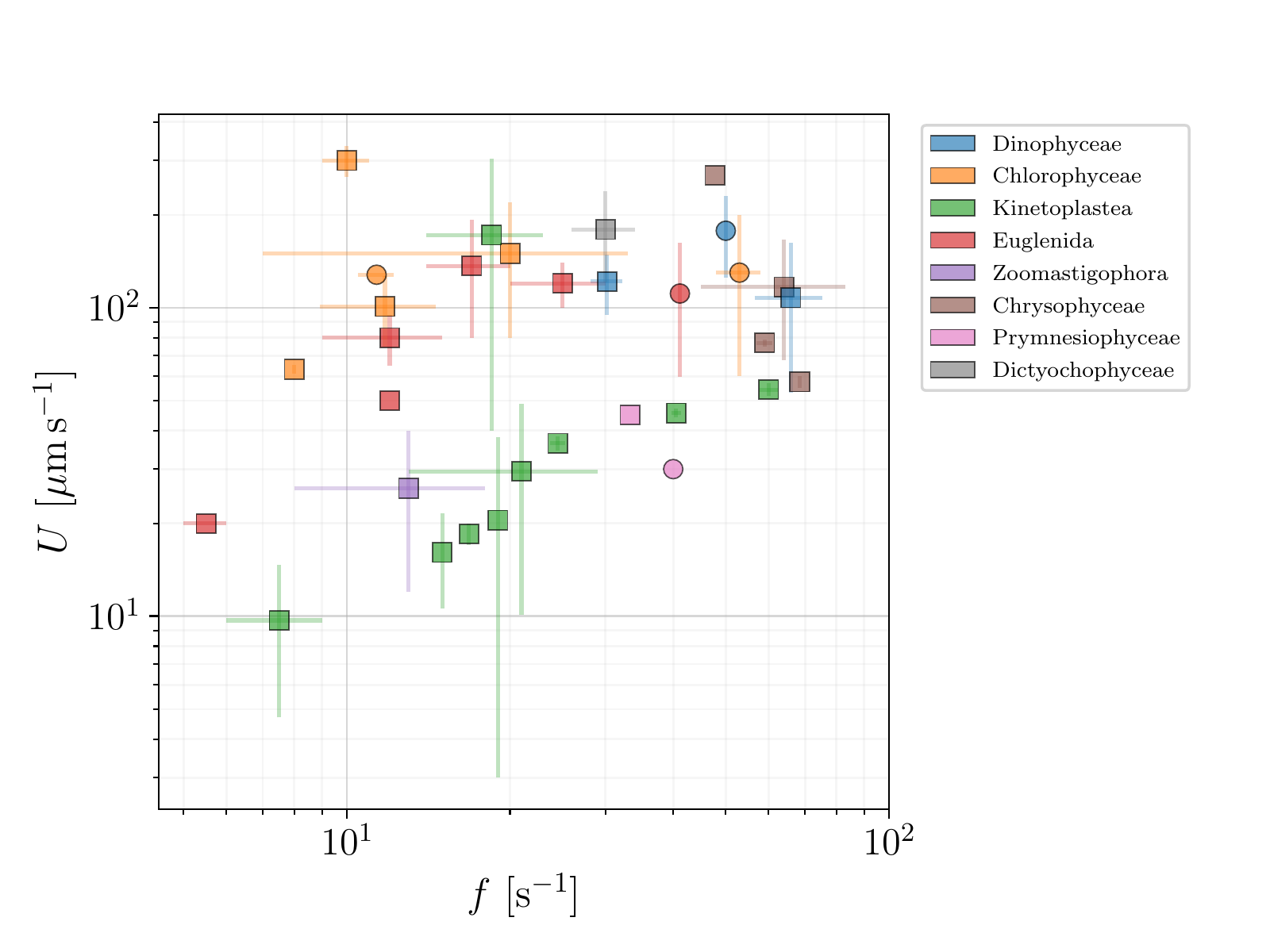}
 \caption{\label{logUvsf_fe}Swimming speed, $U$ ($\mu$m$\,$s$^{-1}$),  plotted versus the frequency of flagellar beat, $f$ (s$^{-1}$), for flagellated eukaryotes {in our dataset} (excluding spermatozoa and ciliates). {Colours mark different classes and sub-classes. Wave-producing organisms are plotted in squares and the remaining flagellated eukaryotes are plotted in circles.}}
 \end{figure}

 \begin{figure}[t]
 \centering
 \includegraphics[width=0.75\columnwidth]{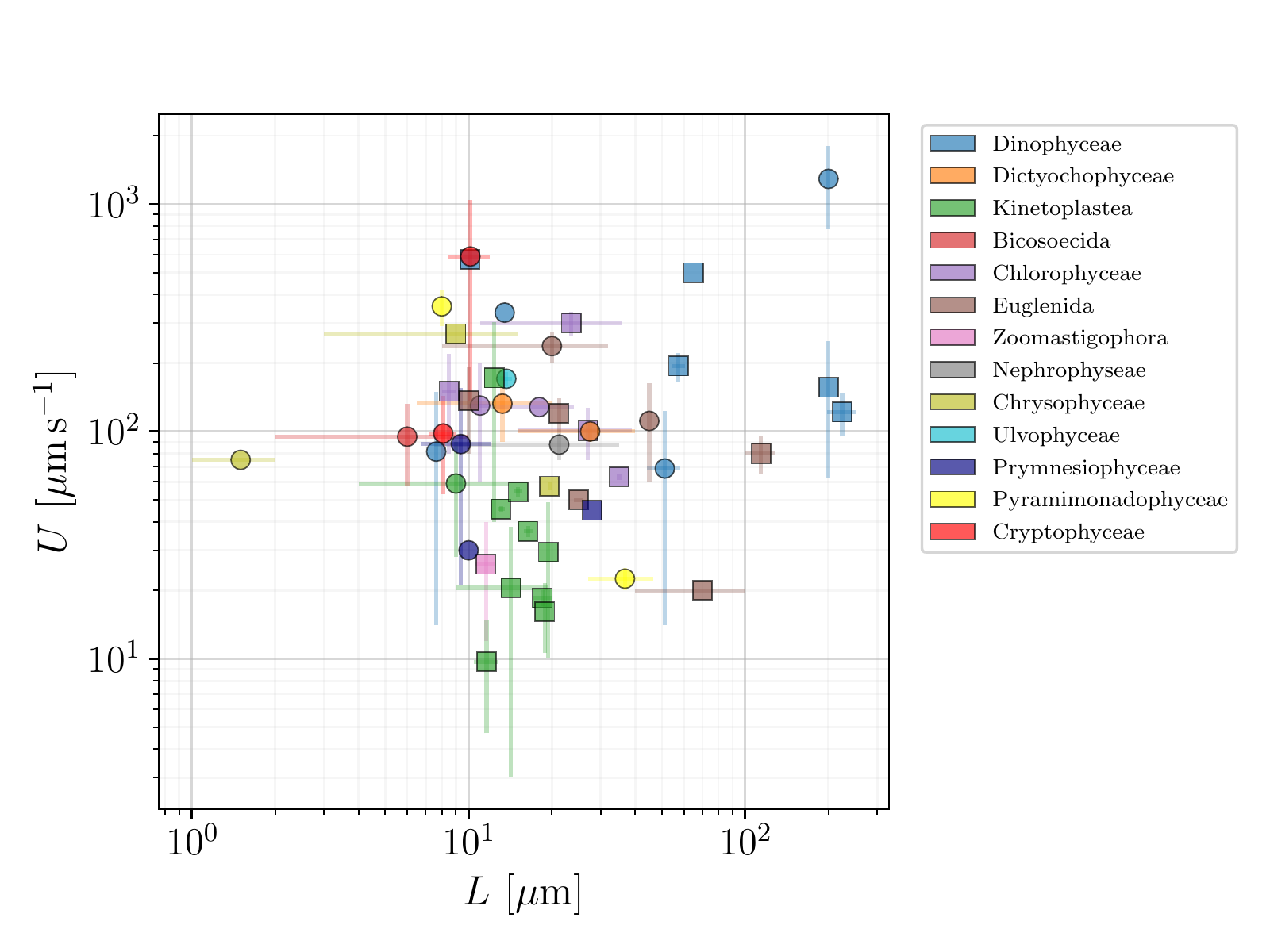}
 \caption{\label{logUvsL_fe}Swimming speed, $U$ ($\mu$m$\,$s$^{-1}$),   vs~length of flagella, $L$ ($\mu$m),   for flagellated eukaryotes {in our database} (excluding spermatozoa and ciliates). {Taxonomic classes are marked by colours. Wave-producers are again plotted in squares, while other flagellates are plotted in circles. }}
 \end{figure}

{In Figs.~\ref{logUvsf_fe} and \ref{logUvsL_fe} we next show how the swimming speeds $U$ {of} the flagellated eukaryotes in our database vary with  the flagellar beat frequencies $f$ and flagellar lengths $L$, respectively.
{Both plots show large variations and no clear trend is evident}. In the next section we will {then} adapt the classical derivation by Gray \& Hancock~\cite{93} as a minimal model for the propulsion of eukaryotic flagellates to see {the role played by} these (and other) parameters in eukaryotic propulsion. }

\subsection{Modelling of swimming for flagellated eukaryotes}

We base the description of the locomotion of flagellated eukaryotes on the assumption that swimming results from planar travelling waves induced in one or more flagella, which push a spheroidal cell body forward.

\begin{figure}[t]
\centering
\includegraphics[width=0.9\columnwidth]{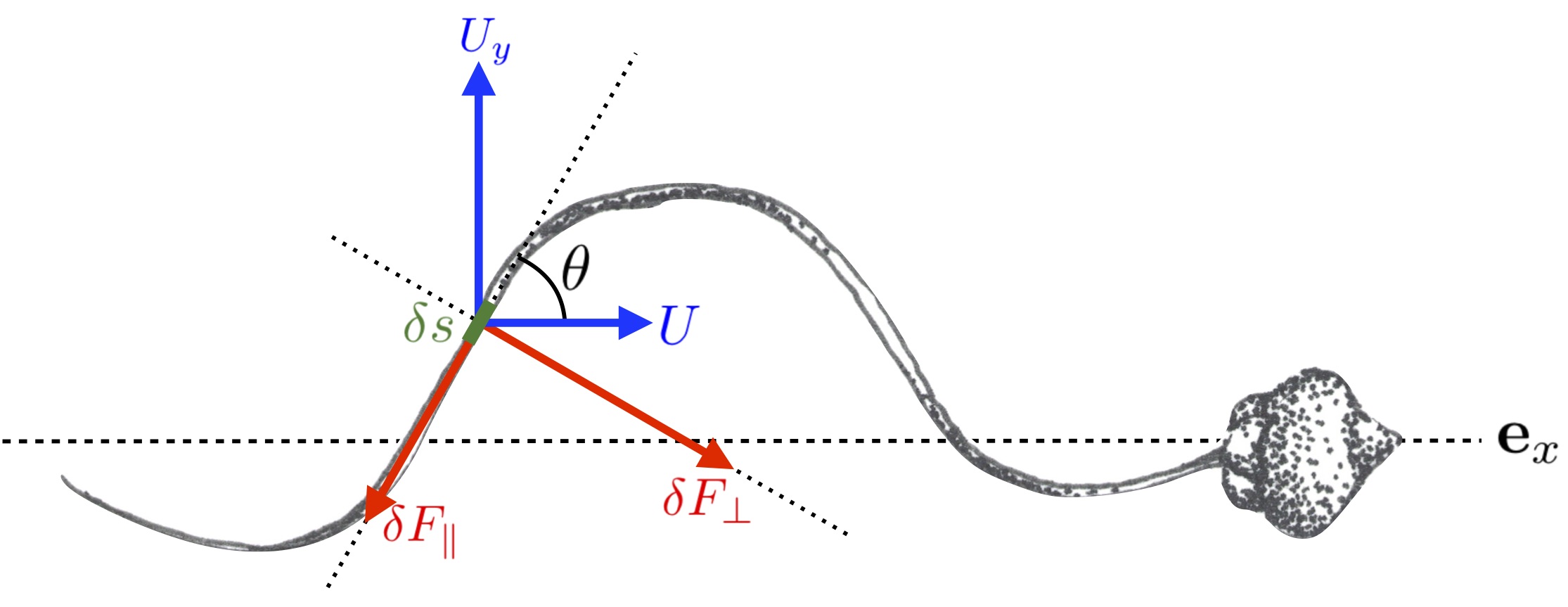}
\caption{Sketch of a swimming eukaryote (spermatozoon of \textit{Chaetopterus}, Annelida) propelled by {a single} flagellum. We distinguish a section of length  $\delta s$ inclined at an angle $\theta$ to the direction of motion $\mathbf{e}_x$, which we use to determine the local hydrodynamic forces exerting on the flagellum. Drawing by Marcos F. Velho Rodrigues.\label{fig:euk_derivation}}
\end{figure}

{The shape of the wave} is described in Cartesian coordinates by $y=y(x,t)$, where $x$ is the direction of cell movement (see Fig.~\ref{fig:euk_derivation}). An infinitesimal segment of the flagellum of length  $\delta s$ inclined at an angle $\theta$ to the axis of movement $\mathbf{e}_x$ is then subjected to a hydrodynamic force perpendicular to its orientation, and given by

 \begin{equation}
 	\delta F_\bot=c_\bot(U_y\cos\theta - U\sin\theta)\ \delta s,
 \end{equation}
and to a force tangential to the segment given by
\begin{equation}
	\delta F_\parallel=c_\parallel(U_y \sin\theta +U \cos\theta)\ \delta s.
\end{equation}
Here $U$ and $U_y(x,t)$ are the local velocities of the flagellum relative to the fluid in the directions along and perpendicular to the overall direction of cell motion, respectively. Furthermore, similarly to the section on prokaryotes,  $c_\bot$ and $c_\parallel$ are the drag coefficients per unit length in the directions normal and tangential to $\delta s$, respectively (see  Eq.~\ref{eq:cbot}).

These two force components produce an infinitesimal net thrust along the $x$ direction, $\delta F=\delta F_\bot \sin\theta -\delta F_\parallel\cos\theta$, which we rewrite as
\begin{equation} \label{eq:dF}
  \delta F = \frac{(c_\bot-c_\parallel)U_y \tan\theta - U (c_\parallel+c_\bot \tan^2\theta)}{1+\tan^2\theta}\ \delta s.
\end{equation}

Taking into account the normal speed to be $U_y=\partial y/\partial t$, using $\tan\theta={\partial y}/{\partial x}$ and $\delta s^2=\delta y^2+\delta x^2$, we transform Eq.~\eqref{eq:dF} into
\begin{equation}\label{eq:dF2}
  \delta F = \frac{(c_\bot-c_\parallel)\displaystyle\frac{\partial y}{\partial t}\displaystyle\frac{\partial y}{\partial x}  - U \left(c_\parallel+c_\bot \left(\displaystyle\frac{\partial y}{\partial x}\right)^2\right)}{\sqrt{1+\left(\displaystyle\frac{\partial y}{\partial x}\right)^2}}\ \delta x.
\end{equation}
{We now} need to specify a particular wave form of the beating pattern. One that is often observed in eukaryotic swimmers is a planar travelling wave~\cite{JahnVotta} which we approximate by a single sine wave of fixed amplitude $h$, wavelength $\lambda$ and beat frequency $f$
\begin{equation}
	y(x,t)=h\sin\left(\frac{2\pi}{\lambda}\left(x+c t\right)\right), \label{eq:sine}
\end{equation}
where $c=\lambda f$ is the speed of the propagating flagellar waves. Substituting the sine wave into Eq.~\eqref{eq:dF2}, and taking the slender limit $c_\bot\approx2c_\parallel$, yields
\begin{equation}
  \delta F = c_\parallel \frac{c A^2-U(1+2A^2)}{\sqrt{1+A^2}}\ \delta x, \label{eq:dFs}
\end{equation}
where $A={\partial y}/{\partial x}=\left({2\pi h}/{\lambda}\right)\cos\left( {2\pi}(x+c t)/{\lambda} \right)$. It is convenient to introduce the number of complete waves $n_w$ in the flagellum of length $L$, defined as
\begin{subequations}
\label{eq:nw}
  \begin{align}
  \frac{1}{n_w}&=\frac{1}{L}\int_{x=0}^\lambda \delta s=\frac{1}{L}\int_{0}^\lambda \sqrt{1+A^2}\ dx. \label{eq:nw1}
\intertext{Because the integrand $\sqrt{1+A^2}$ is a function of period $\lambda$, a simple substitution shows that the number of waves is constant in time, and is given by}
  \frac{1}{n_w}&=\frac{\lambda}{L}\ \Lambda\left(\frac{2\pi h}{\lambda}\right),\label{eq:nw2}
  \end{align}
\end{subequations}
where the auxiliary integral $\Lambda$ is
\begin{equation}
  \Lambda(a)=\frac{1}{2\pi}\int_0^{2\pi}\sqrt{1+a^2\cos^2\alpha}\ d\alpha.\label{eq:LBD}
\end{equation}

With the net thrust $\delta F$ in Eq.~\eqref{eq:dFs} being also of period $\lambda$, a good approximation of the total thrust produced by the entire flagellum independent of time is given by
\begin{equation}\label{eq:fpf}
  n_w\int_{x=0}^{\lambda}\delta F=n_w c_\parallel\lambda\left(c I_1\left(\frac{2\pi h}{\lambda}\right)-U I_2\left(\frac{2\pi h}{\lambda}\right)\right),
\end{equation}
where  we have introduced the two auxiliary integrals
    \begin{equation}\label{integralsFeuk}
  I_1(a)=\frac{1}{2\pi}\int_{0}^{2\pi}\frac{a^2\cos^2\alpha}{\sqrt{1+a^2\cos^2\alpha}}\ d\alpha,\quad
  I_2(a)=\frac{1}{2\pi}\int_{0}^{2\pi} \frac{1+2a^2\cos^2\alpha}{\sqrt{1+a^2\cos^2\alpha}}\ d\alpha.
  \end{equation}
The three functions $\Lambda$, $I_1$ and $I_2$ are easy to evaluate numerically. Alternatively, by writing $\cos^2\alpha = (1+\cos2\alpha)/2$, and neglecting the contributions of the terms in $\cos2\alpha$ in the expressions of Eqs.~\eqref{eq:LBD} and \eqref{integralsFeuk}, one gets explicit approximations
\begin{equation}
 \Lambda(a)\approx\sqrt{1+\frac{a^2}{2}}, \qquad
 I_1(a)\approx\frac{a^2}{2\sqrt{1+\displaystyle\frac{a^2}{2}}}, \qquad  I_2(a)\approx\frac{1+a^2}{\sqrt{1+\displaystyle\frac{a^2}{2}}}.\label{eq:I1I2La}
\end{equation}
Numerical evaluation of the exact expressions for  $\Lambda$, $I_1$ and $I_2$ shows that the approximations above hold to within 13\% accuracy for all values ${h}/{\lambda}<1$.

For the sake of simplicity, we shall suppose that an organism with $N$ beating flagella  is subject to a total thrust equal to $N$ times the thrust generated by each  flagellum and given by Eq.~\eqref{eq:fpf}.
We therefore neglect  hydrodynamic interactions between the flagella, which we  assume    all   beat collinearly along the swimming direction.

Steady swimming requires the thrust produced by the flagella to be balanced by the drag acting on the cell body. The latter is modelled as a prolate spheroid of length $B$ and diameter $W$. The balance of forces acting on the microorganism along $x$ is then
\begin{equation}
	N n_w\int_{x=0}^\lambda\delta F-3\pi\eta U B C_{FB}=0,
\end{equation}
with $C_{FB}(W/B)$ given by Eq.~\eqref{eq:Cfb}. The swimming speed $U$ can thus be written as
\begin{equation}
  \frac{U}{\lambda f}=\frac{I_1\displaystyle\left(\frac{2\pi h}{\lambda}\right)}{I_2\displaystyle\left(\frac{2\pi h}{\lambda}\right)+\displaystyle\frac{3\pi\eta B C_{FB}(W/B)}{N n_w c_\parallel \lambda}}, \label{eq:pswh1}
\end{equation}
or using the definition of $n_w$ in Eq.~\eqref{eq:nw} as
\begin{equation}
  \frac{U}{\lambda f}=\frac{I_1\left(\displaystyle\frac{2\pi h}{\lambda}\right)}{I_2\left(\displaystyle\frac{ 2\pi h}{\lambda}\right)+\displaystyle\frac{3\pi\eta}{N c_\parallel}\displaystyle\frac{B}{L} C_{FB}(W/B)\Lambda\left(\frac{2\pi h}{\lambda}\right)}. \label{eq:pswh2}
\end{equation}
By using $c_\parallel =2\pi\eta\left[\log({L}/{b})-{1}/{2}\right]^{-1}$ as in Eq.~\eqref{eq:cbot}, and approximating integrals $I_1$, $I_2$ and $\Lambda$ with the expressions in Eq.~\eqref{eq:I1I2La}, we arrive at the final expression
\begin{equation}\label{eq:fefit}
  U=\displaystyle\frac{2\pi^2 h^2 f}{\lambda}\left[1+\displaystyle\frac{4\pi^2 h^2}{\lambda^2}+\displaystyle\frac{3 B}{2NL}C_{FB}(W/B)\left(1+\displaystyle\frac{2\pi^2 h^2}{\lambda^2}\right)\left(\log \frac{L}{b}-\frac{1}{2}\right)\right]^{-1}.
\end{equation}

 \subsection{{Insights from data}}

\begin{figure}[t]
\centering
\includegraphics[width=0.75\columnwidth]{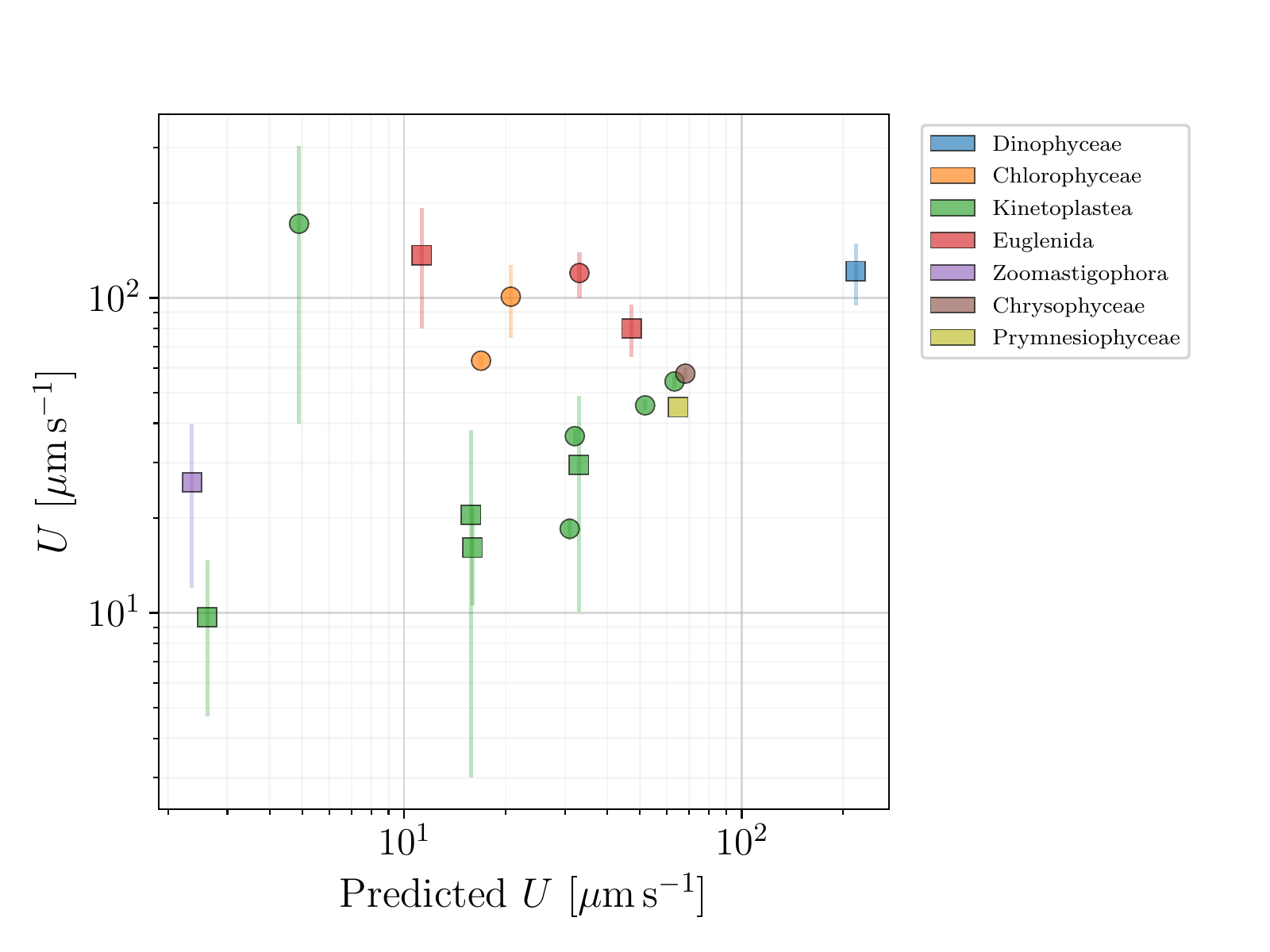}
\caption{{Swimming speeds of flagellated eukaryotes (excluding spermatozoa and ciliates) reported in the database plotted against the theoretical prediction of Eq.~\eqref{eq:pswh2}. Colours mark different classes. Square symbols mark organisms for which the prediction was directly calculated from the available data, while circles represent organisms for which either the body width}
{or one of the flagellar characteristics {has} been estimated (see text for details).}
\label{fig:fe_fit}}
\end{figure}

{We can now use our model to help organise our data on flagellated eukaryotes.}
{In Fig.~\ref{fig:fe_fit}, we compare the swimming speeds from {our dataset} with those predicted by the theoretical model in Eq.~\eqref{eq:pswh2}. Square symbols mark organisms for which all the quantities needed to calculate the predicted speed were available. The species plotted in circles in the figure had their data incomplete. Whenever the body width $W$ was unavailable, we estimated its value using the average aspect ratio $\langle W/B\rangle=0.60$ of Fig.~\ref{histBnL_flageuk_AR}. When one parameter of the flagellar wave was missing, we estimated it with the help of Eq.~\eqref{eq:nw}. The radii of the flagella were all fixed at $b=0.2\ \mu$m.}

{In Fig.~\ref{fig:fe_fit}}, {we see a cluster of data points (mostly {the class} Kinetoplastea) that correlate well with the expected linear dependence. {However,  many of the organisms have  a} swimming speed {that} significantly exceeds the predicted values. This {may} point to other mechanisms being involved, such as different beating patterns}
{and cell body shapes,}
{which   would require a more careful examination. Nevertheless,   the basic {framework} proposed {by} the model provides a useful estimate of the lower bound for the swimming speed, which   can   be exceeded by adopting more effective locomotion strategies suited to the organism and its environment.}

\section{\label{Sec:sperms}Spermatozoa}

The motile behaviour of the spermatozoa of animals has been studied in detail since the beginnings of microscopy due to its   importance for reproductive health. Because a correlation between motility and fertility has been shown to exist~\cite{L111, L224}, {numerous} species of fish~\cite{L222}, birds~\cite{L244}, mammals~\cite{L223,L225,L208}, insects~\cite{L196,L182,L181,L185} and sea urchins~\cite{81} have had their spermatozoa examined. A particular focus {is often} placed on the relation between either the swimming speed or the amplitude of lateral  displacement {of the cell body} and the success in fertilisation by human spermatozoa~\cite{L226}.

{A} remarkable geometrical characteristic of spermatozoa, at least in comparison with other flagellated eukaryotes, is their relatively small heads compared to the length of their flagella.
 Despite this difference, the flagella  of spermatozoa have the same structure detailed above for other eukaryotic cells, and are likewise capable of creating complex waveforms. The mathematical modelling of flagellar locomotion outlined  in the previous section is thus also applicable to the case of swimming spermatozoa.

  Our database of swimming spermatozoa contains 60 {different} species, for which various geometric and dynamic data were found. These include sperm cells of the taxonomic classes  Insecta, Actinopterygii, Mammalia, Amphibia, Polychaeta, Ascidiacea, Echinoidea, Aves, and Bivalvia. {As {mentionned} above, databases of morphological measurements for over 400 spermatozoa, particularly of mammalian species, are available in literature \cite{gage1998,L112,anderson2005sperm} but {since they do not include motility data they} are not included in our database.}

\begin{figure}[t]
\centering
\includegraphics[width=0.5\columnwidth]{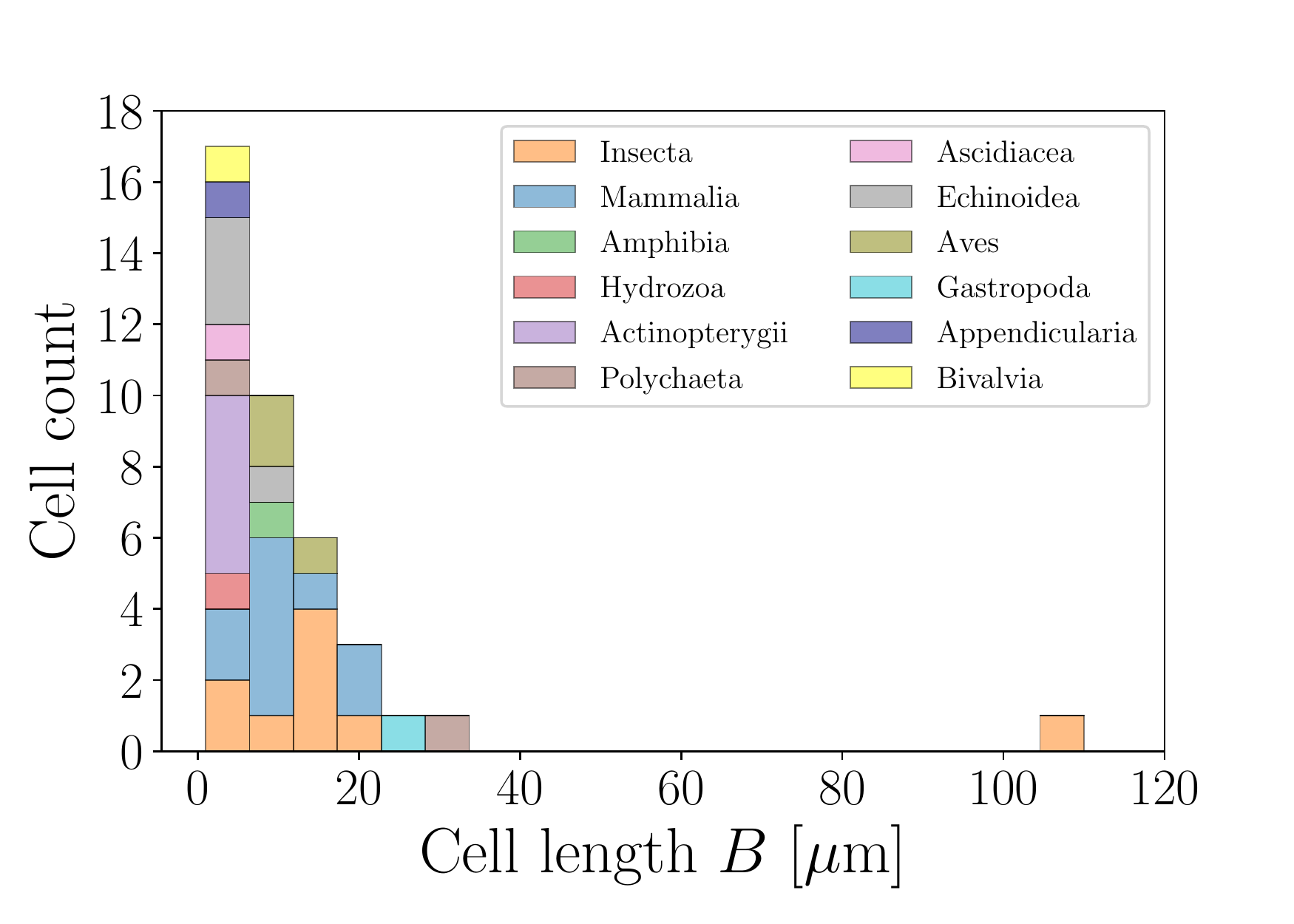}\includegraphics[width=0.5\columnwidth]{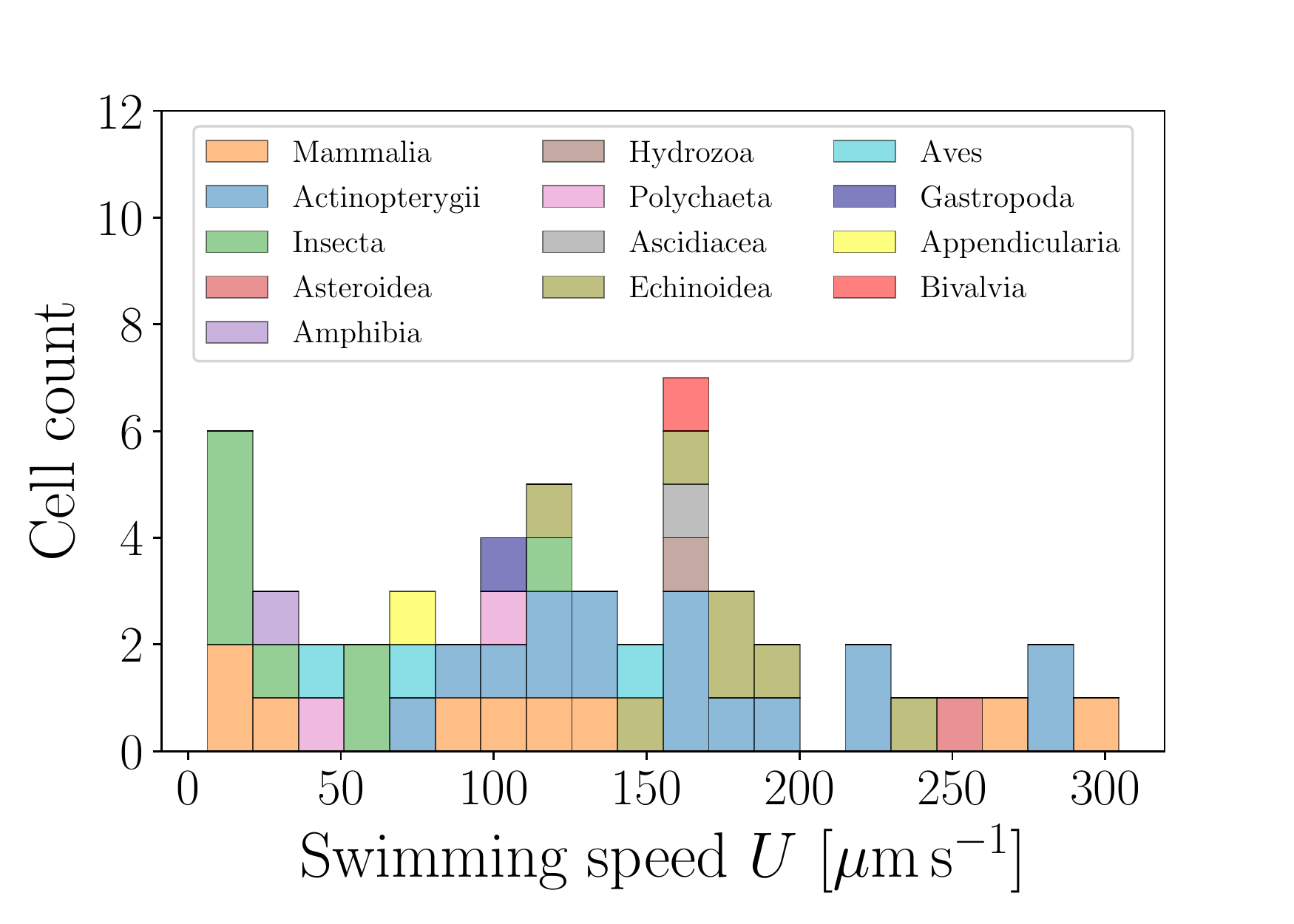}
\caption{Histograms of body lengths, $B$ ($\mu$m, left),  and swimming speeds,  $U$ ($\mu$m$\,$s$^{-1}$, right), for the
spermatozoa in the database. The average cell length is {$\langle B\rangle = 12.21\pm17.25\ \mu\textrm{m}\ (n=39)$}, while the the average swimming speed  is {$\langle U\rangle = 127.23\pm78.49\ \mu$m$\,$s$^{-1}\ (n=52)$} over a wide distribution. {We use} {colours to distinguish between the different taxonomic classes.}
\label{histBnL_sperm}}
\end{figure}

\begin{figure}[t]
\centering
\includegraphics[width=0.5\columnwidth]{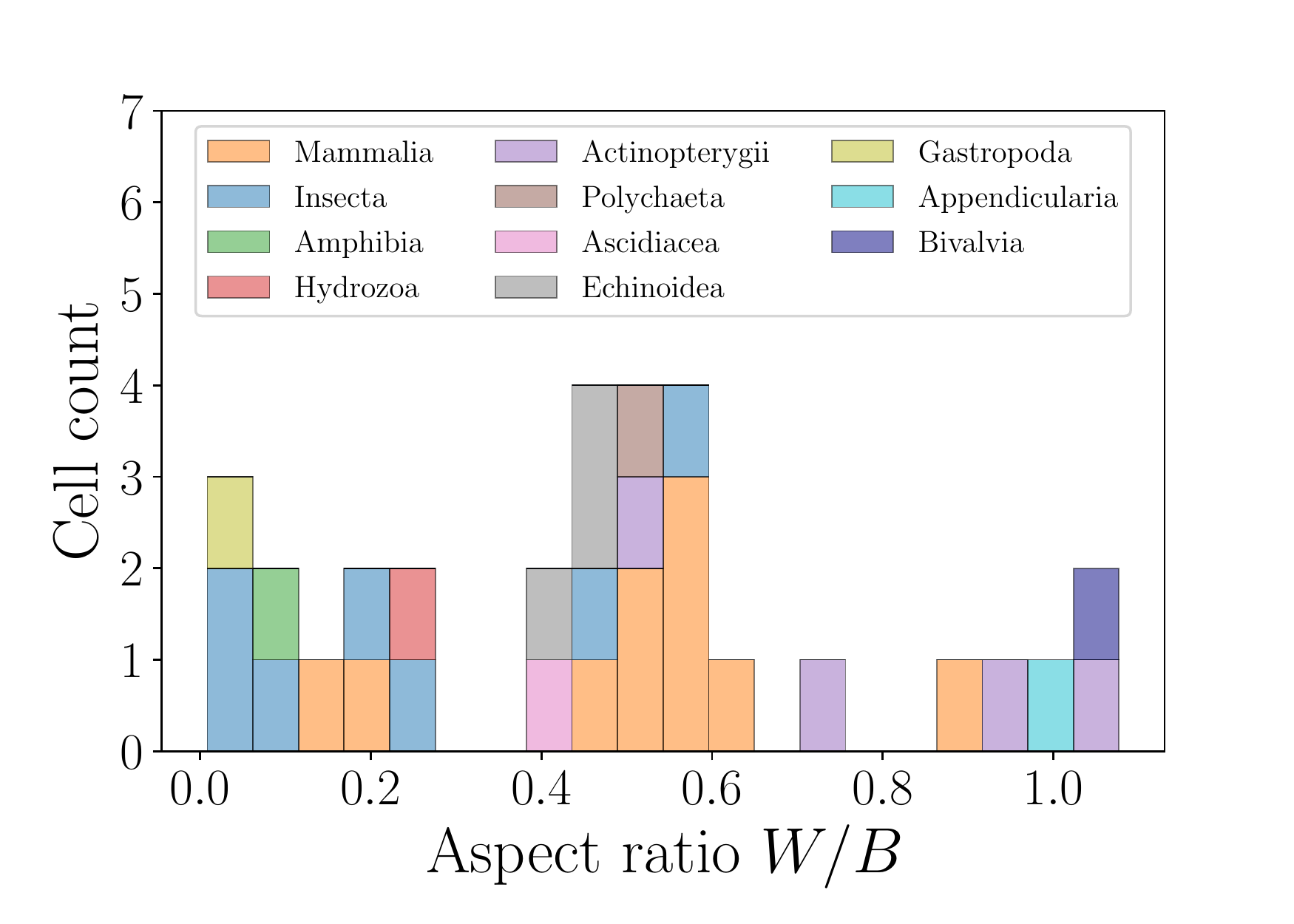}\includegraphics[width=0.5\columnwidth]{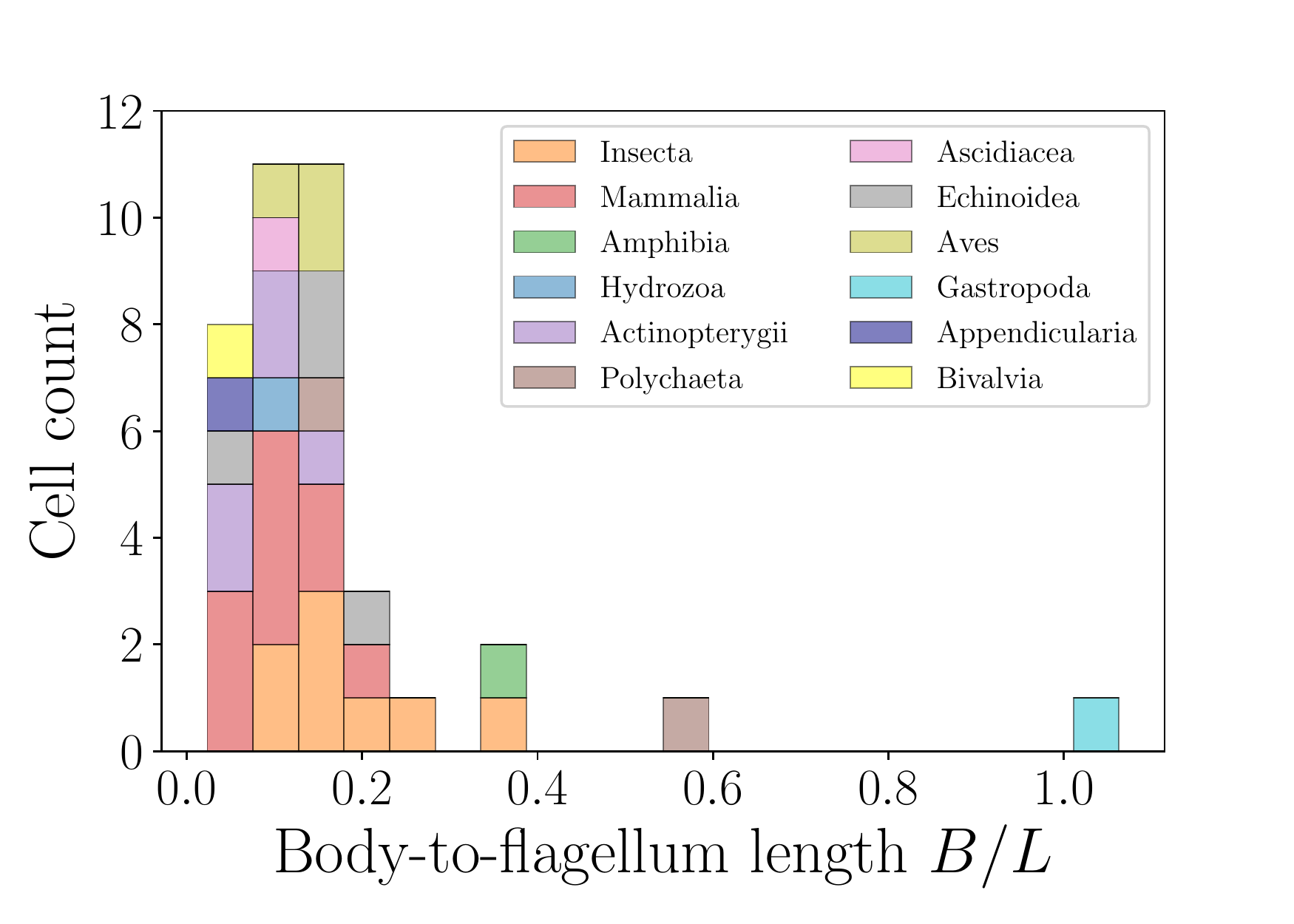}
\caption{Histograms of aspect ratios $W/B$ (left) and body-to-flagellum length $B/L$ (right) for spermatozoa {(colours mark the different taxonomic classes)}. The distribution of cell aspect ratios is rather wide, and yields an average value of {$\langle W/B \rangle = 0.47\pm0.30\ (n=31)$}. The size-to-flagellum length ratios are mostly close to the average {$\langle B/L \rangle = 0.17\pm0.18\ (n=38)$}, showing that in spermatozoa the flagellum length is typically much larger than the {cell} body.
\label{histBnL_sperm_AR}}
\end{figure}

\subsection{Geometry and swimming speeds of the cells}

{The distribution of {cell body} sizes and swimming speeds of spermatozoa are {shown} in Fig.~\ref{histBnL_sperm}, based on the data from Table~\ref{tab:sperm}. With   {body} sizes hardly exceeding 30~$\mu$m (except for one outlier, the cricket spermatozoon, with a size of over 100~$\mu$m), we see that spermatozoa are typically small compared to other eukaryotic cells. The distribution of swimming speeds is relatively uniform, {reaching up} to 300~$\mu$ms$^{-1}$. While their average speeds are close to those of flagellated eukaryotes from Sec.~\ref{Sec:flageuk}, the distribution of speeds is dramatically different, deviating from the log-normal seen for other flagellated eukaryotes~\cite{elife}.}

{A further inspection of the geometry reveals {that} the distribution of sperm cell aspect ratios (Fig.~\ref{histBnL_sperm_AR}, left) {is} widely spread,  {ranging} from elongated to spherical. {A clear} distinguishing feature for spermatozoa is the body-to-flagella length ratio (Fig.~\ref{histBnL_sperm_AR}, right), which is peaked at small values, showing that the spermatozoa of most species have flagella that are over fivefold longer than their body sizes.}

\subsection{Hydrodynamic model for locomotion}

{The locomotion of flagellated spermatozoa follow the same hydrodynamic principles as discussed in detail in Sec.~\ref{Sec:flageuk}. We may {thus use as our starting} point the  the result in Eq.~\eqref{eq:pswh2}, which upon using the drag coefficient $c_\parallel=2\pi\eta [\log({L}/{b})-{1}/{2}]^{-1}$}
{and $N=1$}
{takes the form}
\begin{equation}\label{eq:pswh3}
  \frac{U}{\lambda f}=\displaystyle\frac{I_1(2\pi{h}/{\lambda})}{I_2(2\pi{h}/{\lambda})+\displaystyle\frac{3B}{2L}C_{FB}({W}/{B})\Lambda(2\pi{h}/{\lambda})[\log({L}/{b})-{1}/{2}]}.
\end{equation}
Note that the  second term in the denominator of the right-hand side of Eq.~\eqref{eq:pswh3} is the  hydrodynamic load of the dragging cell body, {which we include although the flagella are notably longer than  cell bodies for spermatozoa.}

\subsection{{Insights from data}}

\begin{figure}[t]
\centering
\includegraphics[width=0.75\columnwidth]{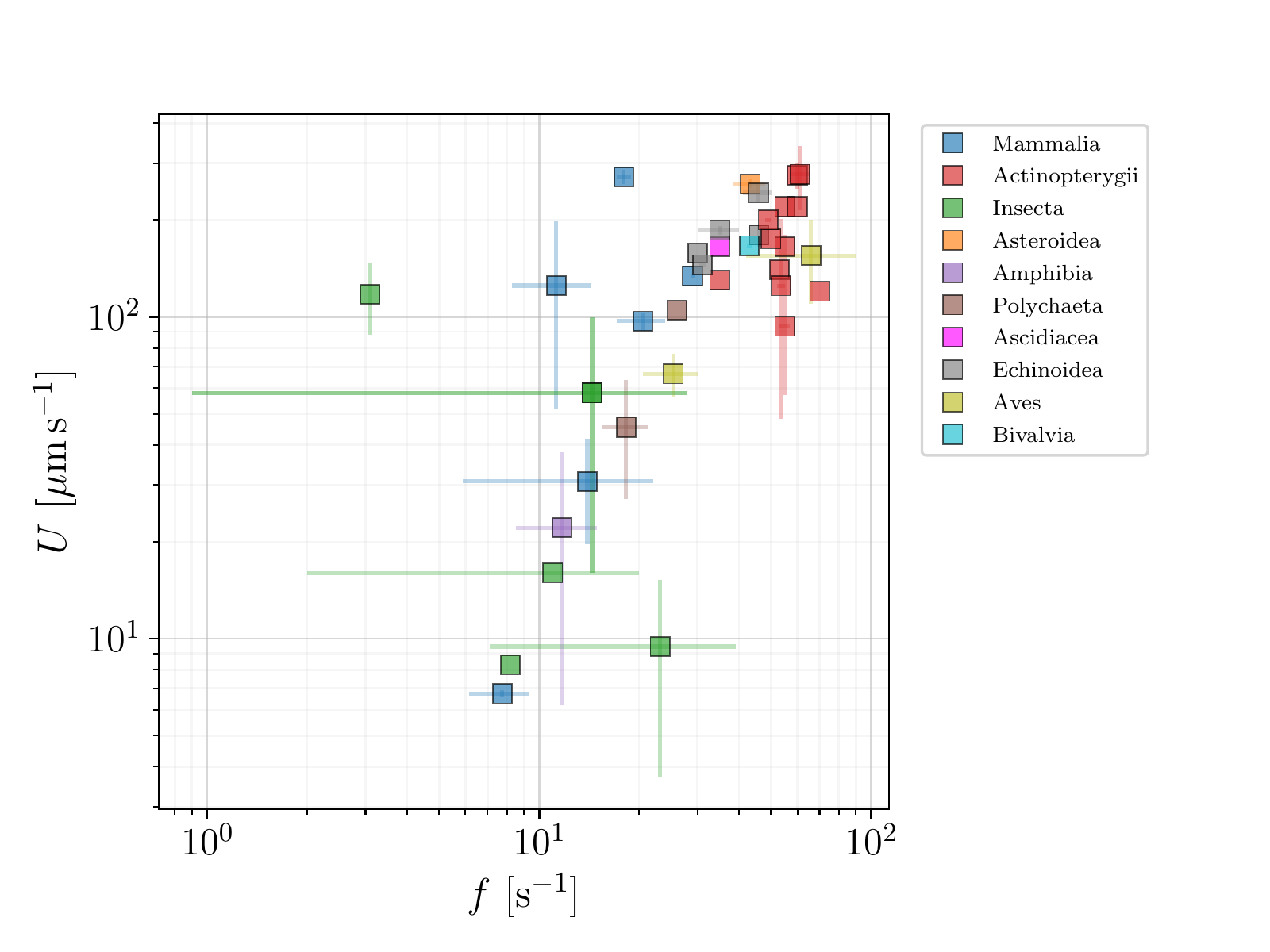}
\caption{Swimming speeds, $U$ ($\mu$m$\,$s$^{-1}$), as function of flagellar beat frequency $f$ (s$^{-1}$), for spermatozoa. A {strong} correlation between $U$ and $f$ is apparent on the figure.
\label{Uvsf_spmtz}}
\end{figure}

\begin{figure}[t]
\centering
\includegraphics[width=0.75\columnwidth]{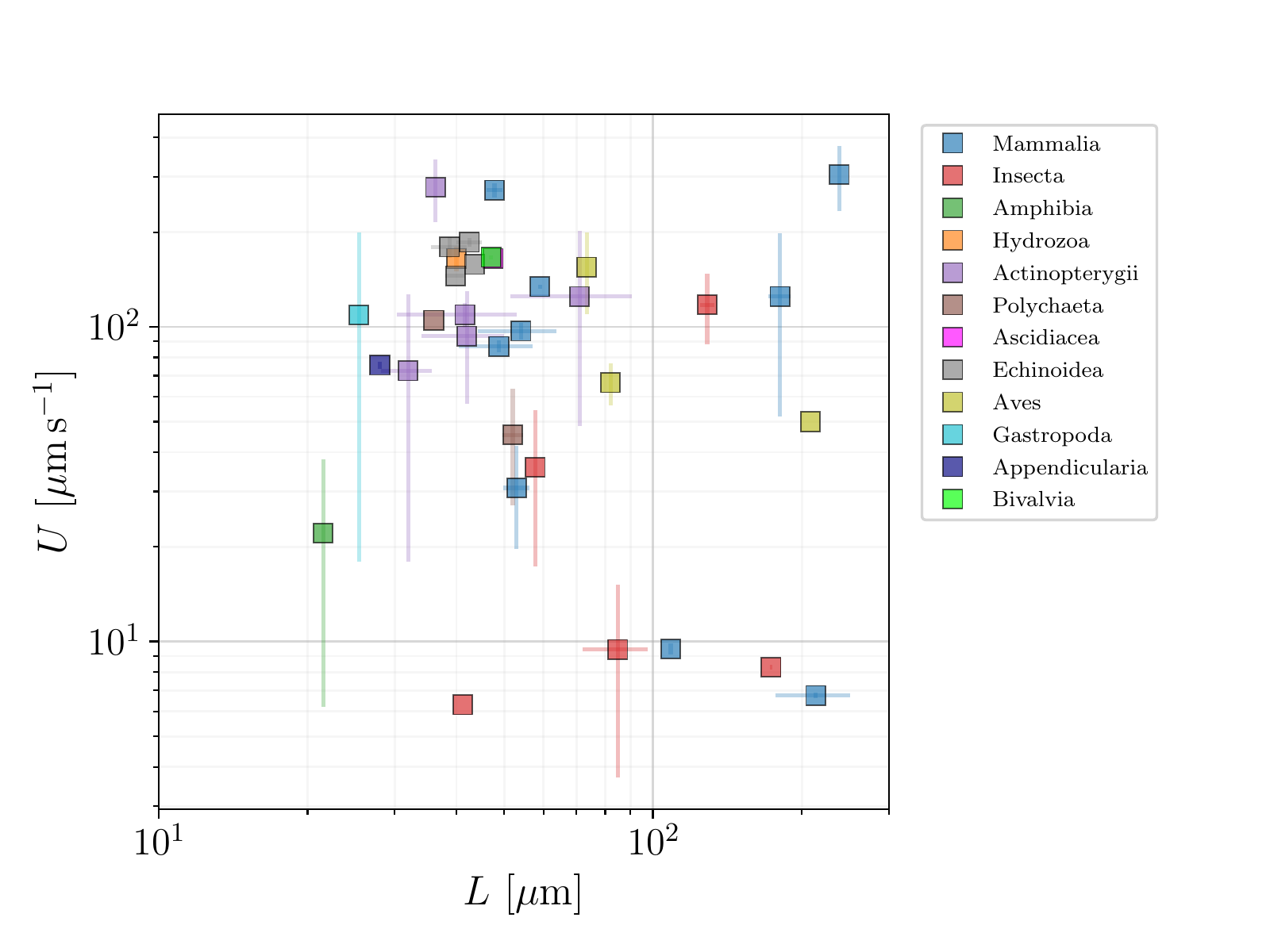}
\caption{Swimming speeds, $U$ ($\mu$m$\,$s$^{-1}$), as function of flagellar lengths, $L$ ($\mu$m), for spermatozoa.
In contrast with the {result} in Fig.~\ref{Uvsf_spmtz}, no clear correlation between $U$ and $L$ is observed here. \label{UvsL_spmtz}}
\end{figure}

{We again turn our attention to the {behaviour} of the swimming speeds for the cells.} In Fig.~\ref{Uvsf_spmtz}, we examine the dependence of the spermatozoa swimming speed $U$ on the flagellar beat frequency, $f$. With most spermatozoa operating in the frequency range between 10 and 100~Hz, and swimming speeds of up to 300~$\mu$m$\,$s$^{-1}$, we observe a {pronounced} correlation
{between these two variables across   our database.} In Fig.~\ref{UvsL_spmtz}, we also show the dependence of the swimming speed  $U$ on the flagellar length $L$, which  ranges from about 20 to 120~$\mu$m. Here, in contrast, no direct or apparent correlation is seen between $U$ and $L$.

\begin{figure}[t]
\centering
\includegraphics[width=0.75\columnwidth]{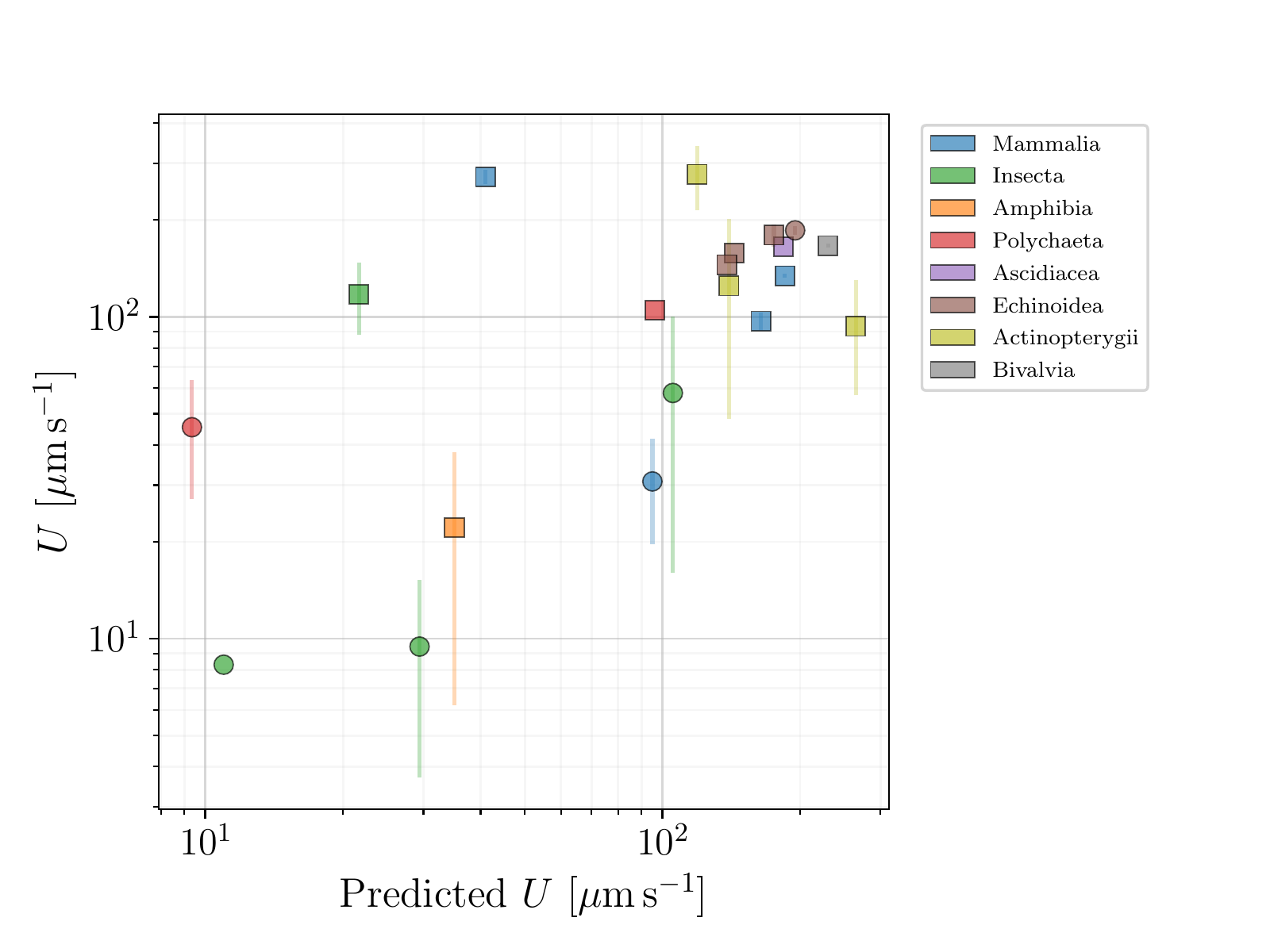}
\caption{
Reported propulsion speed of spermatozoa compared with the values predicted by the theoretical model in Eq.~\eqref{eq:pswh3}. {Colour scheme distinguishes between the different taxonomic classes.} Squares represent spermatozoa that had all parameters available in the literature{, while the circles mark cases where at least one parameter had to be estimated (via $\langle W/B\rangle=0.47$ from Fig.~\ref{histBnL_sperm_AR} or through Eq.~\eqref{eq:nw}).}
\label{fig:spmt_fit}}
\end{figure}

To {help organise the information} on the locomotion of sperm cells {in our database}, we   resort to the   model from Eq.~\eqref{eq:pswh3}, which we compare with the collected data in   Fig.~\ref{fig:spmt_fit}. {Circles represent organisms for which either {the body width} $W$ was unavailable (in which case we assumed $W/B=0.47$, the average value from Fig.~\ref{histBnL_sperm_AR}), or for which one parameter of the flagellar wave was missing (and was thus estimated using Eq.~\eqref{eq:nw}). The thickness of the flagella was fixed at $2b=0.4~\mu$m.}
{We see that the model of Eq.~\eqref{eq:pswh3} is able to capture the essence of spermatozoan swimming,    and  better   than it did for   flagellated eukaryotes in the previous section. The outliers can  likely be explained by the {use of more complex wave patterns}  in some species.}

 \section{\label{Sec:cil}Ciliates}

Within the diverse group of flagellated eukaryotes, the {final} family  of organisms is
 distinguished by their remarkably large number of flagella,  ranging from  hundreds to tens of thousands (see {the distribution in} Fig.~\ref{fig:Nflag}). These flagella are   short compared to the size of the cell body and are called {in {this} case} cilia -- hence the name of ciliates given to the whole group.  Ciliates have developed a locomotive strategy relying on the phased beating of their many cilia. Typically, a single cilium  beats using  a two-stroke motion with a power stroke of an extended  cilium followed by a recovery stroke where the cilium is curved, generating a polarised beat \cite{lauga2009}.
 From the phased beat of neighbouring cilia, collective motion is induced that pumps the surrounding fluid~\cite{BandW}, thus creating the {hydrodynamic} forces necessary for locomotion.  This collective sequential movement of cilia is often observable through the {so-called} metachronal waves of deformation travelling over the surfaces of ciliated cells, resembling  spectator  waves in stadiums.
 Yet, the underlying ciliary structure is not easily observed and only a few studies  report successfully the wavelengths of metachronal waves and ciliary beat frequencies. In particular, for the model organisms in the genus \textit{Paramecium} the frequencies of ciliary beat of all the different regions of the cell have been accessed~\cite{L206}.

The mathematical modelling of metachronal waves can be undertaken at various levels of complexity \cite{childress81,lauga2009}, starting with coarse-grained {continuum} models, such as the squirmer model~\cite{lighthill1952,blake1971}, up to detailed simulations of the deformations of individual cilia interacting hydrodynamically~\cite{liron_mochon_1976,Elgeti2013emergence}. Non-hydrodynamic interactions  via  intra-cellular  coupling mediated by the cell body are also important~\cite{quaranta2015,wan2016}.  Independently of the specific coordination mechanism, ciliates  all swim by transporting the surrounding fluid along their surfaces, and  move in the  direction opposite to the fluid motion. By using different models for this effective transport mechanism, we can now test several hypotheses across our database of ciliates, which involves data for {93} species. Note that the distribution of swimming speeds across species from this dataset has been published in our earlier contribution~\cite{elife}.

 \subsection{Geometry and swimming speeds of the cells}

\begin{figure}[t]
\centering
\includegraphics[width=0.5\columnwidth]{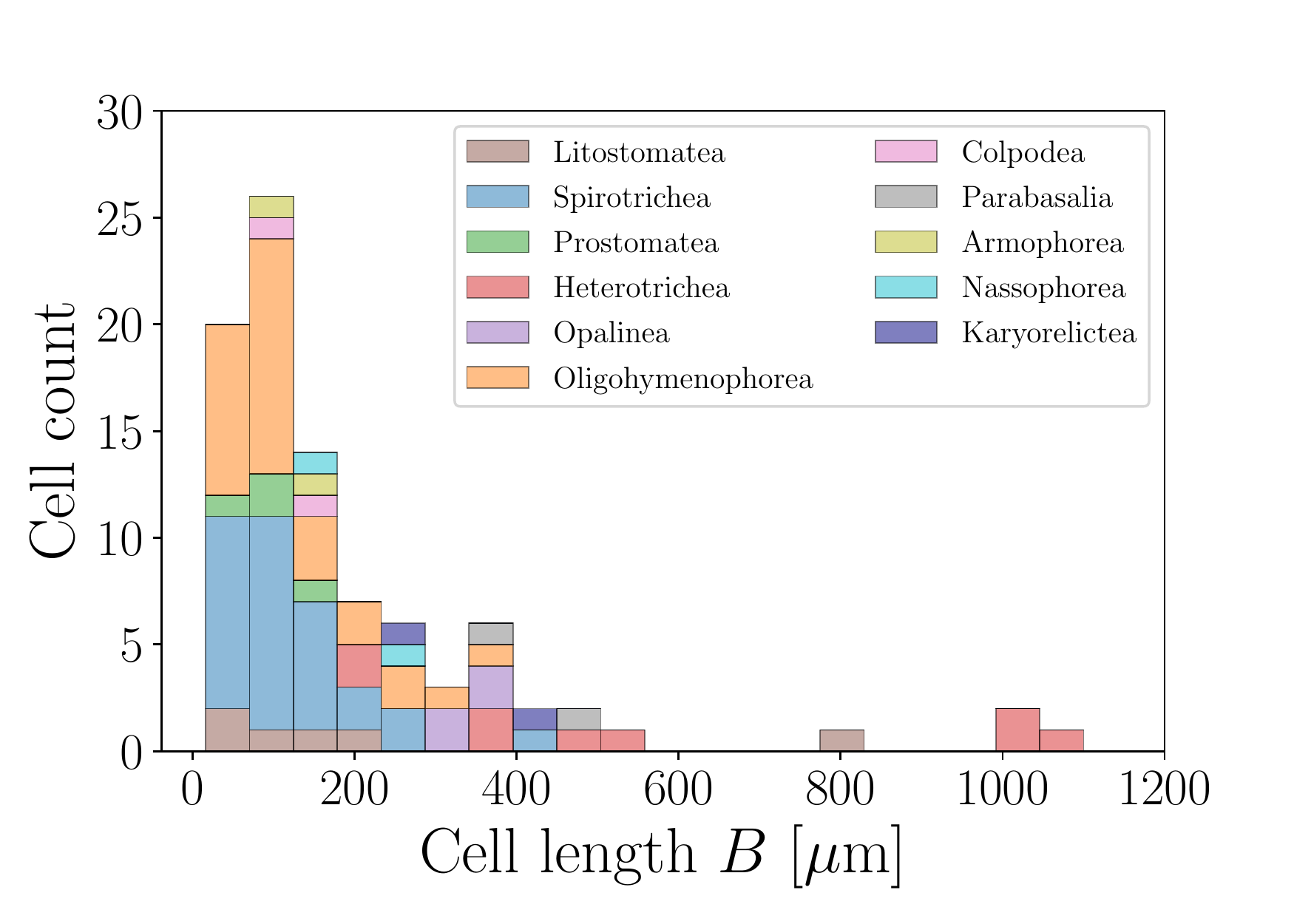}\includegraphics[width=0.5\columnwidth]{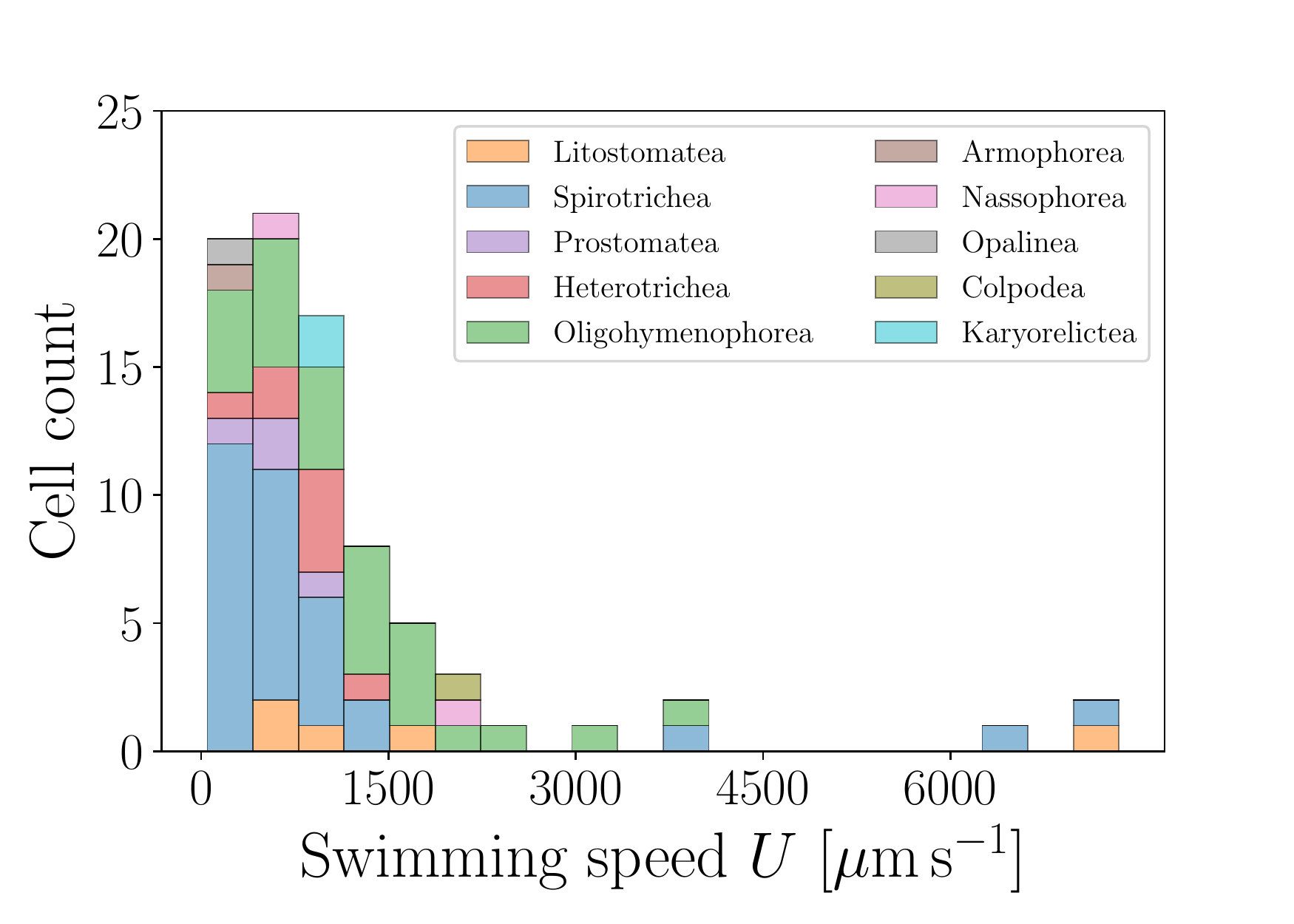}
\caption{Histograms of body lengths, $B$ ($\mu$m, left),  and swimming speeds,  $U$ ($\mu$m$\,$s$^{-1}$, right), for the {{93}} ciliates in the database. Ciliates are by far the largest organisms in our database, with the average cell length of {$\langle B\rangle = 194.87\pm207.45\ \mu\textrm{m}\ (n=91)$}, and an average swimming speed {$\langle U\rangle = 1147.57\pm1375.64\ \mu\textrm{m}\,\textrm{s}^{-1}\ (n=81)$}.
\label{histBnL_cil}}
\end{figure}

In Fig.~\ref{histBnL_cil}, we present histograms of sizes and swimming speeds for {the} ciliates in our database. Most of the organisms are close to, or slightly below, average values, which is highlighted by the skewness of the distributions~\cite{elife}. {The cells are} notably larger {(average length about 200~$\mu\textrm{m}$)} and faster {(average speed of over one millimetre per second)} tha{n} any other group {in our database}.   {As a result}, the {dimensionless} P\'eclet number for relevant molecular solutes (such as ions) around the ciliates is {of the order of} 100 which means that, in contrast to bacteria {and} flagellates, ciliates live in a physical environment where advection and thus the ability to stir the surrounding fluid may be the life-driving mechanism~\cite{elife}.

\begin{figure}[t]
\centering
\includegraphics[width=0.5\columnwidth]{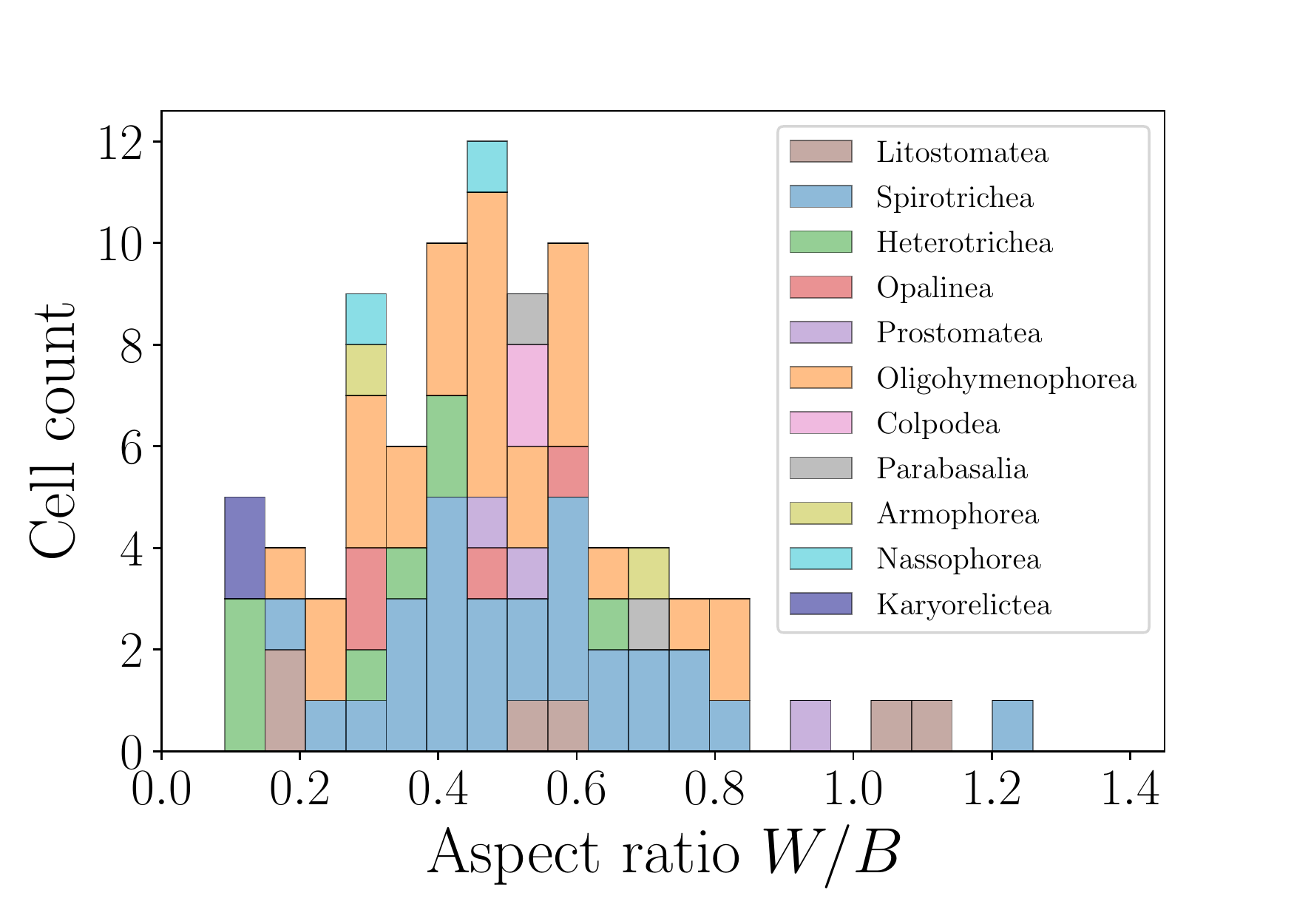}\includegraphics[width=0.5\columnwidth]{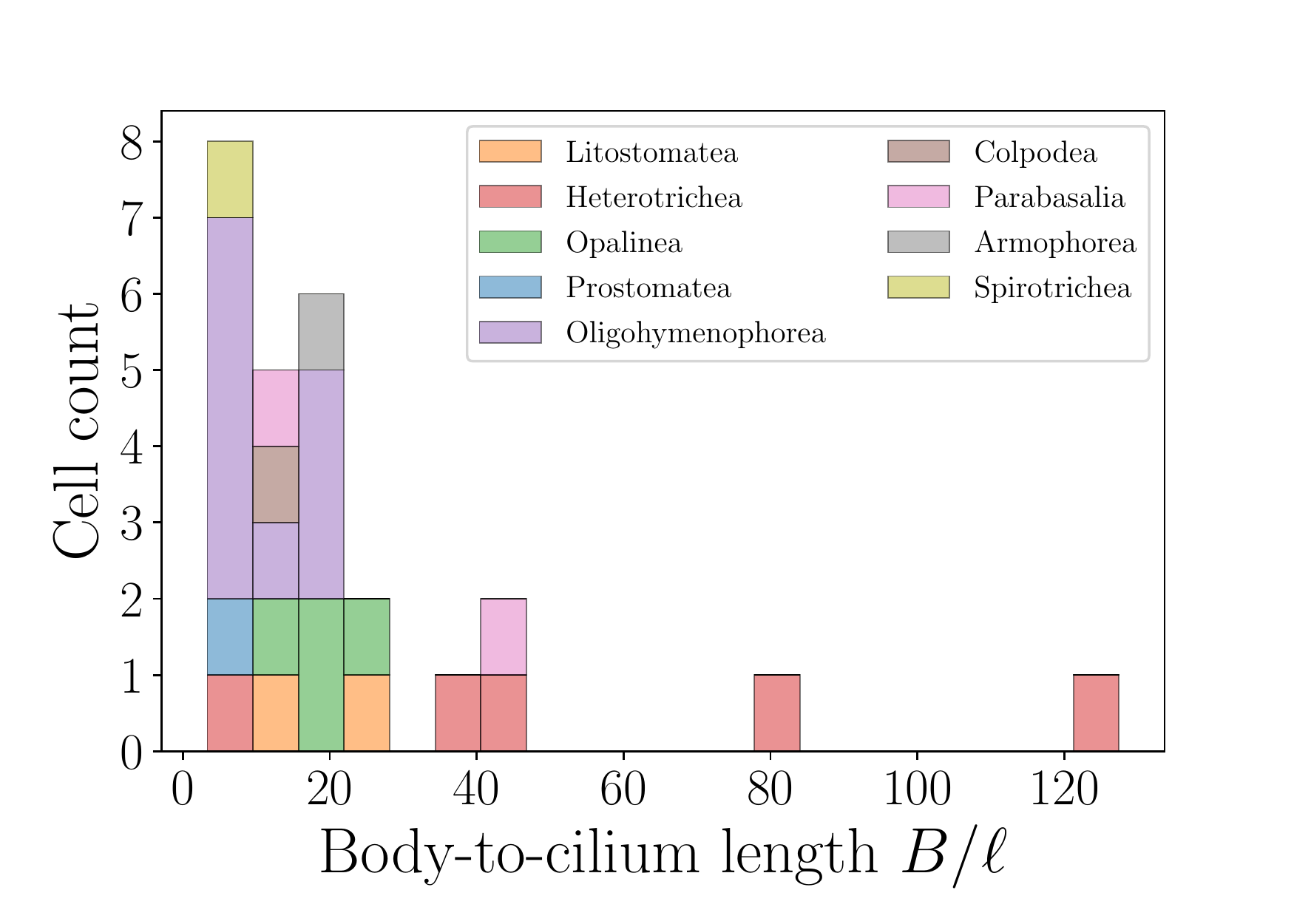}
\caption{Histograms of aspect ratios $W/B$ (left) and body-to-cilium length $B/\ell$ (right) for ciliates. Most of the cells are prolate, with the mean aspect ratio {$\langle W/B \rangle = 0.49\pm0.22\ (n=86)$}. The size-to-flagellum length ratios have average values {$\langle B/\ell \rangle = 23.13\pm 27.03\ (n=26)$}.
\label{histBnL_cil_AR}}
\end{figure}

The distribution of aspect ratios of the cells, along with the body-to-cilia lengths, are shown in Fig.~\ref{histBnL_cil_AR}. The former peaks at the mean value of about 0.5, indicating prolate cell bodies. The  large   values of the body-to-cilium length ratios confirm that cilia {take the form of} tiny hairs covering the cell body, much smaller than the body itself. This  {in turn} justifies coarse-grained  modelling approaches representing the cell body as a continuous surface capable of exerting stress, thereby locally averaging the collective motion of  individual cilia.

 \subsection{Models for ciliary propulsion}

{In search of means to {organise our data} on the locomotion of ciliates,}
{we {propose} below three {distinct} ciliary propulsion models  that {each} assume a different property to be constant {among the cells}}
{during forward swimming.}
{These {three   approaches model  the swimming of the cells as  induced by:} {\bf (A)}  a constant tangential stress exerted  on the cell surface by the cilia array; {\bf(B)} a  constant force exerted by each individual cilium on the fluid;  {\bf(C)} a constant effective fluid speed induced near the cell surface by the cilia.}

We model a ciliate cell as a prolate spheroid of length $B$ and diameter $W$. We set the $x$-axis along the {long axis of the cell},  taken to also be the direction of movement. The ciliate swimming with speed $U$ along $x$ is then subject to a viscous drag of  magnitude
\begin{equation}
  D=3\pi\eta B C_{FB} U,
\label{eq:Dcil}\end{equation}
with the geometry-dependent coefficient $C_{FB}$   in Eq.\eqref{eq:Cfb}. Balancing this drag    with the propulsive force generated by the collective action of the cilia yields different models for the swimming speed $U$, according to how one exactly models the propulsive force.

{Some aspects of the  mathematical description of the cell will be useful in what follows}.
A cross-section of the spheroid containing $\mathbf{e}_x$ is an ellipse of eccentricity $e=\sqrt{1-(W/B)^2}$. Every point of the ellipse can be parametrised in polar coordinates by
\begin{equation}
  r(\theta)=\frac{W}{2\sqrt{1-(e\cos\theta)^2}},
\label{eq:rell}\end{equation}
with the origin placed at the centre between its foci. Every point on the surface of the spheroidal body can then be written using spherical coordinates as
\begin{equation}
  \mathbf{x}(\theta, \varphi)=r(\theta)\left[ \cos\theta\,\mathbf{e}_x + \sin\theta \left(\cos\varphi\,\mathbf{e}_y + \sin\varphi\,\mathbf{e}_z\right)\right],\ \theta \in [0,\pi],\ \varphi \in [0,2\pi[.
\label{eq:xell}\end{equation}
 One may thus write the axisymmetric, unit vector tangential to the spheroidal surface and pointing along the polar angle as
\begin{equation}
  \mathbf{t}(\theta)=\frac{1}{\sqrt{r(\theta)^2+r'(\theta)^2}}\left[r'(\theta)\mathbf{e}_r + r(\theta)\mathbf{e}_\theta\right],
\label{eq:tell}\end{equation}
where $r'(\theta)=\mathrm{d}r(\theta)/\mathrm{d}\theta$. Finally, an infinitesimal surface element on the spheroidal surface is given by
\begin{equation}
  \mathrm{d}S=r(\theta)\sin\theta\sqrt{r(\theta)^2 + r'(\theta)^2}\, \mathrm{d}\theta\ \mathrm{d}\varphi.
\label{eq:dS}\end{equation}

Let then  $\mathbf{x}$ be a given point on the spheroidal surface, Eq.~\eqref{eq:xell}. The probability of having a cilium in an area $\mathrm{d}S$ around $\mathbf{x}$ is denoted by $p(\mathbf{x})\mathrm{d}S$, and we take the probability density to be uniform by setting $p(\mathbf{x})=1/\mathscr{S}$ for every $\mathbf{x}$ of Eq.~\eqref{eq:xell}, where
\begin{equation}
  \mathscr{S}=\int_0^{2\pi} \int_0^\pi r(\theta)\sin\theta\sqrt{r(\theta)^2 + r'(\theta)^2} \mathrm{d}\theta\ \mathrm{d}\varphi=\frac{\pi W^2}{2}\left[1+\frac{\arcsin e}{e\sqrt{1-e^2}}\right]
\label{eq:Sell}\end{equation}
is the surface area of the spheroid.

In order to proceed,  we {now} need to balance  the drag force with ciliary propulsion, and thus need to specify the details of the propulsion mechanism.

\paragraph{(A) Constant tangential stresses.}

The simplest model for ciliary propulsion assumes that the array of cilia  exerts a constant, axisymmetric stress of magnitude $\tau$ along the tangent vector $\mathbf{t}$. Using Eqs.~\eqref{eq:tell} and \eqref{eq:dS}, the total propulsive force can then be written as
\begin{equation}
  P_\tau=\int_S \tau (-\mathbf{t}\cdot\mathbf{e}_x) \mathrm{d}S=\tau\,\mathcal{I}_\mathbf{t}(B,W),
\label{eq:Ptau}\end{equation}
{with a purely geometric} factor given by
\begin{equation}
  \mathcal{I}_\mathbf{t}(B,W)=2\pi\int_0^\pi \left[(r(\theta)\sin\theta)^2 - r(\theta)r'(\theta)\sin\theta\cos\theta\right]\mathrm{d}\theta.
\label{eq:Itcil}\end{equation}

Balancing the propulsion  $P_\tau$ from Eq.~\eqref{eq:Ptau} with the drag $D$ given by Eq.~\eqref{eq:Dcil} and solving for the swimming speed $U_\tau$  leads then to the theoretical model
\begin{equation}
  U_\tau=\tau \frac{\mathcal{I}_\mathbf{t}}{3\pi \eta B C_{FB}}.
\label{eq:Utau}\end{equation}

\paragraph{(B) Constant force per cilium.}

{In the second modelling approach}, one may imagine  that each cilium, whose base lies at the point $\mathbf{x}(\theta,\varphi)$, {exerts} a constant force $F$ {along the} tangent vector $\mathbf{t}$. One cilium then contributes a local thrust along $x$ of magnitude
\begin{equation}
  F (-\mathbf{t})\cdot\mathbf{e}_x=\frac{F}{\sqrt{r(\theta)^2+r'(\theta)^2}}\left[r(\theta)\sin\theta-r'(\theta)\cos\theta\right].
\label{eq:Tcil}\end{equation}
{If the ciliated cell possesses $N$ such cilia, uniformly distributed over its surface,}
{the central limit theorem establishes the total propulsive force to be}
\begin{equation}
  P_F=F\frac{N}{\mathscr{S}} \mathcal{I}_\mathbf{t}(B,W).
\label{eq:PF}\end{equation}
After balancing with the drag, this leads to the ciliary swimming speed $ U_F$ predicted by this model as
\begin{equation}
  U_F=F\frac{N\mathcal{I}_\mathbf{t}}{3\pi \eta B\mathscr{S}C_{FB}}.
\label{eq:UF}\end{equation}

\paragraph{(C) Constant surface velocity.}

The third modelling approach assumes that the local speed of the fluid induced by ciliary motion is {(almost)} constant. To quantify this hypothesis, consider a spheroidal cell with a prescribed tangential surface velocity distribution
{$\mathbf{u}_{\rm s}=u_{\rm s}(\zeta)\,\mathbf{t}$, where
$\zeta=\cos\theta$ and $\mathbf{t}$ is given by Eq.~\eqref{eq:tell}.
}

In this case, the Lorentz reciprocal theorem may be used to relate the propulsion speed $U_\text{s}$ of a squirming organism to the surface velocity distribution \cite{Leshansky2007,pohnl2020axisymmetric} by
\begin{equation}
	U_\text{s} = -\frac{\tau_0}{2} \int_{-1}^{1}\left( \frac{1-\zeta^2}{\tau_0^2 - \zeta^2}\right)^{1/2} u_\text{s}(\zeta)\,\mathrm{d}\zeta,
\end{equation}
{where $\tau_0 = 1/e =  1/\sqrt{1-(W/B)^2}$ is fixed by the morphology of the swimmer.}

Following past work~\cite{Leshansky2007}, if we take an almost uniform surface velocity distribution of the form
\begin{equation}
	u_\text{s}(\zeta) = - \tau_0 \hat u_\text{s} \left( \frac{1-\zeta^2}{\tau_0^2 - \zeta^2}\right)^{1/2},
\end{equation}
where the constant $\hat u_\text{s}$ sets the characteristic surface velocity scale, {we obtain a model for} the swimming speed {as given by} $U_\text{s}= \hat u_\text{s}\left[\tau_0^2 - \tau_0 (\tau_0^2-1)\coth^{-1}\tau_0\right]$, which may also be written in terms of the eccentricity $e$  as
\begin{equation}
	U_\text{s}(e) = \frac{\hat u_\text{s}}{e^2}\left(1-\frac{1-e^2}{e} \tanh^{-1} e\right).
\label{eq:Ucilu}\end{equation}
With this particular flow assumption, for a very slender cell body ($e\to 1$), $U_\text{s}\to \hat u_\text{s}$, while for a spherical cell ($e\to 0$) we get $U_\text{s}\to 2\hat u_\text{s}/3$, in agreement with the classical result~\cite{stone1996propulsion}.

\begin{figure}[t]
\centering
\includegraphics[width=0.85\columnwidth]{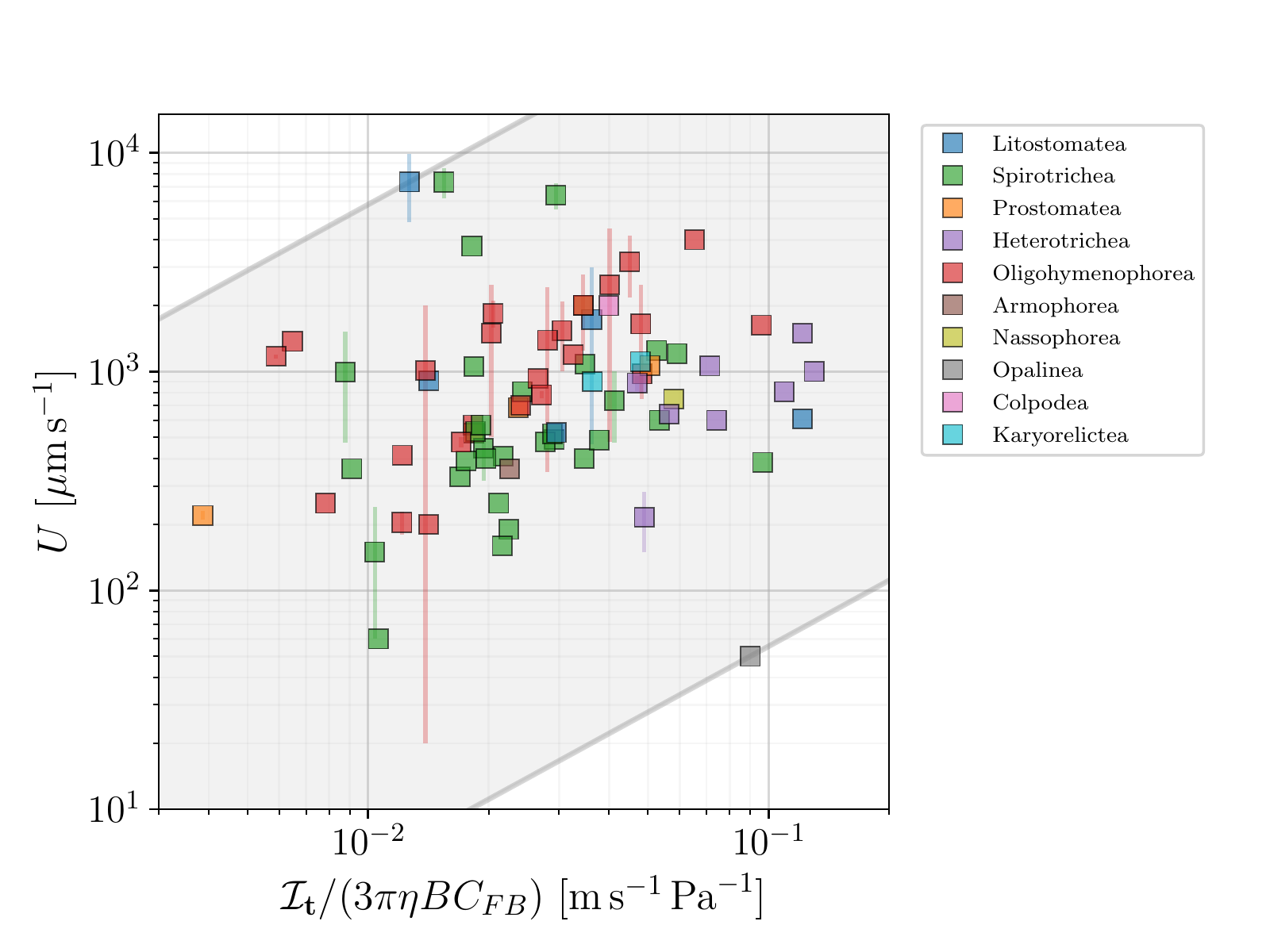}\
\caption{{The swimming speed $U$ for ciliates plotted versus the numerical factor accompanying the constant tangential stress assumed in model {\bf (A)} and Eq.~\eqref{eq:Utau}. The shaded area encloses all organisms and serves as an estimate of the average {effective} tangential stress for all organisms, with the lower bound of $\tau_\text{min}=0.55$ mPa, and the upper bound of $\tau_\text{max}=580$ mPa. Colours distinguish between classes of ciliated organisms. The scatter of data suggests that only a {large} range of values for the stress of individual organisms can be inferred.}
\label{fig:UcilSig}}
\end{figure}

\subsection{{Insights from  data}}

We begin with the constant tangential stress model {\bf (A)}, where the swimming speed is given by Eq.~\eqref{eq:Utau}.
{In Fig.~\ref{fig:UcilSig} we plot the measured speed for all ciliated species in our database versus the factors accompanying the tangential stress $\tau$ in Eq.~\eqref{eq:Utau}. The scatter of the data points {clearly} does not support the hypothesis of universal surface stress for all organisms. The model can however be used to estimate}
{the {effective} stress $\tau$ on the surface of each ciliate in the database. The shaded area represents the bounds for $\tau$, and fall in the {wide} range $0.55-580$~mPa.}
{These {values are consistent} with the estimate of $\tau\approx10$~mPa for \textit{Volvox} colonies~\cite{Short8315}.}

\begin{figure}[t!]
\centering
\includegraphics[width=0.85\columnwidth]{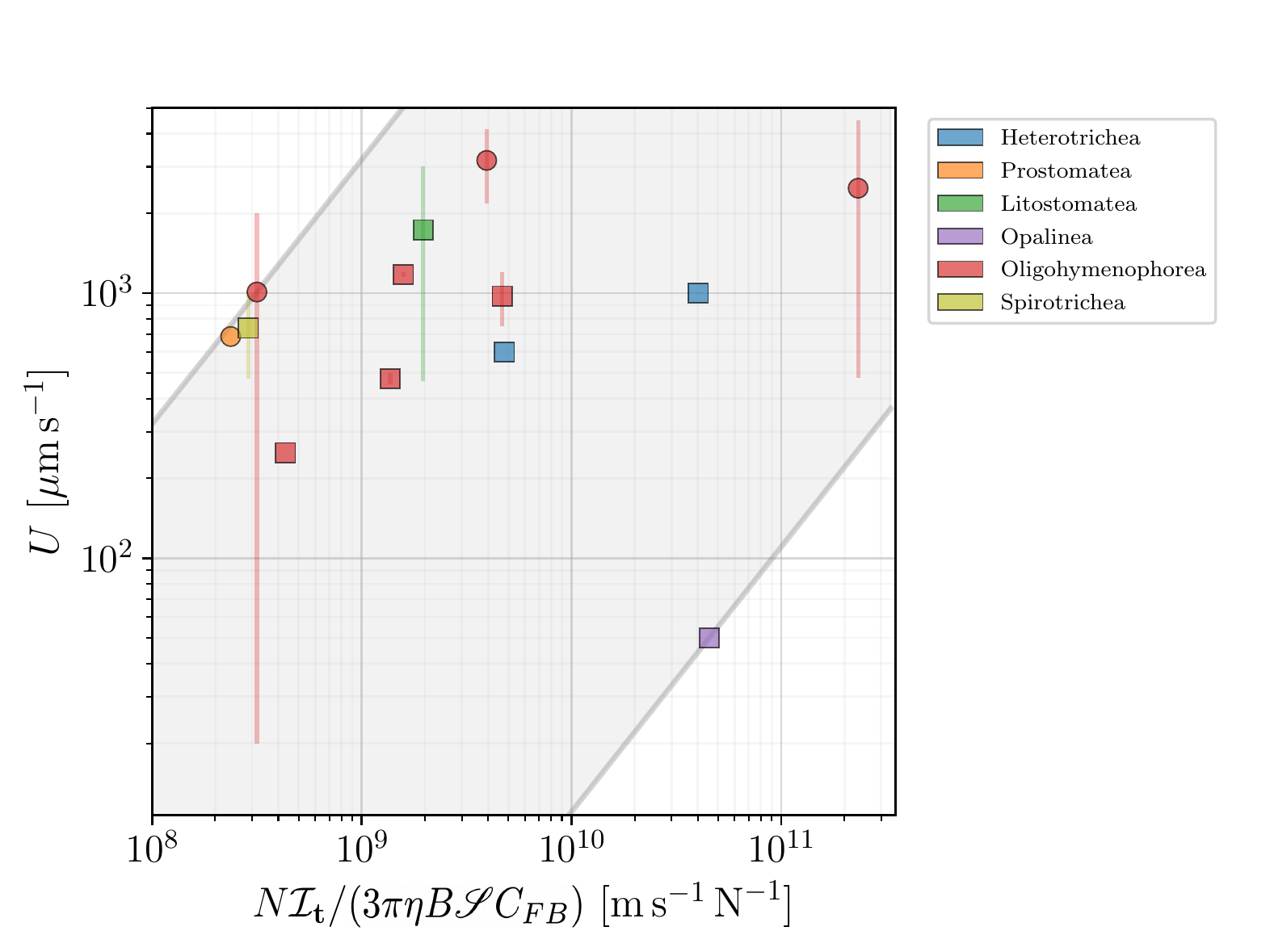}
\caption{ {Reported swimming speed $U$ plotted against the numerical prefactor of Eq.~\eqref{eq:UF}, assuming a constant {effective} force per cilium in the propulsion model {\bf (B)}. Square symbols mark organisms for which the prediction was directly calculated from the available data, while circles represent those for which we estimated the number $N$ of cilia. Colours distinguish the different taxonomic classes. The visible large scatter of data sets the bounds for the {effective} force per cilium to be in the range of $1.10\ 10^{-3}$ to 3.19~pN, represented by the shaded area in grey.}
}
\label{fig:UcilFN}
\end{figure}

{ A similar comparison for the `constant force per cilium' model {\bf (B)}, quantified by Eq.~\eqref{eq:UF}, requires the knowledge of the number of cilia $N$ for each swimmer. This number is, however, scarcely reported in literature, with only   9 values registered in our database. For some species, {however}, measurements report the number of cilia per unit area $\kappa$, or the distance between neighbouring cilia $d$. Using the latter, we can estimate the number of cilia per unit area to be  $\kappa_d\approx 1/d^{2}$. Using Eq.~\eqref{eq:Sell}, $\kappa$ and $N$ can be easily related via $N=\kappa\mathscr{S}$. By doing so, we determined $N$ (equivalently $\kappa$) for a total of 15 ciliated species out of 93, 13 of which had information about the cell swimming speed. In Fig.~\ref{fig:UcilFN} we plot the reported swimming speed versus the right-hand side of Eq.~\eqref{eq:UF} to estimate the {effective} force per cilium {$F$}. We report our estimated values of $F$ for each species in Table \ref{tab:F_cil}.}
{Our data encloses previous estimates in the range $0.3-1.0$~pN~\cite{Solari2006}, and show that the {effective} tangential forces {exerted} by each cilium may even be two orders of magnitude lower for species like {\it Opalina ranarum}. }

\begin{table}[h]
    \centering
    \begin{tabular}{l|l}
      \textbf{Species}  	&	$F$ [pN]	\\\hline\hline
      \textit{Blepharisma} sp. & $1.25\ 10^{-1}$\\
      \textit{Coleps hirtus} & 2.89\\
      \textit{Didinium nasutum} & $8.82\ 10^{-1}$\\
     \textit{Opalina ranarum} &$1.10\ 10^{-3}$\\
      \textit{Paramecium caudatum} & $1.07\ 10^{-2}$\\
      \textit{Paramecium multimicronucleatum} & $8.04\ 10^{-1}$\\
      \textit{Paramecium} spp. & $2.09\ 10^{-1}$\\
      \textit{Spirostomum} sp. & $2.49\ 10^{-2}$\\
      \textit{Stylonichia} sp. & 2.57\\
      \textit{Tetrahymena pyriformis} & $3.48\ 10^{-1}$\\
      \textit{Uronema marinum} & 3.19 \\
      \textit{Uronema} sp. & $7.43\ 10^{-1}$ \\
      \textit{Uronemella} spp. & $5.78\ 10^{-1}$\\\hline
    \end{tabular}
    \caption{Estimated values of the {effective} tangential force $F$ {exerted} by each cilium for the species in Fig.~\ref{fig:UcilFN}.    \label{tab:F_cil}}
\end{table}

\begin{figure}[h!]
\centering
\includegraphics[width=0.85\columnwidth]{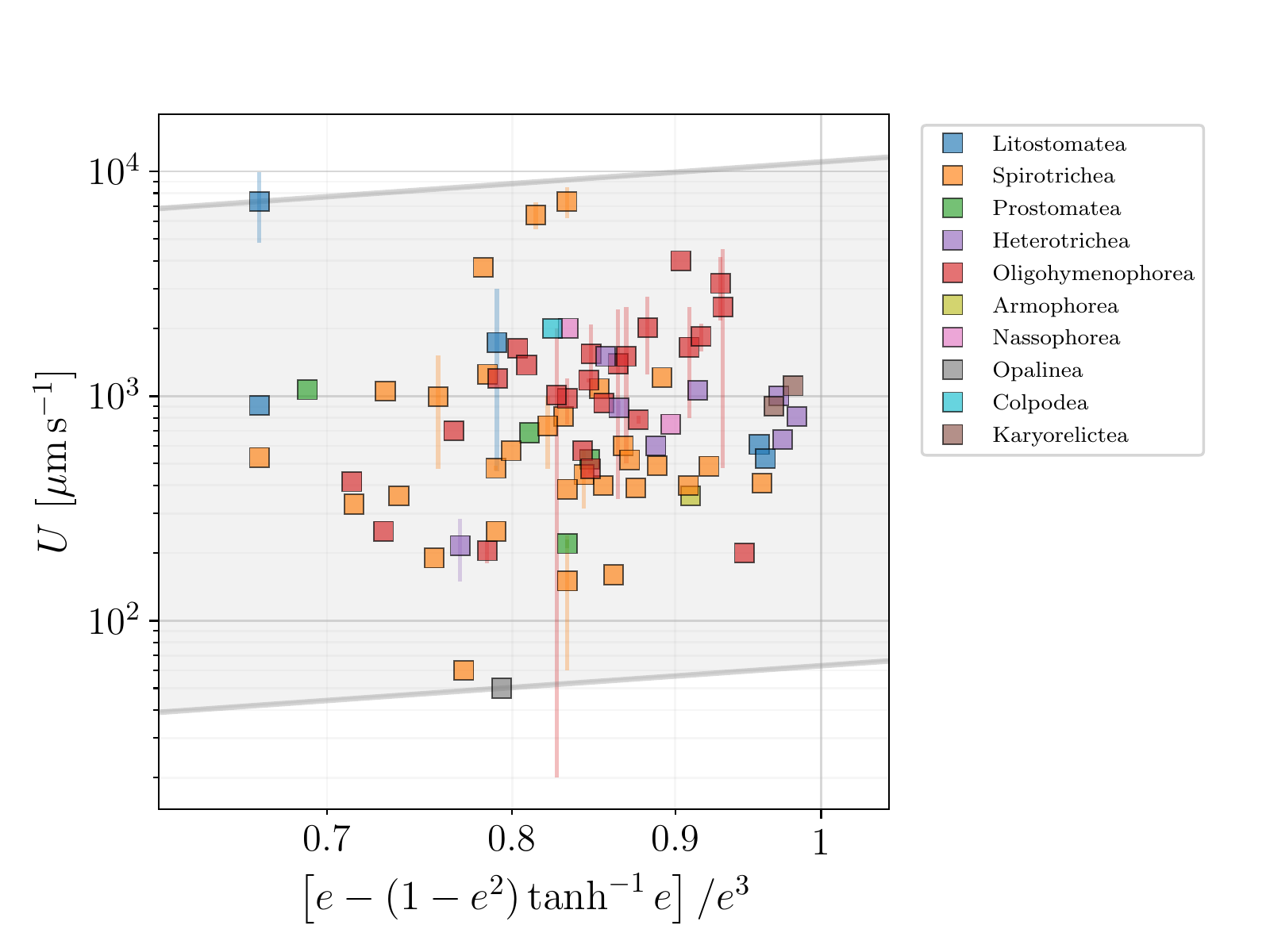}\
\caption{ {Reported swimming speeds {from our database} $U$ plotted against the geometric factor from Eq.~\eqref{eq:Ucilu} for the constant-flow model {\bf (C)}.
The data can be used to estimate the range of {effective} surface velocities to be in the range from $63.0\ \mu\textrm{m}\,\textrm{s}^{-1}$ to $1.10\ 10^4\, \mu\textrm{m}\,\textrm{s}^{-1}$, in the grey shaded area. Colours {allow to} distinguish different taxonomic classes.}
\label{fig:Ucilus}}
\end{figure}

{The third model {\bf (C)} assumes the creation of local flows by a}{n almost} {constant surface velocity,} {whose order of magnitude is fixed by $\hat{u}_s$.}
{The predictions of Eq.~\eqref{eq:Ucilu} suggest that}
{the swimming speed and the surface velocity are related by a simple geometric parameter,} {namely a function of the cell body eccentricity, $e$.
In Fig.~\ref{fig:Ucilus}  we plot the  measured ciliate velocities against the theoretical geometric factor determined for each species from our data.}
{The model can be used to estimate the magnitude of the {effective} average surface speed for each species. The resulting values span from a few tens of $\mu\text{m}\,\text{s}^{-1}$ to about $10^4$~$\mu\text{m}\,\text{s}^{-1}$. The average value  {of the effective}
surface velocity, calculated for all species, $\langle\hat u_\text{s}\rangle=1.42\ 10^{3}\,\mu$m$\,s^{-1}$, is about $2$ to $3$ times the average metachronal wave speed we estimate from our data, $\lambda_{MW} f$, where $\lambda_{MW}$ is the wavelength of the metachronal wave created by the collective ciliary beating at frequency $f$. }
{Here also our data confirm and extend previous estimates. For example in Ref.~}{\cite{L204},   tracking microscopy and fluid velocimetry were used to determine with precision the flow field of a freely swimming \textit{Volvox} colony, resulting in estimates of the surface speed   $\hat{u}_\text{s}\approx 200-250\ \mu$m$\,$s$^{-1}$ for species swimming at $U\approx 100-150\ \mu$m$\,$s$^{-1}$.}

 \section{\label{sec:disc} Conclusion}

\subsection{Summary and perspective}

In this paper, {based on an initial selection of six {seminal} papers in the field of biological fluid dynamics and  physics}, we  assembled a summary of the   experimental data produced to date on the characterisation of motile behaviour of unicellular microswimmers. {The material  gathered provides a convenient and practical reference point for future studies.} Our database includes empirical data on the motility of four categories of organisms, namely bacteria (and archaea), flagellated eukaryotes, spermatozoa and ciliates. Whenever possible, we reported the following {biological, morphological, kinematic and dynamical  parameters}: species, geometry and size of the organisms, swimming speeds, actuation frequencies, actuation amplitudes, number of flagella and properties of the surrounding fluid. {In all cases, we also give the}  appropriate references to {the} publications reporting the measurements. {We then analysed this information   by characterising  {some of the statistical properties} of the cells {in our database} and by introducing   theoretical models for each {main species in order} to establish guiding principles for the presentation of the data. We particularly focused on the dependence of the swimming speed on the characteristics of the swimmers and environmental properties. The analysis shows that qualitative trends established in the  theoretical framework {based on motility in} Stokes flows agree{s broadly} with the reported data but {that the large} degree of variability among species   precludes drawing general conclusions from the dataset.  {The modelling approaches can however}  be helpful} {in  rationalising  the data, pointing out the relevant dynamic quantities governing the locomotion of each individual group. {In particular, our} data confirm and extend estimates of these parameters previously reported in the literature.}

An important  result highlighted by our study is that a tremendous statistical variability exists in the available data, not only within domains~\cite{elife} but also within smaller taxonomic groups. Little is known about the variability of motility within individual species in Nature, neither in terms of their morphological characteristics (e.g.~size  and shape distribution), nor in the details of their propulsion (flagellar or ciliary  motion). In fact, for every single set in our database, it is not clear at  all  how representative any particular measurement  is of a group of similar organisms in the same environment?  How sensitive  are the propulsion characteristics of  these cells to  changes in the  environmental stimuli and how do they adapt to new conditions?  With the enhanced capacity to process large datasets and with new developments regarding automation of image analysis, the task of gathering and processing statistical data is becoming increasingly feasible, and  new works  will be able to  discover the fundamental principles dictating the locomotion of similar species within the same taxonomic group.

{The database in its current form, which is stored on the OSF repository~{\cite{Rodrigues_Lisicki_Lauga_2020}}, would benefit from the collaborative effort of the community. By growing further, it would help provide {up-to-date} information on the dynamics of {a variety of} organisms and   populations, {hopefully further encouraging collaborations between cell biologists and physical scientists}. To aid this process, our database is available on GitHub~\cite{BOSOgit}, where it can be extended and enriched.}

We gave our database the  appellation of BOSO-Micro. The first term stands for Bank Of Swimming Organisms while the second is  there to emphasise that we have focused our  work on microscopic unicellular organisms. We hope that new  versions of the  database, BOSO-X, will be built by focusing for X on different organisms. An obvious suggestion  would   be to assemble a  BOSO-Fish database, given the  large amount  of experimental, computational and theoretical knowledge on the swimming of fish. We hope that building exhaustive databases of this sort will further facilitate the work of physical scientists on biological problems related to locomotion.

\subsection{Caveats and limitations}

{The  collection of data {gathered in our database}  is inevitably incomplete and biased, in particular due to the way the initial set of literature sources, {focused on biophysical studies,} was chosen. Despite {our efforts to carry out} a broad search for swimming data, it is possible that important references were left out. The mitigation strategy in this case relies on making the database public~{\cite{Rodrigues_Lisicki_Lauga_2020}} and expandable}{~\cite{BOSOgit}.}

{Regarding the presented data, a major limitation is {of course} their sparseness. The relevant parameters in the description of motility are incomplete for many species, especially the variables related to the beating of cilia and
flagella, which
hinders direct comparison with theoretical models.}

{Furthermore}, {the database} {was populated using data} {presented across different papers, books, registers and reports, and the multiplicity of sources introduces a significant and inherent noise. For many species, reported measurements of one or more characteristics refer to different experimental environments. Even if those are reported, different strains of the same species may behave differently under slightly modified physical and chemical conditions.}

{It is also important to highlight the limitations and assumptions of the models used in our paper.} {The {models were}  designed to assist the presentation of data in the context of established   ideas regarding microscale locomotion, and to provide quick estimates of the }{relevant} {dynamic characteristics of microswimmers.} {Common to all the models is our assumption that the cell bodies are spheroidal. A look at Fig.~\ref{fig:Tree} quickly reveals that this hypothesis is a crude approximation for many species in our analyses (e.g. {\it Caulobacter crescentus}, {\it Ceratium tripos}, {\it Stentor}). We have made this choice in modelling in order to account for the influence of both the cell body length and width {in an analytical way}. The diversity of form, which might be crucial for certain locomotion strategies, has no reflection in the considered simplistic models, yet it must be incorporated into specific models describing particular organisms.} {Similarly, in the case of swimming eukaryotic cells, several of our hypotheses  on the flagellar beat ought to be {examined} carefully. For spermatozoa and flagellated eukaryotes,
we assumed the  form of a simple sinusoidal wave, whereas many species display more complex  {flagellar} beating patterns   (e.g.~complex waves displayed in {\it Columba livia} and {\it Sturnus vulgaris} spermatozoa). For flagellated eukaryotes, {we have neglected hydrodynamic interactions between flagella, which is a simplified approximation}. In the case of ciliates, the three models we have introduced also do not take into account   hydrodynamic interactions between neighbouring cilia, nor   the effect of the polarised beating of cilia and their recovery stroke.}  {Despite these limitations, we hope that the use of modelling may also prove useful in rationalising and organising future data on swimming organisms along similar lines.}

 \subsection*{Acknowledgments}
We deeply appreciate the help and suggestions of Dr.~Derek Scales. We also thank Masha Dvoriashyna, Christian Esparza-L\'opez, Ivan Tanasijevic, Maria T\u{a}tulea-Codrean and Albane Th\'ery for useful feedback.

 This project has received funding from the European Research Council (ERC) under the European Union's Horizon 2020 research and innovation program (grant agreement 682754 to EL), from the National Science Centre of Poland (grant Sonata no. 2018/31/D/ST3/02408 to ML) and from Campus France (Eiffel Scholarship no. 812884G to MFVR).

\newpage

\appendix
\section{\label{sec:data} The database of swimming microorganisms}
\nolinenumbers

In Table \ref{tab:symbols} we present a short glossary with the main symbols used in the database.

\begin{table}[h!]
    \centering
    \begin{tabular}{cp{9cm}c}
         \textbf{Symbol} & \textbf{Meaning} & \textbf{Unit}  \\\hline\hline
         $B$ & Body length & \m\\
         $W$ & Body width & \m\\
         $N$ & Number of flagella or cilia & - \\
         $L$ & Lengths (mostly flagella, otherwise specified) & \m \\
         $n_w$ & Number of waves (full periods, or crests) produced by flagellar beat & - \\
         $\lambda$ & Wavelength of flagellar waves (of helicoidal body and of metachronal waves indicated by a subscript $B$ and ${MW}$, respectively) & \m \\
         $\Lambda$ & Length of a complete wave along the flagellum (or path, indicated by subscript) & \m \\
         $h$ & Amplitude of waves (for helicoidal bodies, a subscript $B$ added) & \m \\
         $U$ & Swimming velocity & $\mu\text{m}\,\text{s}^{-1}$ \\
         $\omega$ & Flagellar beat frequency & s$^{-1}$\\
         $\Omega$ & Frequency of the rotation of cell body & s$^{-1}$ \\
         $c$ & Wave speed of flagellar beat (or metachronal wave) & $\mu\text{m}\,\text{s}^{-1}$ \\
         $V$ & Volume of cell body & \m$^3$ \\
         $\ell$ & Length of cilia & \m \\
         $d$ & Distance between cilia & \m \\
         $b$ & Radius of flagella & \m \\
         $\kappa$ & Number of cilia per unit area &  \m$^{-2}$ \\
         $f$ & Beating frequency of cilia & s$^{-1}$ \\
         $G$ & Gyration (frequency at which organisms revolve around the axis of movement) & s$^{-1}$ \\
         $\eta$ & Viscosity of the swimming medium & $\text{mPa}\,\text{s}$ \\\hline
    \end{tabular}
    \caption{List  of symbols used in the database, together with their  explanation and    units. }
    \label{tab:symbols}

\end{table}

For every entry in the database, in the case when more than one measurement was available, we report the average value and the standard deviation using the $\pm$ notation.  Values inside parentheses specify the range of the values measured,  e.g.~$(x_{\min}-x_{\max})$. Sometimes only the upper boundary was available, indicated by a preceding `$\max$'. When the information was not available in the texts of the articles, the figures or the graphics were analysed with the GNU Image Manipulation Program (GIMP) software in order to extract data. This is indicated in the tables by a superscript \f\ or \g\ respectively, if figures or graphics were used.

The various tables of data are organised as follows. Table \ref{tab:rod} contains the data for {78} organisms in the branch of bacteria (with 5 spiral-shaped bacteria included). Spirochaetes (18 species) and \textit{Spiroplasma} (2 species) were separated from the other bacteria because of their distinct mode of locomotion and are presented in Table \ref{tab:spiro}. The data for the 10 species of archaea are contained in Table \ref{tab:archaea}.

Eukaryotes have also been divided into three groups. The data for flagellated eukaryotes ({121} species) are presented first in Table \ref{tab:flageuk}, followed by spermatozoa (60 species) in Table \ref{tab:sperm} and finally ciliates ({93} species) in Table \ref{tab:cilia}.

\newgeometry{top=0.85in,left=0.5in,right=0.5in, footskip=0.35in, bottom=1in}
\pagestyle{tab}
\begin{sidewaystable}[H]
 \caption{Data for swimming bacteria (Spirochaetes and \textit{Spiroplasma} excluded)}
 \label{tab:rod}
 \scriptsize
\begin{tabular}{|p{2.15cm}|p{2cm}p{2cm}p{5cm}|p{2.15cm}p{5cm}|p{2.1cm}|}
    \hline \hline
       \textbf{Species} &\multicolumn{3}{c|}{\centering \textbf{Geometry}}&\multicolumn{2}{c|}{\centering\textbf{Kinematics}}&\textbf{References} \\[3pt]
       & \centering$B$ &\centering $W$ & Flagella & \centering$U$ & Notes & \\

\hline	\textit{Agrobacterium sanguineum}	&\centering &\centering &							& \centering\parbox[t]{2.15cm}{\centering $25.2$ \\ $(\max 35)$} &	\parbox[t]{5cm}{Mean run time $=0.11$ s with\\ acceleration $=138\ \mu$m$\,$s$^{-2}$.} &	\cite{marine}	\\

\hline	\parbox[t]{2.15cm}{$\alpha$-proteobacte-\\rium AB015520\nu}	&\centering &\centering &	\nu GenBank closest matching organism.	& \centering\parbox[t]{2.15cm}{\centering $17.25\pm4.05$ \\ $(\max 55)$} &	\parbox[t]{5cm}{Mean run time $=(0.19-0.21)$ s with\\ acceleration $=(96-124)\ \mu$m$\,$s$^{-2}$.} &	\cite{marine}	\\

\hline	\parbox[t]{2.15cm}{$\alpha$-proteobacte-\\rium KAT8 }	&\centering$4$ &\centering$1$ &		& \centering\parbox[t]{2.15cm}{\centering $22\ (\max28)$} &	\parbox[t]{5cm}{GenBank AF025321. $V=3.1$.\\Mean run time $=3.05$\,s. } &	\cite{Kiorboe2002}	\\

\hline	\textit{Alteromonas macleodii}	&\centering$(2-7)$ &\centering$0.4$ &	Monopolar flagellum.						& \centering\parbox[t]{2.15cm}{\centering $19\pm2.9$ \\ $(\max 55)$} &	\parbox[t]{5cm}{Mean run time $=(0.13-0.2)$ s with\\ acceleration $=(112-139)\ \mu$m$\,$s$^{-2}$.} &	\cite{marine,gauthier1977alteromonas,vantrappen2004alteromonas}	\\

\hline	\parbox[t]{2.15cm}{\textit{Arthrobacter}\\ \textit{histidinolovorans}}	&\centering &\centering $0.26$ &							& \centering\parbox[t]{2.15cm}{\centering $23.3$ \\ $(\max 55)$} &	\parbox[t]{5cm}{Mean run time $=0.19$ s with\\ acceleration $=166\ \mu$m$\,$s$^{-2}$.} &	\cite{marine}	\\
\hline
    \end{tabular}
\end{sidewaystable}

\begin{sidewaystable}[H]
 \scriptsize
    \begin{tabular}{|p{2.15cm}|p{2cm}p{2cm}p{5cm}|p{2.15cm}p{5cm}|p{2.1cm}|}
    \multicolumn{7}{c}{\normalsize{\textbf{Table \ref{tab:rod} continued:} Data for swimming bacteria (Spirochaetes and \textit{Spiroplasma} excluded)}}\\
    \hline
       \textbf{Species} &\multicolumn{3}{c|}{\centering \textbf{Geometry}}&\multicolumn{2}{c|}{\centering\textbf{Kinematics}}&\textbf{References} \\[3pt]
       & \centering$B$ &\centering $W$ & Flagella & \centering$U$ & Notes & \\
\hline	\textit{Curacaobacter baltica}	&\centering$2.5$ &\centering $1$ &							& \centering\parbox[t]{2.15cm}{\centering $21\ (\max30)$} &	\parbox[t]{5cm}{Average run length $=6.2\,$s. $V=2$.} &	\cite{Kiorboe2002}	\\

       \hline	\textit{Escherichia coli}	&\centering$	2.5\pm0.6	$&\centering$	0.88\pm0.09	$&	\parbox[t]{5cm}{Peritrichous flagella. Usually $N=6$\\ ($N=3.37\pm1.59$\nu\nd),	$L=8.3\pm2.0$\nu\nt  \\($L=7.3\pm2.4$\nu\nd),	$n_w\approx2$,	\\$\lambda=2.366\pm0.121$\nq,	$h=0.38$\nu\nd \f.} 		&\centering\parbox[t]{2.15cm}{\centering $	24.1\pm10$ \\ $(14.2-60)$\nc}&	$\omega=131\pm31$\nu\nd (also $\omega=270$\ns$)$, $\Omega=23\pm8$\nu\nd, $\omega_{\text{motor}}=154\pm30$\nu\nd. Additional parameters measured\nu: twiddle length $=0.14\pm0.19$ s, run length $=0.86\pm1.18$ s, change in direction from run to run $=68\pm36^\circ$, change in direction while running $=23\pm23^\circ$. \nu For wild-type strain AW405. \nd Observed in the presence of Alexa Fluor 532. \nt In the presence of motility buffer, at 23\dc. \nq Average from 4 strains. \nc Average from values of 10 articles. \ns For strain HCB437 at 32\dc.	&	\cite{139,183,32,L73,L74,L75,L123,L100,158,L115,L117,L129,L130,L136,L101,L218,L251, L227,L74,L165}\\

       \hline	\textit{Flavobacterium uliginosum}	&\centering &\centering $0.225\pm0.035$ &							& \centering\parbox[t]{2.15cm}{\centering $22.45\pm9.75$ \\ $(\max 65)$} &	\parbox[t]{5cm}{Mean run time $=(0.16-0.17)$ s with\\ acceleration $=(80-117)\,\mu$m$\,$s$^{-2}$.} &	\cite{marine,konishi1994}	\\
\hline
    \end{tabular}
    \end{sidewaystable}

    \begin{sidewaystable}[H]
    \scriptsize
    \begin{tabular}{|p{2.15cm}|p{2cm}p{2cm}p{5cm}|p{2.15cm}p{5cm}|p{2.1cm}|}
    \multicolumn{7}{c}{\normalsize{\textbf{Table \ref{tab:rod} continued:} Data for swimming bacteria (Spirochaetes and \textit{Spiroplasma} excluded)}}\\
    \hline
    \textbf{Species} &\multicolumn{3}{c|}{\centering \textbf{Geometry}}&\multicolumn{2}{c|}{\centering\textbf{Kinematics}}&\textbf{References} \\[3pt]
    & \centering$B$ &\centering $W$ & Flagella & \centering$U$ & Notes & \\

    \hline	\textit{Photobacterium profundum}	&\centering$	2.94	$\f&\centering$	1.76	$\f&	\parbox[t]{5cm}{Probably monopolar.	$L=9$\f,\\	$n_w\approx3.25$\f, $\lambda=2.22 $\f, $h=0.39$\f.}		&\centering$	(25.8$\nu$-28.2$\nd)&			Data for strain SS9. This species' motile behaviour was observed as a function of the pressure in the observation chamber $p$. \nu At 20\dc\ and $p=0.1$ MPa. \nd At 20\dc\ and $p=30$ MPa. Motility ceased when $p$ was superior to 170 MPa. For strain 3TCK: $U_{\max}=21.7\,$\m$\,s^{-1}$ at 20\dc\ and $p=0.1$MPa, no motility observed if $p>150$ MPa.	&	\cite{L136}	\\

\hline	\textit{Polaribacter irgensii}	&\centering$(0.8-48)$ &\centering $(0.25-0.5)$ &	Polar flagella.						& \centering\parbox[t]{2.15cm}{\centering $24.6$ \\ $(\max 55)$} &	\parbox[t]{5cm}{Mean run time $=0.23$ s with\\ acceleration $=178\,\mu$m$\,$s$^{-2}$.} &	\cite{marine,gosink1998polaribacter}	\\

\hline	\parbox[t]{2.15cm}{\textit{Pseudoalteromo}-\\\textit{nas citrea}}	&\centering$(1.5-4.5)$ &\centering $(0.41-1.5)$ &	Monopolar flagellum.						& \centering\parbox[t]{2.15cm}{\centering $32.2$ \\ $(\max 75)$} &	\parbox[t]{5cm}{Mean run time $=0.17$ s with\\ acceleration $=145\,\mu$m$\,$s$^{-2}$.} &	\cite{marine,gauthier1977alteromonas}	\\
\hline
    \end{tabular}
\end{sidewaystable}

\begin{sidewaystable}[H]
 \scriptsize
    \begin{tabular}{|p{2.15cm}|p{2cm}p{2cm}p{5cm}|p{2.15cm}p{5cm}|p{2.1cm}|}
    \multicolumn{7}{c}{\normalsize{\textbf{Table \ref{tab:rod} continued:} Data for swimming bacteria (Spirochaetes and \textit{Spiroplasma} excluded)}}\\
    \hline
       \textbf{Species} &\multicolumn{3}{c|}{\centering \textbf{Geometry}}&\multicolumn{2}{c|}{\centering\textbf{Kinematics}}&\textbf{References} \\[3pt]
       & \centering$B$ &\centering $W$ & Flagella & \centering$U$ & Notes & \\

       \hline	\textit{Sarcina ureae}	&\centering$	(1.97$\f$-4)	$&\centering$		$&	Peritrichous flagella. 	$L=(13.3-15.7)$\f,	$n_w\approx(2-5)$,	$\lambda_{\text{short}}=1.639\pm0.0054$, $\lambda_{\text{long}}=3.193\pm0.0048$,	$h=0.37$\f.		&\centering$	(18.75$\nu$-28.1$\nd$)	$&			Occurs in packets of 8 cocci with one flagellum per cell. \nu $\eta = 1.16$. \nd Strain ATCC 13881 at 20\dc. 	&	\cite{139, 183, L117, L217, L218}	\\

       \hline \parbox[t]{2.15cm}{\textit{Selenomonas}\\{\it ruminantium}}	&\centering$	3.98	$\f&\centering$	1.17	$\f&	$N=6\pm1.4$,					Three configurations of flagella were observed in function of pH and salt concentration: \underline{Coiled form}: Left-handed helix with $\lambda \approx0$, $h=0.965$, when pH $=(5-8)$ in the absence of salt. \underline{Normal}: Left-handed helix with $\lambda=4.7$, $h=0.965$, when pH $>8$, for every salt concentration. \underline{Large Curly}: Right-handed helix with $\lambda=4.84$, $h=0.93$, when pH $<5$. For pH $<3$ the flagella were disintegrated.	&\centering$	16\pm6	$&			\textit{S. ruminantium} subsp. lactilytica TAM6421 (NBRC103574) grown anaerobically at 37\dc.	&	\cite{L135}	\\

    \hline \parbox[t]{2.1cm}{\textit{Serpens}\\\textit{flexibilis}}	&\centering$	(8-12)	$&\centering$	(0.3-0.4)	$&	Bipolar and lateral. $N= (4-10)$ per tuft,	$L=(15-30)$,		$\lambda=0.88$,	$h=0.29$.		&\centering$	(1.11$\nu$-16$\nd$)	$&			\nu Strain PFR-1 at 30\dc\ with 0.5\% agar. \nd Strain PFR-3, $\eta=6.5$. 	&	\cite{L130, L132}	\\
\hline
    \end{tabular}
\end{sidewaystable}

\begin{sidewaystable}[H]
 \scriptsize
    \begin{tabular}{|p{2.15cm}|p{2cm}p{2cm}p{5cm}|p{2.15cm}p{5cm}|p{2.1cm}|}
    \multicolumn{7}{c}{\normalsize{\textbf{Table \ref{tab:rod} continued:} Data for swimming bacteria (Spirochaetes and \textit{Spiroplasma} excluded)}}\\
    \hline
       \textbf{Species} &\multicolumn{3}{c|}{\centering \textbf{Geometry}}&\multicolumn{2}{c|}{\centering\textbf{Kinematics}}&\textbf{References} \\[3pt]
       & \centering$B$ &\centering $W$ & Flagella & \centering$U$ & Notes & \\

       \hline \textit{Thiovulum majus}        &\centering\parbox[t]{2cm}{\centering$7\pm 3$\\$(5-25)$\f\nu}&         &        $N\approx100$ peritrichous flagella.        $L=2.3$\f.                                        &\centering\parbox[t]{2.15cm}{\centering$        330$\\ $(150-600)        $}&                        $\omega = (20-63)$, $\Omega= 6.6\pm 3.4$, cells were attached and exerted a force of $\approx40$ pN in the surrounding water. Cells swam in helical paths of $h_{\rm path}=(5-40)$, $\lambda_{\rm path}=(40-250)$, in periods of $(0.2-1)$ s. \nu Different sizes have been reported for different populations of the same species.   &        \cite{L121,L69v2C,N9,M12}       \\

       \hline	\parbox[t]{2cm}{\textit{Vibrio}\\\textit{alginolyticus}}&\centering$	1.92\pm0.46	$&\centering$	0.8\pm0.49	$&	\parbox[t]{5cm}{	$L=5.02\pm1.15$,	$n_w=2.76 (=L/\Lambda)$,\\	$\lambda=1.58\pm0.14$, $\Lambda=1.82\pm0.16$, 	\\$h=0.14\pm0.02$.}		&\centering$	(77$\nu$-147$\nd$)	$&			$\omega = (690$\nu$-1660$\nd$)$. Mutant YM42 grown in HI broth at 37\dc. \nu At 25\dc. \nd At 35\dc.	&	\cite{L76,L69v2B,L101,L118,L139}	\\

       \hline	\parbox[t]{2cm}{\textit{Vibrio}\\\textit{anguillarum}}	&\centering$	(1-3.28$\f$)	$&\centering$	(0.5-1$\f$)	$&\parbox[t]{5cm}{	Sheathed polar flagella.	$L>4.5$\f\nu, \\	$n_w>1$,	$\lambda=2.73$\f,	$h=0.3$\f.}		&\centering$	(25$\nd$-40$\nt$)	$&			Strain NB10. \nu Flagellum not entirely in the picture. \nd At 5\dc. \nt At 25\dc	&	\cite{L141,L148,L253}	\\
\hline \hline
    \end{tabular}

    \bigskip\bigskip

    \caption{Data for swimming Spirochaetes and \textit{Spiroplasma}}
    \label{tab:spiro}
    \begin{tabular}{|p{2.15cm}|p{2cm}p{2cm}p{5cm}|p{2.15cm}p{5cm}|p{2.1cm}|}
    \hline \hline
    \textbf{Species} &\multicolumn{3}{c|}{\textbf{Geometry}}&\multicolumn{2}{c|}{\textbf{Kinematics}}&\textbf{References}\\[3pt] & \centering$B$ & \centering$W$ & Helicoidal body & \centering$U$ & Notes & \\

    \hline        \parbox[t]{2.15cm}{\textit{Borrelia}\\{\it burgdorferi}}        &\centering\parbox[t]{2cm}{\centering$        (4-30)        $}&\centering$        (0.2-0.33)        $& \parbox[t]{5cm}{$N=2\times (7-11)$,                $n_w=(3-10)$,\\        $\lambda_B=(3.01\pm0.18-3.29\pm0.07)$, \\       $h_B=(0.41\pm0.02-0.77\pm0.03)$.}                &\centering$(4.25-6.8\pm2.4$\nu$)        $&                        $\omega=10.2$, $c=34.24$. \nu In BSK-II medium with no Ficoll ($\eta\approx\eta_{\rm water}$). The torque was estimated to be $2700$ pN nm.     &        \cite{L104,L105,L106,L69v4,N13,M8}       \\

    \hline \parbox[t]{2.15cm}{\textit{Brachyspira}\\{\it hyodysenteriae}} &  & 	&		$h=0.43\pm0.03$, $\lambda=2.84\pm0.1$. &	\centering $40\pm4$ &	Wild type. &	\cite{N4} \\

    \hline	\parbox[t]{2.1cm}{\textit{Brachyspira}\\\textit{pilosicoli}}	&\centering$	5.36	$\f&\centering$	0.24	$\f&				$\lambda_B=2.2$,	$h_B=0.32$.		&\centering\parbox[t]{2.15cm}{\centering$	5.9\pm1.7	$}&			$\omega=8.83\pm3.35$. Swine intestinal strain NKf1 grown at 37\dc.	&	\cite{L269,L279}	\\

    \hline	\parbox[t]{2.1cm}{\textit{Cristispira}\\\textit{balbianii}}	&\centering$	(22.5$\f$-80)	$&\centering$	(0.65$\f$-2)	$&	$N>100$,		$n_w=(3$\f$-3.5)$.	Parameters out of picture of the veil-like crista (the body loses its regular spiral form on fixation): $\lambda_B=6$ and	$h_B=1.52$.	(\cite{32} mentions irregular spiral shapes with $n_w\approx(2-5)$, amplitude of 8 \m\ and depth of 1.6 \m).	&\centering$		$&		 		&	\cite{31,32,120,L69v4,BandW}	\\
\hline
    \end{tabular}
\end{sidewaystable}

\begin{sidewaystable}[H]
    \scriptsize
    \begin{tabular}{|p{2.15cm}|p{2cm}p{2cm}p{5cm}|p{2.15cm}p{5cm}|p{2.1cm}|}
    \multicolumn{7}{c}{\normalsize{\textbf{Table \ref{tab:spiro} continued:} Data for swimming Spirochaetes and \textit{Spiroplasma}}}\\
    \hline
       \textbf{Species} &\multicolumn{3}{c|}{\centering \textbf{Geometry}}&\multicolumn{2}{c|}{\centering\textbf{Kinematics}}&\textbf{References} \\[3pt]
       & \centering$B$ &\centering $W$ & Helicoidal body & \centering$U$ & Notes & \\

       \hline	\parbox[t]{2.1cm}{\textit{Leptospira}\\\textit{biflexa}}	&\centering$	14	$\nu&\centering$	(0.14-0.15)	$&						When a cell translates, its anterior end is spiral-shaped ($h_B=0.3$, $\lambda_B=2.7$, $W=0.18$, ``S-end'') and the posterior end is hook-shaped (H-end). The anterior S-end is gyrated by the rotation of Periplasmatic Flagella (PF), and the coiled protoplasmic cylinder (PC) ($h_B=0.09$, $\lambda_B=0.7$, $W=0.14$) rotates in the opposite direction. The S- and H-shapes of the two ends are thought to be determined by the shape of PF.	&\centering$	14.2\pm2.9	$\nd&			The anterior S-end is left-handed and gyrates counter-clockwise ($\omega=74.4\pm33.6$\g\nd), which produces backward motion of the spiral wave. In contrast, the protoplasmic cylinder is right-handed and rotates clockwise ($\omega=135.6\pm22.8$\g\nd). The posterior H-end is approximately planar and rotates in the same direction as the S-end to allow the cell to translate without twisting ($\omega=20.97\pm12.58$\g\nd). \nu $L_{\text{H-end}} \approx L_{\text{S-end}}=3$ and $L_{PC}=8$. \nd $\eta = 0.86$. 	&	\cite{L230,31}	\\

    \hline	\parbox[t]{2.15cm}{\textit{Leptospira icte-}\\{\it rohemorrhagiae}}	&\centering\parbox[t]{2cm}{\centering$	7.5$\\ $(4-10)	$}&\centering$	(0.07-0.27)	$&	\parbox[t]{5cm}{$N=0$,			$\lambda_B=(0.3-0.6)$,\\	$h_B=(0.25-0.45)$.}		&\centering$		$&			At 19-25\dc.	&	\cite{32,L69v4,L248}	\\
\hline
    \end{tabular}
\end{sidewaystable}

\begin{sidewaystable}[H]
    \scriptsize
    \begin{tabular}{|p{2.15cm}|p{2cm}p{2cm}p{5cm}|p{2.15cm}p{5cm}|p{2.1cm}|}
    \multicolumn{7}{c}{\normalsize{\textbf{Table \ref{tab:spiro} continued:} Data for swimming Spirochaetes and \textit{Spiroplasma}}}\\
    \hline
       \textbf{Species} &\multicolumn{3}{c|}{\centering \textbf{Geometry}}&\multicolumn{2}{c|}{\centering\textbf{Kinematics}}&\textbf{References} \\[3pt]
       & \centering$B$ &\centering $W$ & Helicoidal body & \centering$U$ & Notes & \\

\hline	\parbox[t]{2.15cm}{\textit{Treponema}\\{\it denticola}}	&\centering$	7.7\pm0.94	$&\centering$	0.2\pm0.02	$&	\parbox[t]{5cm}{Flagellar arrangement 2:4:2,\\			$\lambda_B=1.23\pm0.15$,	$h_B=0.5\pm0.05$.	}	&\centering\parbox[t]{2.15cm}{\centering$	0$\nu\\ $(0.015-28.1)	$}&		\parbox[t]{5cm}{Strains ATCC 33520, ATCC35405 and ATCC35404 were examined. \nu \textit{T. denticola} are unable to translate unless suspended in a gel-like medium. Swimming speed is strongly dependent on viscosity and temperature, e.g.~for ATCC 33520: $U(\eta = 9.2, 25$\dc$)=2.53\pm0.34\, (1.65-$\\$4.85)$, $U(\eta =216, 25$\dc$)=6.31\pm1.25$\\$(3.33-8.63)$, $U(\eta=9.2, 35$\dc$)=5.48\pm$\\$1.25\,(3.63-8.64),\ U(\eta=140, 35$\dc$)=$\\$19.31\pm4.46\,(11.82-28.1)$.}	&	\cite{L108,L69v4,L128}	\\

\hline \parbox[t]{2.15cm}{\textit{Treponema}\\{\it pallidum}} &	\centering $(18.46-23.65)$\f &	\centering $0.59$\f &	\parbox[t]{5cm}{$N=2\times(2-4)$, $n_w\approx(8-10)$\f,\\ $\lambda_B=1.56\pm0.04$, $h_B=0.28\pm0.01$.}  &	\centering $1.9\pm0.2$ &	In CMRL medium with no Ficoll ($\eta=\eta_{\rm water}$). The torque was estimated to be $800$\,pN\,nm. &	\cite{M8}\\

\hline \parbox[t]{2.15cm}{\textit{Treponema}\\{\it primitia}} &	\centering $32.31$\nu &	\centering $0.35$ &	\parbox[t]{5cm}{$N=2$, $n_w\approx 2$, $\lambda_B=2.5$, $h_B=0.6$.\\ The cells looked like rigid helices at all times.} &	\centering $12$ &	$\omega=16.67$. \nu Estimated using the pitch and radius of a helix with two complete turns. &	\cite{M9} \\
\hline \hline
    \end{tabular}

\bigskip\bigskip

\scriptsize
    \caption{Data for swimming Archaea}
    \label{tab:archaea}
    \begin{tabular}{|p{2.15cm}|p{2cm}p{2cm}p{5cm}|p{2.15cm}p{5cm}|p{2.1cm}|}
    \hline \hline
    \multirow{2}{*}{\textbf{Species}} &\multicolumn{3}{c|}{\textbf{Geometry}}&\multicolumn{2}{c|}{\textbf{Kinematics}}&\multirow{2}{*}{\textbf{References}}\\[3pt] & \centering$B$ & \centering$W$ & Flagella &\centering $U$ & Notes & \\

    \hline	\textit{Halobacterium halobium}	&\centering$	(2$\nu$-5$\nd$) $&  &	Cells can have either a mono or bipolar tuft of flagella and display ``Super flagella".	$L=3.85\pm0.1$ ($L_\text{super}=22\pm2$),		$\lambda=1\pm0.02$,	$h=0.36\pm0.09$,		&\centering\parbox[t]{2.15cm}{\centering$	(1.59\pm0.39$\nt$-$\\$2.94\pm0.34$\nq$) 	$}&			Strains wild-type NRL, M407, M416 and Flx37 were examined. \nu Monopolar flagellated cells. \nd Bipolar flagellated cells. \nt Counterclockwise rotation of flagellar bundle. \nq Clockwise rotation of flagellar bundle. 	&	\cite{L213,L239}	\\

    \hline	\textit{Halobacterium salinarum}	&\centering\parbox[t]{2cm}{\centering$2.6\pm0.5$\\$(1.6$\f$-10)	$}&\centering$	0.43\pm0.07$\\$(\max1)	$&	\parbox[t]{5cm}{Cells can show either a mono- or bipolar tuft of flagella. 	$L=4.3\pm1$,		$\lambda=2.1\pm0.2$,\\	$h=0.22\pm0.03$.}	&\centering\parbox[t]{2.15cm}{\centering$	3.3\pm0.9$\\ $(\max 10)	$}&	\parbox[t]{5cm}{$\omega=23\pm5$, $\Omega=2.9\pm2.5$. Estimated\\power and torque required to rotate the archaellum are $7.7\,10^{-18}$\,W and $50$ pN\,nm, respectively. Temperature range of swimming: 20$-$65\dc.}	&	\cite{L73,32,L256,Kinosita2016}	\\

    \hline	\parbox[t]{2.15cm}{\textit{Methanocaldococ-} \textit{cus jannaschii}}	&\centering$	1.5			$&  &	Polar bundle of flagella.	$L<5$,	$n_w=2$.				&\centering\parbox[t]{2.15cm}{\centering$	380\pm40$\\$ (\max 589) 	$}&			Temperature range of swimming: 20-90\dc\ (optimal growth at 85\dc). In zigzag movement, $U=(50-100)$.	&	\cite{L73,L69v1,L214}	\\
\hline\hline
    \end{tabular}
    \end{sidewaystable}

\begin{sidewaystable}[H]
    \scriptsize
    \caption{Data for swimming flagellated eukaryotes}
    \label{tab:flageuk}
    \begin{tabular}{|p{2.15cm}|p{2cm}p{2cm}p{5cm}|p{2.15cm}p{5cm}|p{2.1cm}|}
    \hline\hline
       \textbf{Species} &\multicolumn{3}{c|}{\centering \textbf{Geometry}}&\multicolumn{2}{c|}{\centering\textbf{Kinematics}}&\textbf{References} \\[3pt]
       & \centering$B$ &\centering $W$ & Flagella & \centering$U$ & Notes & \\

\hline	\textit{Actinomonas mirabilis}	&\centering$	$&\centering $ $ &		&\centering\parbox[t]{2.15cm}{\centering$240$}&		$V=75$.	&	\cite{Hansen1997}	\\

       \hline	\textit{Alexandrium minutum}	&\centering$	21.7\pm2.2	$&\centering$		$&							&\centering\parbox[t]{2.15cm}{\centering$	(64\pm23$\nu$-$\\$320\pm84$\nd$)	$}&			\nu At 12\dc. \nd At 24\dc. Swimming speeds also available at 8, 16 and 20\dc.	&	\cite{L270}	\\

    \hline	\parbox[t]{2.15cm}{\textit{Alexandrium}\\{\it ostenfeldii}}	&\centering$	41.1\pm4.5	$&\centering$		$&							&\centering\parbox[t]{2.15cm}{\centering$	(66\pm19$\nu$-$\\$150\pm34$\nd$)	$}&			\nu At 12\dc. \nd At 16\dc. Swimming speeds also available at 8 and 20\dc	&	\cite{L270}	\\

    \hline	\textit{Alexandrium tamarense}	&\centering$	26.7\pm2.6	$&\centering$		$&							&\centering\parbox[t]{2.15cm}{\centering$	(108\pm44$\nu$-$\\$255\pm81$\nd$)	$}&			A. tamarense isolate SB50 appeared in doublet configuration and swam at $U=344\pm52\,(\max472)$ (compare: single cells $U=238\pm64\,(\max360)$). \nu At 12\dc. \nd At 24\dc. Swimming speeds also available at 8, 16 and 20\dc.	&	\cite{L270}	\\

    \hline	\textit{Amphidinium britannicum}	&\centering\parbox[t]{2cm}{\centering $	51.2\pm7$ \\ $ (39.5-69.9)	$} & \parbox[t]{2cm}{\centering $	36.3\pm4.6$ \\ $(30-53.9)	$} &	\parbox[t]{5cm}{$N=2$ (longitudinal and transverse):\\	$L_L\approx B$,		$\lambda_T=3.375$\f, 	$h_T=0.74$\f.}		&\centering\parbox[t]{2.15cm}{\centering$	72.85$\\$(14.1-123.3)	$}&			Algae from Biologische Anstalt Helgoland, ME76, (Herdman) Lebour, measured with Laser Doppler Spectroscopy. 	&	\cite{L154,L245}	\\
\hline

    \end{tabular}
\end{sidewaystable}

\begin{sidewaystable}[H]
    \scriptsize
    \begin{tabular}{|p{2.15cm}|p{2cm}p{2cm}p{5cm}|p{2.15cm}p{5cm}|p{2.1cm}|}
    \multicolumn{7}{c}{\normalsize{\textbf{Table \ref{tab:flageuk} continued:} Data for swimming flagellated eukaryotes}}\\
    \hline
       \textbf{Species} &\multicolumn{3}{c|}{\centering \textbf{Geometry}}&\multicolumn{2}{c|}{\centering\textbf{Kinematics}}&\textbf{References} \\[3pt]
       & \centering$B$ &\centering $W$ & Flagella & \centering$U$ & Notes & \\

       \hline	\textit{Chlamydomonas reinhardtii}&\centering$	(7-13)	$&\centering$		$&	$N=2$,	$L=(10-12)$.				Breaststroke beat.	&\centering\parbox[t]{2.15cm}{\centering$	92.27$\\ $(60-200)	$}&		$\Omega=(0.13-0.382)$, $\omega=53\pm5$.	Wild-type \textit{C. reinhardtii} (cc1690).	&	\cite{83,L161,L202,L203,L204,L250}	\\

    \hline	\textit{Chlamydomonas} sp. &\centering$	13	$&\centering$		$&\parbox[t]{5cm}{	$N=2$,	$L=35\,(\approx2.5\times B)$,	$n_w=1$,	$\lambda=$\\$6.3$,		planar DDW or rowing breaststroke.}	&\centering$	(61-65.4)	$&		$\omega=8$.		&	\cite{116,143,L96,83,L154,BandW}\\

    \hline	\textit{Codonosiga botrytis}&\centering$	15	$&\centering$	5	$&	$N=1$,	$L=30$.					&\centering$		$&	$c=500$,	$\omega=28$.		&	\cite{194,198,ws7}	\\

    \hline	\textit{Crithidia deanei}&\centering$	7.4\pm0.2	$&\centering$		$&		\parbox[t]{5cm}{$L=13.1\pm0.4$,	$n_w=0.77\,(=L/\Lambda)$,\\	$\lambda=11.7\pm0.2$,	$h=2.2\pm0.05$.}		&\centering$	45.6\pm1.5	$&	$c=466\pm12$,	$\omega=40.5\pm0.8$.	Cells were cultured at 28\dc, and examined at room temperature, 22\dc.	&	\cite{L170}	\\

    \hline	\parbox[t]{2.3cm}{\textit{Crithidia}\\\textit{fasciculata}}&\centering$	11.1\pm0.3	$&\centering$		$&		\parbox[t]{5cm}{$L=15.1\pm0.5$,	$n_w=0.94\,(=L/\Lambda)$,	\\$\lambda=11.6\pm0.2$,	$h=2.2\pm0.07$.}		&\centering$	54.3\pm2.6	$&	$c=680\pm28$,	$\omega=60\pm2.3$.	Cells were cultured at 28\dc, and examined at room temperature = 22\dc.	&	\cite{L170}	\\
\hline
    \end{tabular}
 \end{sidewaystable}

\begin{sidewaystable}[H]
    \scriptsize
    \begin{tabular}{|p{2.15cm}|p{2cm}p{2cm}p{5cm}|p{2.15cm}p{5cm}|p{2.1cm}|}
    \multicolumn{7}{c}{\normalsize{\textbf{Table \ref{tab:flageuk} continued:} Data for swimming flagellated eukaryotes}}\\
    \hline
       \textbf{Species} &\multicolumn{3}{c|}{\centering \textbf{Geometry}}&\multicolumn{2}{c|}{\centering\textbf{Kinematics}}&\textbf{References} \\[3pt]
       & \centering$B$ &\centering $W$ & Flagella & \centering$U$ & Notes & \\

    \hline	\parbox[t]{2.2cm}{\textit{Gyrodinium}\\\textit{aureolum}}&\centering$	(27-34)	$&\centering$		$&							&\centering$	(33-245) 	$&	 $G=(2-3)$.	Algae from University of Oslo, Hulburt, measured with Laser Doppler Spectroscopy. $V=250$.	&	\cite{L154,L160}	\\

    \hline	\parbox[t]{2.15cm}{\textit{Gyrodinium}\\{\it dorsum}}&\centering$	37.5\pm4.1	$&\centering$	31.3\pm3	$&	$N=2$,					either helical or planar beat.	&\centering\parbox[t]{2.15cm}{\centering$	324\pm43.8$\\ $(254-454)	$}&	\parbox[t]{5cm}{	$\Omega=(1.5-2.32)$.	At 20\dc. Swimming speed for cells with: short flagella $=240$\\$\pm47\,(120-316)$\,\m\,s$^{-1}$, no longitudinal flagella $=147\pm28.5\,(93-224)$\,\m\,s$^{-1}$.}	&	\cite{99,97,83,L64,L158,BandW}	\\

    \hline	\parbox[t]{2.15cm}{\textit{Gyrodinium}\\{\it dorsum}}&\centering$	34.5	$&\centering$	24.5	$&	$N=1$.					Planar beat.	&\centering$	148.35	$&		$\Omega=8.2$	(This \textit{Gyrodinium} has no helical flagella).	&	\cite{99,BandW}	\\

    \hline	\parbox[t]{2.15cm}{\textit{Hemidinium}\\{\it nasutum}}&\centering\parbox[t]{2cm}{\centering$	26.8$\\ $(24.4$\f$-30)	$}&\centering\parbox[t]{2.15cm}{\centering$	17.1$\\ $(15.2$\f$-20) 	$}&							&\centering$	105.6	$&		$\Omega=1$.		&	\cite{L158,L171,L233}	\\

    \hline	\parbox[t]{2.15cm}{\textit{Hemiselmis}\\{\it simplex}}&\centering$	(4-6.5)	$&\centering$	3	$&	$N=2$.						&\centering\parbox[t]{2.15cm}{\centering$	260$\\ $(200-450)	$}&		$G=(7-10)$,	$V=10$.	&	\cite{L160,L165,L167}	\\
\hline
    \end{tabular}
 \end{sidewaystable}

\begin{sidewaystable}[H]
    \scriptsize
    \begin{tabular}{|p{2.15cm}|p{2cm}p{2cm}p{5cm}|p{2.15cm}p{5cm}|p{2.1cm}|}
    \multicolumn{7}{c}{\normalsize{\textbf{Table \ref{tab:flageuk} continued:} Data for swimming flagellated eukaryotes}}\\
    \hline
       \textbf{Species} &\multicolumn{3}{c|}{\centering \textbf{Geometry}}&\multicolumn{2}{c|}{\centering\textbf{Kinematics}}&\textbf{References} \\[3pt]
       & \centering$B$ &\centering $W$ & Flagella & \centering$U$ & Notes & \\

    \hline	\textit{Ochromonas malhamensis}&\centering$	3	$&\centering$		$&	\parbox[t]{5cm}{$N=1$,		$n_w=2.8$,	$\lambda=7$,	$h=1$.\\	Mastigonemes. Planar DDW.}	&\centering$	(55-60)	$&		$\omega=68.44$.	At 18\dc.	&	\cite{107,116,BandW}\\

    \hline	\textit{Ochromonas minima}&\centering$	(3.5-6.5)	$&\centering$		$&	$N=2$,	$L_1=(1-2)$ and $L_2=B/3$.					&\centering$	75	$&		 $G=(0.25-1.25)$, $V=25$.	&	\cite{L160,L167}	\\

    \hline	\parbox[t]{2.15cm}{\textit{Olisthodiscus}\\{\it luteus}}&\centering$	(15-30)	$&\centering$		$&							&\centering\parbox[t]{2.15cm}{\centering$	140$\\ $(20-160)	$}&		 $G=(0.5-1)$. Algae from the University of Oslo, N. Carter, measured with Laser Doppler Spectroscopy. $V=600$.	&	\cite{L154,L160}	\\

    \hline	\parbox[t]{2.2cm}{\textit{Oxyrrhis}\\ \textit{marina}}&\centering$	(28.2-50.8) 	$&\centering$		$&	$N=2$ (longitudinal and transverse).						&\centering\parbox[t]{2.15cm}{\centering$	300\pm134$\\ $(90-700)	$\nu}&		$\Omega=9.3$.	Cells swim in a helicoidal path with $h_{\text{path}}=18$ and $\lambda_{\text{path}} = 108$, with $\omega_{\text{normal}} = 9.8$. Speed increased slightly in the presence of food cells. \nu Average of 7 registered values.	&	\cite{L143,L166,L179,L173}	\\
\hline
    \end{tabular}
 \end{sidewaystable}

\begin{sidewaystable}[H]
    \scriptsize
    \begin{tabular}{|p{2.15cm}|p{2cm}p{2cm}p{5cm}|p{2.15cm}p{5cm}|p{2.1cm}|}
    \multicolumn{7}{c}{\normalsize{\textbf{Table \ref{tab:flageuk} continued:} Data for swimming flagellated eukaryotes}}\\
    \hline
       \textbf{Species} &\multicolumn{3}{c|}{\centering \textbf{Geometry}}&\multicolumn{2}{c|}{\centering\textbf{Kinematics}}&\textbf{References} \\[3pt]
       & \centering$B$ &\centering $W$ & Flagella & \centering$U$ & Notes & \\

    \hline	\parbox[t]{2.3cm}{\textit{Peridinium}\\\textit{umbonatum}}&\centering$	28\,(25-35)	$&\centering$	23\,(21-30)	$&							&\centering$	250	$&		$\Omega=1.67$.		&	\cite{L158,L172,L233}	\\

    \hline	\textit{Phaeocystis pouchetii}&\centering$	 (4.5-8)	$&\centering$		$&	$N=2$,	$L=1.5\times B$.					&\centering$	(21-155)	$&			Algae from Biologische Anstalt Helgoland, ME64, (Harlot) Lagerh., measured with Laser Doppler Spectroscopy. 	&	\cite{L154,L167}	\\

    \hline	\textit{Polytoma uvella} &\centering\parbox[t]{2cm}{\centering$	18.25$\nu \\$(15-30)	$}&\centering$	(9-20)	$&	$N=2$,	$L\approx B$,		$\lambda=15$,	$h=2.9$\f.	Planar DDW and rowing breaststroke beating.	&\centering\parbox[t]{2.15cm}{\centering$	103.7$\nu \\$(74.8-127)	$}&	$c>312$,	$\Omega=(3-4)$, $\omega=11.7$\nu\,$(7-20)$.	At 20-22\dc. \nu Average of the different values registered.	&	\cite{40,85,143,83,L96,116,133,BandW}\\

    \hline	\parbox[t]{2.3cm}{\textit{Polytomella}\\\textit{agilis}}&\centering$	(9.8-15)	$&\centering$	(4.9-9)	$&	$N=4$,	$L=(8-9)$,				Planar DDW and rowing breaststroke.	&\centering$	(80-220)	$&	$c=(90-450)$,	$\omega=(7-33)$.	At 20-22\dc.	&	\cite{84,85,116,83,L161,BandW}\\

    \hline	\textit{Poteriodendron} sp. &\centering$		$&\centering$		$&		$L=35$,		$\lambda=4$,	$h\approx2$.	Planar beat.	&	\centering Sessile	&		$\omega=40$.	At 20\dc.	&	\cite{L65}\\
\hline
    \end{tabular}
\end{sidewaystable}

\begin{sidewaystable}[H]
    \scriptsize
    \begin{tabular}{|p{2.15cm}|p{2cm}p{2cm}p{5cm}|p{2.15cm}p{5cm}|p{2.1cm}|}
    \multicolumn{7}{c}{\normalsize{\textbf{Table \ref{tab:flageuk} continued:} Data for swimming flagellated eukaryotes}}\\
    \hline
       \textbf{Species} &\multicolumn{3}{c|}{\centering \textbf{Geometry}}&\multicolumn{2}{c|}{\centering\textbf{Kinematics}}&\textbf{References} \\[3pt]
       & \centering$B$ &\centering $W$ & Flagella & \centering$U$ & Notes & \\

    \hline	\parbox[t]{2.3cm}{\textit{Scrippsiella}\\\textit{trochoidea}}&\centering\parbox[t]{2cm}{\centering$	25.3 \pm 2.4$\\ $(\max 35)	$}&\centering$	19.9 \pm 2.1	$&		&\centering\parbox[t]{2.15cm}{\centering$	82$\\ $(22.2-153)	$}&			$V=3600$. Algae from Biologische Anstalt Helgoland, ME64, (Stein) Loeblich, measured with Laser Doppler Spectroscopy. 	&	\cite{L64,L154,L155,Hansen1997}	\\

    \hline	\textit{Spumella} sp. &\centering$	10	$\nu&\centering$		$&							&\centering$	25\pm2	$& \nu Equivalent Spherical diameter. (Synonym of \textit{Monas} O. F. M{\"u}ller 1773 and of \textit{Heterochromonas} Pascher 1912)		&	\cite{L180,L167}	\\

    \hline	\textit{Teleaulax} sp. &\centering$	(12-15)	$&\centering$		$&	$N=2$,	$L\approx0.6\times B$\f\nu.					&\centering$	(53-56)	$&		\parbox[t]{5cm}{Behaviour in the presence of the predator \textit{Oxyrrhis marina}. Prior to encounter with predator: $U=(61-76)$,\\ post-encounter: $U=(133-143)$. \\\nu For \textit{T. acuta}}	&	\cite{L173,L167}	\\
\hline \hline
    \end{tabular}
\end{sidewaystable}

\begin{sidewaystable}[H]
    \scriptsize

    \caption{Data for spermatozoa}
    \label{tab:sperm}
    \begin{tabular}{|p{2.15cm}|p{2cm}p{2cm}p{5cm}|p{2.15cm}p{5cm}|p{2.1cm}|}
    \hline \hline
    \textbf{Species} &\multicolumn{3}{c|}{\textbf{Geometry}}&\multicolumn{2}{c|}{\textbf{Kinematics}}&\multirow{2}{*}{\textbf{References}}\\[3pt] & \centering$B$ & \centering$W$ & Flagella &\centering $U$ & Notes & \\

    \hline	(Cricket)	&\centering$	110\pm10	$&\centering$		$&		$L=870\pm31.6$,		$\lambda\approx20$,	$h=0.9$\nu. 		&\centering$		$&		$\omega=13.3\pm3.4$\nu.	\nu In basic suspension medium at $18.5\pm0.5$\dc\ and $\eta=1.2$.	&	\cite{L186}	\\

    \hline        (Guinea Pig)        &\centering$        10.86$\f&\centering$        9.68 $\f&                $L=108.55$ ($L_\text{midpiece}=11.5$).                                        & \centering $9.48\pm0.40$ &                                &        \cite{L223,N3}      \\

    \hline        (Rabbit)        &\centering$         (8.06-8.51)        $&\centering$        (4.59-4.98)        $&         \parbox[t]{5cm}{       $L=(46-49.51)$ ($L_\text{midpiece}=8.81$),\\ $\lambda=41.6\pm4.2$\nu, $h=3.3\pm0.3$\nu.}                                       &\centering\parbox[t]{2.15cm}{\centering$        (101\pm7$\nd$-$\\$272\pm14$\nt$)         $}&        $\omega_{\text{shallow}}=17\pm0.9$, $\omega_{\text{deep}}=18\pm0.9$. New Zealand white rabbit spermatozoa at 37\dc. \nu From tracings, using the values of $B,\ W$ and $L$. \nd Average path velocity in shallow slide (25 \m) with ampullar fluid. \nt Head velocity in deep slide (100 \m) with ampullar fluid. Cells diluted in BO medium were also studied.        &        \cite{164,L112,L174,M37}        \\
\hline
    \end{tabular}
\end{sidewaystable}

\begin{sidewaystable}[H]
    \scriptsize
    \begin{tabular}{|p{2.15cm}|p{2cm}p{2cm}p{5cm}|p{2.15cm}p{5cm}|p{2.1cm}|}
    \multicolumn{7}{c}{\normalsize{\textbf{Table \ref{tab:sperm} continued:} Data for spermatozoa}}\\
    \hline
       \textbf{Species} &\multicolumn{3}{c|}{\centering \textbf{Geometry}}&\multicolumn{2}{c|}{\centering\textbf{Kinematics}}&\textbf{References} \\[3pt]
       & \centering$B$ &\centering $W$ & Flagella & \centering$U$ & Notes & \\

       \hline	\textit{Colobocentrotus} (sea urchin)	&\centering$	7.17\pm0.13	$\f&\centering$	3.1\pm0.36	$\f&	\parbox[t]{5cm}{	$L=(35.5$\f$-42)$, 	$n_w=(1.25-1.5)$,\\	$\lambda=20.9\pm3$\f,	$h=3.94\pm0.95$\f.\\	2D beat.}	&\centering$	(165.6-193.2) 	$&		\parbox[t]{5cm}{$\omega=46$.	At 23-26\dc, $\eta = 1.8$. The authors also studied the movement of ATP-reactivated sperm: $U=(73.6-83.2)$,\\ $\omega=32$.}	&	\cite{46,81,BandW}	\\

       \hline	\parbox[t]{2.15cm}{\textit{Columba livia}\\ (pigeon)}	&\centering$	16	$&\centering$		$&		\parbox[t]{5cm}{$L=132\pm11.1$ ($L_\text{midpiece}= 98.1\pm11.2$).\\				More complex than a helical wave, with consistent angular velocity always CW.}	&\centering$		$&				&	\cite{L244}	\\

       \hline	\parbox[t]{2.15cm}{\textit{Coturnix}\\\textit{coturnix} var.\\ \textit{japonica} (quail)}&\centering$		$&\centering$		$&	\parbox[t]{5cm}{	$L=208$ ($L_\text{midpiece}=161$).				Irregular\\ beat, decaying towards the end of the midpiece.}	&\centering$	(\max 50)	$&		$\Omega= \max 4$ (CW).	At 20-23\dc.	&	\cite{L244}	\\

       \hline	\textit{Culex} (mosquito)	&\centering$	(13.7$\nu$-14.6$\f$)	$&\centering$	0.48	$\nu&		$L>41.2$\nu\nd,	$n_w=3.3$,	$\lambda=15.5$,	$h=2.6$.		&\centering$	6.3	$\f&			\nu \textit{C. pipiens quinquefasciatus}. \nd Not entirely in the picture.	&	\cite{165,L229,BandW}	\\
\hline
    \end{tabular}
\end{sidewaystable}

\begin{sidewaystable}[H]
    \scriptsize
    \begin{tabular}{|p{2.15cm}|p{2cm}p{2cm}p{5cm}|p{2.15cm}p{5cm}|p{2.1cm}|}
    \multicolumn{7}{c}{\normalsize{\textbf{Table \ref{tab:sperm} continued:} Data for spermatozoa}}\\
    \hline
       \textbf{Species} &\multicolumn{3}{c|}{\centering \textbf{Geometry}}&\multicolumn{2}{c|}{\centering\textbf{Kinematics}}&\textbf{References} \\[3pt]
       & \centering$B$ &\centering $W$ & Flagella & \centering$U$ & Notes & \\

       \hline	\parbox[t]{2.15cm}{\textit{Lygaeus}\\ (milkweed bug)}	&\centering$	(4.8-5.24$\f$)	$&\centering$	(0.7$\f$-1)	$&		\parbox[t]{5cm}{$L>29.8$\nu,	$n_w=2.3$,	$\lambda=(13-14.5$\f$)$,\\	$h=(1.3$\f$-2.1)$.	}	&\centering$		$&		$\omega=(110-130)$.	\nu Not entirely in the picture and estimating the end of the head and beginning of the tail.	&	\cite{165,L183,BandW}	\\

       \hline	\parbox[t]{2.15cm}{\textit{Lytechinus}\\ (sea urchin)}	&\centering$	(5.1-7.55$\f$)	$&\centering$	(2.9-2.97$\f$) 	$&		\parbox[t]{5cm}{$L=43.5$,	$n_w=1.45$, $\Lambda=29.9$,\\	$\lambda=24.8$\nu\,$(22.6-30)$,	$h=(4.6-4.7)$.\\	2D beat.}	&\centering$	158	$&	\parbox[t]{5cm}{$c=(854-900)$,	$\omega=30$.	At 16\dc, \\$\eta = 1.4$. \nu Average of all registered values.}	&	\cite{41,46,70,L184,BandW}	\\

       \hline	\parbox[t]{2.15cm}{\textit{Megaselia}\\{\it scalaris} (fly)}	&\centering$	18.7\pm0.54	$&\centering$	0.16\pm0.01	$&	\parbox[t]{5cm}{	$L=128.7\pm4.09$. Characteristic\\ large and small waves:	$n_{w\text{large}}\approx1.1$,\\	$\lambda_{\text{large}}=(68-75$\f$)$, $\lambda_{\text{small}}=7$,\\	$h_{\text{large}}=(9.3-10.27$\f$)$, $h_{\text{small}}=0.5$.}		&\centering$	117.6\pm29.6	$\nu&		$\omega_{\text{large}}=3.1$.	\nu For straight cells as they move in natural fluid (rounded and linear cells could be observed). Rounded cells moved with $U=12.7\pm6$. Cells in methyl cellulose tended to be linear and move at $U=35.5\pm10.3$.	&	\cite{L185,L187,L188}	\\
\hline
    \end{tabular}
\end{sidewaystable}

\begin{sidewaystable}[H]
    \scriptsize
    \begin{tabular}{|p{2.15cm}|p{2cm}p{2cm}p{5cm}|p{2.15cm}p{5cm}|p{2.1cm}|}
    \multicolumn{7}{c}{\normalsize{\textbf{Table \ref{tab:sperm} continued:} Data for spermatozoa}}\\
    \hline
       \textbf{Species} &\multicolumn{3}{c|}{\centering \textbf{Geometry}}&\multicolumn{2}{c|}{\centering\textbf{Kinematics}}&\textbf{References} \\[3pt]
       & \centering$B$ &\centering $W$ & Flagella & \centering$U$ & Notes & \\

    \hline	\parbox[t]{2.15cm}{\textit{Oncorhynchus}\\{\it mykiss} (trout)}	&\centering$		$&\centering$		$&							&\centering$	220	$\nu&		$\omega=55$.	\nu Activity lasted 30 s.	&	\cite{L222}	\\

    \hline	\textit{Ostrea} (oyster)	&\centering$	2.6	$&\centering$	2.8	$&		\parbox[t]{5cm}{$L=47$,		$\lambda=25.6$,	$h=4.7$.	2D and 3D\\ beat.}	&\centering$	(163.8-169)	$&		$\omega=43$.	At 23\dc.	&	\cite{70,L194,BandW}	\\

    \hline	\textit{Ovis} (ram)	&\centering$	10.6	$&\centering$	6.2	$&		\parbox[t]{5cm}{$L=59$,		$\lambda=36.5$,	$h=7.3$.	2D and 3D\\ beat.}	&\centering$	(132.3-136)	$&		$\omega=29$.	At 35.5\dc.	&	\cite{70,L194,BandW}	\\

    \hline	\parbox[t]{2.15cm}{\textit{Periplaneta}\\{\it americana}\\ (cockroach)}	&\centering$	14.85	$\f\nu&\centering$	0.95	$\f&		$L=57.75$\f,			$h=6.43$\f		&\centering\parbox[t]{2.15cm}{\centering$	(16.1\pm1.22$\nd$-$\\$53.6\pm3.1$\nt$)	$}&		\parbox[t]{5cm}{	\nu The acrosome measured 2.08. \nd At $15-$\\16.6\dc. \nt At $37-$39\dc}	&	\cite{L190,L183,L229}	\\

    \hline	\parbox[t]{2.15cm}{\textit{Polyodon}\\{\it spathula}\\ (paddlefish)}	&\centering$		$&\centering$		$&							&\centering$	175	$&		$\omega=50$.	Activity lasted 50 s.	&	\cite{L222}	\\
\hline\hline
    \end{tabular}
\end{sidewaystable}

\begin{sidewaystable}[H]
    \scriptsize
    \caption{Data for ciliates}
    \label{tab:cilia}
    \begin{tabular}{|p{2.15cm}|p{2cm}p{2cm}p{5cm}|p{2.15cm}p{5cm}|p{2.1cm}|}
    \hline \hline
    \textbf{Species} &\multicolumn{3}{c|}{\centering \textbf{Geometry}}&\multicolumn{2}{c|}{\centering\textbf{Kinematics}}&\textbf{References} \\[3pt]
       & \centering$B$ &\centering $W$ & Cilia & \centering$U$ & Notes & \\

    \hline \textit{Amphileptus gigas}&\centering$	808	$&\centering$	136	$&							&\centering$	608	$&				&	\cite{47}	\\

    \hline \textit{Amphorides quadrilineata}&\centering$	138	$&\centering$	47	$&							&\centering$	490	$&				&	\cite{L144}	\\

    \hline \parbox[t]{2.15cm}{\textit{Balanion}\\{\it comatum}}&\centering$	16	$\nu&\centering$		$& This species has one caudal cilium.							&\centering$	220\pm10	$&			 \nu Equivalent spherical diameter.	&	\cite{L180}	\\

    \hline \parbox[t]{2.15cm}{\textit{Balantidium}\\{\it entozoon}}&\centering$	(84.5$\f$-106)	$&\centering$	(43.26$\f$-55.6)	$&	Cells can swim either with dexio-symplectic (slow swimming, right handed spiral path, $\lambda_{MW}=3.83$\f) or dexio-antiplectic metachrony (fast swimming, left handed spiral path, $\lambda_{MW}=6.32$\f), $\ell=(3.71-4.72)$. 				&\centering$		$&				&	\cite{149,162}	\\

    \hline \textit{Blepharisma} sp. &\centering$	350	$&\centering$	120	$&	$\ell=7.5$,	$N=7000$ (excluding compound cilia), $\kappa=0.1$.					&\centering$	600	$&	$V=1830000$.			&	\cite{L157,L161,Hansen1997}	\\

    \hline \textit{Cepedea} sp. &\centering$	333	$\f&\centering$	148.5	$\f&	\parbox[t]{5cm}{		Symplectic metachrony,	$\lambda_{MW}=37$,\\ $\ell\approx25$\f.}			&\centering$		$&			Considering $250\times$ magnification.	&	\cite{162}	\\
\hline
    \end{tabular}
\end{sidewaystable}

\begin{sidewaystable}[H]
    \scriptsize
    \begin{tabular}{|p{2.25cm}|p{2cm}p{2cm}p{5cm}|p{2.15cm}p{5cm}|p{2.1cm}|}
    \multicolumn{7}{c}{\normalsize{\textbf{Table \ref{tab:cilia} continued:} Data for ciliates}}\\
    \hline
       \textbf{Species} &\multicolumn{3}{c|}{\centering \textbf{Geometry}}&\multicolumn{2}{c|}{\centering\textbf{Kinematics}}&\textbf{References} \\[3pt]
       & \centering$B$ &\centering $W$ & Cilia & \centering$U$ & Notes & \\

    \hline \textit{Koruga} sp. &\centering$	(300-400)	$&\centering$	(200-300)	$&\parbox[t]{5cm}{	$\ell=(20-30)$,		symplectic metachrony,\\	$\lambda_{MW}=(22-40)$.	}		&\centering$	\mathcal{O}(100)	$&				&	\cite{149}\\

    \hline \textit{Laboea strobila}&\centering$	100	$&\centering$	49	$&							&\centering$	810	$&				&	\cite{L144}	\\

    \hline \parbox[t]{2.1cm}{\textit{Lacrymaria}\\\textit{lagenula}}&\centering$	42	$&\centering$	45	$&							&\centering$	909	$&			At 26\dc.	&	\cite{47}	\\

    \hline \textit{Lembadion bullinum}&\centering$	43	$&\centering$	36	$&							&\centering$	415	$&				&	\cite{47}	\\

    \hline \textit{Lembus velifer}&\centering$	87	$&\centering$	17	$&							&\centering$	200	$&				&	\cite{47}	\\
\hline
    \end{tabular}
\end{sidewaystable}

\begin{sidewaystable}[H]
    \scriptsize
    \begin{tabular}{|p{2.15cm}|p{2cm}p{2cm}p{5cm}|p{2.15cm}p{5cm}|p{2.1cm}|}
    \multicolumn{7}{c}{\normalsize{\textbf{Table \ref{tab:cilia} continued:} Data for ciliates}}\\
    \hline
       \textbf{Species} &\multicolumn{3}{c|}{\centering \textbf{Geometry}}&\multicolumn{2}{c|}{\centering\textbf{Kinematics}}&\textbf{References} \\[3pt]
       & \centering$B$ &\centering $W$ & Cilia & \centering$U$ & Notes & \\

    \hline \textit{Paramecium tetraurelia}&\centering$	124\pm20	$&\centering$	46\pm5	$&							&\centering\parbox[t]{2.15cm}{\centering$	784\pm31$\\$(\max 1376)	$}&	\parbox[t]{5cm}{$f_{\text{cortex}}=35\pm4$, $f_{\text{mouth}}=66\pm8$,	$c_{MW}=$\\$(461-1596)$.	Wild-type cells of stock d4-2 grown at 27\dc\ swimming in 0.2 ml in depression slides maintained at a temperature between 20 and 25\dc. There is data available for some mutants too.}	&	\cite{L200}	\\

    \hline \textit{Paramecium woodruffi}&\centering\parbox[t]{2cm}{\centering$	169$\\ $(98-222)	$}&\centering\parbox[t]{2cm}{\centering$	62$\\ $(42-72)	$}&			Antiplectic metachrony.				&\centering\parbox[t]{2.15cm}{\centering$	2000$\\$ (1250-2777)	$}&			This species could also swim in right-handed spirals with $U=609\,(581-666)$.	&	\cite{48}	\\

    \hline \parbox[t]{2.15cm}{\textit{Porpostoma}\\{\it notatum}}&\centering$	107.7	$\f&\centering$	29	$\f&							&\centering$	(1583.4-2101)	$\f&				&	\cite{L176}	\\

    \hline \textit{Prorodon teres}&\centering$	175	$&\centering$	160	$&							&\centering$	1066	$&				&	\cite{47}	\\

    \hline \textit{Protoopalina} sp. &\centering$	315	$\f&\centering$	92.4	$\f&	\parbox[t]{5cm}{$\ell=15.1$, symplectic metachrony,\\ $\lambda_{MW}=(20.54-27.6)$.}			&\centering$		$&				&	\cite{162}	\\
\hline
    \end{tabular}
\end{sidewaystable}

\begin{sidewaystable}[H]
    \scriptsize
    \begin{tabular}{|p{2.15cm}|p{2cm}p{2cm}p{5cm}|p{2.15cm}p{5cm}|p{2.1cm}|}
    \multicolumn{7}{c}{\normalsize{\textbf{Table \ref{tab:cilia} continued:} Data for ciliates}}\\
    \hline
       \textbf{Species} &\multicolumn{3}{c|}{\centering \textbf{Geometry}}&\multicolumn{2}{c|}{\centering\textbf{Kinematics}}&\textbf{References} \\[3pt]
       & \centering$B$ &\centering $W$ & Cilia & \centering$U$ & Notes & \\

    \hline \textit{Tetrahymena pyriformis}&\centering\parbox[t]{2cm}{\centering$	70$\\ $(55.7-89.77$\f$) 	$}&\centering\parbox[t]{2cm}{\centering$	27.5$\f\\ $(20-45)	$}&	\parbox[t]{5cm}{$\ell=(7-14.35$\f$)$,	$N=500$ (excluding\\ compound cilia) divided in $(17-23)$\\ columns, $\kappa=0.2$, $d=(2.84-6.16)$\f,\\ dexio-antiplectic metachrony (similar to \textit{Paramecium}),	$\lambda_{MW}\approx16.2$.}			&\centering\parbox[t]{2.15cm}{\centering$	480$\\ $(451.2-500)	$}&	$f=20$.			&	\cite{149,162,231,L157,L161,BandW}	\\

    \hline \textit{Tetrahymena thermophila} &	\centering $(46.2-47.1)$\f &	\centering $(28.3-28.8)$\f &	$\ell=5.30\pm0.95$\f. &	\centering $204.5\pm24.2$\g &	$f=15.9\pm3.7$. Values for wild-type cell (CU427.4) &	\cite{M39} \\

    \hline \textit{Tillina magna}&\centering\parbox[t]{2cm}{\centering$	162$\\ $(150-175)	$}&\centering\parbox[t]{2cm}{\centering$	82$\\ $(75-90)	$}&							&\centering$	2000	$&			At 25\dc.	&	\cite{47,149}	\\

    \hline \parbox[t]{2.15cm}{\textit{Tintinnopsis}\\{\it kofoidi}}&\centering$	100	$&\centering$	29	$&							&\centering$	400	$&				&	\cite{L144}	\\

    \hline \textit{Tintinnopsis minuta}&\centering$	40	$&\centering$	26	$&							&\centering$	60	$&				&	\cite{L144}	\\
\hline \hline
    \end{tabular}
\end{sidewaystable}


\end{document}